\documentclass[useAMS,usenatbib]{mnras}
\usepackage{graphicx}
\usepackage{hyperref}
\usepackage{amsfonts,amsmath}
\usepackage{xspace}
\usepackage{gensymb}
\usepackage{subfigure}
\usepackage{color}

\usepackage{times}
\newcommand{\co}{CO}
\newcommand{\cotwoone}{CO(2-1)\xspace}

\newcommand{\kks}{${\rm K\,km\,s^{-1}}$\xspace}

\newcommand{\msol}{M$_{\odot}$\xspace}
\newcommand{\solmass}{M$_{\odot}$\xspace}
\newcommand{\kms}{${\rm km\,s^{-1}}$\xspace}
\newcommand{\htwo}{H$_{2}$\xspace}
\newcommand{\msolpcsq}{M$_{\odot}\,{\rm pc}^{-2}$\xspace}

\newcommand{\hi}{{\sc H\,i}\xspace}
\newcommand{\hib}{{\rm{\sc H\,i}}\xspace}
\newcommand{\hibold}{{\bf H\,{\sc i}}\xspace}
\newcommand{\hii}{{\sc H\,ii}\xspace}

\newcommand{\vpeak}{$v_{\rm peak}$\xspace}
\newcommand{\vcent}{$v_{\rm cent}$\xspace}
\newcommand{\vrot}{$v_{\rm rot}$\xspace}

\newcommand{\diskfit}{{\sc diskfit}}

\graphicspath{{./}{figures/}} % Force your latex program to look into Images folder for pictures (reduce clutter)

\title[Atomic Gas in M33]{Kinematics of the Atomic ISM in M33 on 80 pc scales}

\author[Koch et al.]{Eric W. Koch$^{1}$\thanks{E-mail:
ekoch@ualberta.ca (EWK); rosolowsky@ualberta.ca (EWR)}, Erik W. Rosolowsky$^{1}$, Felix J. Lockman$^{2}$, Amanda A. Kepley$^{3}$, \newauthor Adam Leroy$^{4}$, Andreas Schruba$^{5}$, Jonathan Braine$^{6}$, Julianne Dalcanton$^{7}$, \newauthor
Megan C. Johnson$^{8}$, Sne\v{z}ana Stanimirovi\'{c}$^{9}$\\
$^{1}$University of Alberta, Department of Physics, 4-183 CCIS, Edmonton AB T6G 2E1, Canada \\
$^{2}$Green Bank Observatory, P.O. Box 2, Green Bank, WV 24944, USA\\
$^{3}$National Radio Astronomy Observatory, 520 Edgemont Road, Charlottesville, VA 22903-2475, USA\\
$^{4}$The Ohio State University, Department of Astronomy, 140 West 18th Avenue, Columbus, OH 43210, USA\\
$^{5}$Max-Planck-Institut f\"{u}r extraterrestrische Physik, Giessenbachstra\ss e 1, D-85748 Garching, Germany\\
$^{6}$Laboratoire d'Astrophysique de Bordeaux, Univ. Bordeaux, CNRS, B18N, all\'ee Geoffroy Saint-Hilaire, 33615 Pessac, France\\
$^{7}$Department of Astronomy, Box 351580 University of Washington, Seattle, WA 98195\\
$^{8}$United States Naval Observatory, 3450 Massachusetts Ave NW, Washington, D.C., 20392, USA\\
$^{9}$University of Wisconsin, Department of Astronomy, 475 N Charter St., Madison, WI 53706, USA}

\begin{document}

\date{Draft date: \today}

\pagerange{\pageref{firstpage}--\pageref{lastpage}} \pubyear{2017}

\maketitle

\label{firstpage}

\begin{abstract}
We present new L-band ($1\mbox{--}2$ GHz) observations of the nearby spiral galaxy M33 with 80~pc resolution obtained with the Karl G.~Jansky Very Large Array. The \hi observations, combined with \hi measurements from the Green Bank Telescope, improve the spectral resolution and sensitivity ($2.8$~K rms noise in a $0.2$~\kms channel) compared to previous observations.  We find individual profiles are usually non-Gaussian, harbouring line wings, multiple components, and asymmetries.  Given this spectral complexity, we quantify the motions in the atomic ISM through moment analysis of the spectra and fits to aligned, stacked profiles. The measured value of the \hi line width depends strongly on the method used, with the velocity stacked profiles aligned to the peak velocity giving the minimum value of $\sigma = 7$ \kms and all other methods giving higher values ($\sigma\sim10$ \kms). All measurements of the line width show a shallow radial trend, with $\sigma$ decreasing by $\sim2$ \kms from $R_{\rm gal}=0$ to $R_{\rm gal}=8$ kpc.  We consider a number of energy sources that might maintain the line width against turbulent dissipation, but no single source is adequate.  We find excess emission relative to a Gaussian in the stacked profile line wings, ranging from 9\% to 26\% depending on how the spectra are aligned.  By splitting the line wings into symmetric and asymmetric components, we find that the lagging rotational disk accounts for one-third of the line wing flux. We also find emission far from the rotation-axis of the galaxy in multiple discrete \hi clouds, including a filament with a projected length of $\sim8$ kpc.
\end{abstract}


\section{Introduction}
\label{sec:intro}

\hi 21-cm emission is an ideal tracer of kinematics in the atomic interstellar medium (ISM) because of its ubiquitous distribution that extends beyond the optical extent of galactic disks. On large-scales, \hi emission can be used to study disk kinematics and rotation, while small-scale variations trace the kinematic and turbulent nature of the ISM.

Modeling the kinematics of the atomic ISM remains a significant challenge. The \hi detected within the Milky Way exhibits kinematic features related to the evolution of clouds on small scales. Our limited perspective within the Galaxy makes it difficult to place these small-scale motions in the context of large-scale motions and the galactic potential.  However, since the kinematics on both scales are linked, a consistent understanding of ISM kinematics requires sampling scales from molecular clouds to the entire disk \citep{Dobbs2014prpl.conf....3D}.  The external view of nearby galaxies can overcome these limitations in Galactic observations. However, these observations often lack spatial resolution and high sensitivity, yielding studies of 21-cm emission that blend the small-scale motions seen in Milky Way studies.

This trade-off of resolution and scale between Galactic and extragalactic observations is also important for discerning the multiple thermal states in the atomic medium. While superseded by later work, the \citet{Field1969ApJ...155L.149F} model predicts the presence of a warm neutral medium (WNM, $n\sim1$ cm$^{-3}$, $T\sim1000$ K) and cool neutral medium (CNM, $n\sim10$ cm$^{-3}$, $T\sim 100$ K) over a range of pressure. Subsequent studies of the atomic medium qualitatively support the two-phase picture \citep{Wolfire1995ApJ...443..152W,Wolfire2003ApJ...587..278W}.  However, in observational studies, it remains difficult to separate these two gas phases. The optically-thin WNM is typically measured in \hi emission, while the optically-thick CNM can only be unambiguously separated through absorption studies.

Decomposing \hi spectra into thermal components has best been studied in the Milky Way through absorption towards bright extragalactic sources \citep{Heiles2003ApJ...586.1067H,Murray2015ApJ...804...89M}, or by tracing \hi self-absorption \citep{Gibson2005ApJ...626..195G}.  These studies find that the atomic ISM is found in both the WNM and CNM states, though a non-negligible fraction of components are in an unstable intermediate state \citep{Heiles2003ApJ...586.1067H,Murray2015ApJ...804...89M}. Extragalactic absorption studies toward $\sim 50$ lines-of-sight in Local Group galaxies suggest that the fraction of the atomic medium in the CNM varies between local systems \citep{Dickey1990ApJ...352..522D,Dickey1994A&A...289..357D,Dickey2000ApJ...536..756D}.

Most extragalactic studies focus on interpreting the \hi emission line profiles at the telescope resolution (typically $200\mbox{--}1000$ pc), though different modelling approaches are used.  Fitting individual spectra with a Gaussian is the most straight-forward approach \citep{Boulanger1992A&A...266...37B,Mogotsi2016AJ....151...15M}, however non-Gaussian line features or multiple components require a more sophisticated model, such as a two-Gaussian model \citep{Young1996ApJ...462..203Y,Warren2012ApJ...757...84W}. A simpler approach is to only estimate the line width of a spectrum using the second-moment \citep{Tamburro2009AJ....137.4424T}, though this approach has been found to overestimate the line width in many cases \citep{Mogotsi2016AJ....151...15M}.

To study kinematics in in faint regions, several studies stack velocity-aligned spectra to increase the signal-to-noise (S/N) in the data.  Nearly all studies that utilize this method find a common line shape of a central Gaussian peak with enhanced line wings \citep{Young1996ApJ...462..203Y,Braun1997ApJ...484..637B,Petric2007AJ....134.1952P,Ianjasm2012AJ....144...96I,Stilp2013ApJ...765..136S}.  To characterize this shape, the stacked spectra are either fit with a two-Gaussian model \citep{Young1996ApJ...462..203Y,Ianjasm2012AJ....144...96I} or as a Gaussian peak with a non-parametric measure of the enhanced line wings \citep{Stilp2013ApJ...765..136S,Stilp2013bApJ...773...88S}. The \hi line widths from a single Gaussian range from $6\mbox{--}10$ \kms, while the two-Gaussian models have a narrow component width of 3 to 6 \kms, and 9 to 25 \kms in the wider component.

The results of stacked profile studies, and the models chosen to explain the profile shape, have led to diverging physical interpretations of the atomic medium.  Studies using a two-Gaussian model argue that the narrow and broad components are naturally explained by the CNM and WNM, respectively \citep{Young1996ApJ...462..203Y,Ianjasm2012AJ....144...96I}. On the other hand, \citet{Stilp2013ApJ...765..136S} argues that the central Gaussian peak represents emission from a turbulent mixture of CNM and WNM, and the enhanced line wings result from stellar feedback.  Alternatively, \citet{Braun1997ApJ...484..637B} proposes that the brightest emission arises from a ``high-brightness network'' (HBN) of narrow filamentary structure across the disk.  The HBN then arises from a dense and optically thick CNM component, where the central peak becomes flattened in individual spectra due to the higher optical depth \citep{Braun2009ApJ...695..937B,Braun2012ApJ...749...87B}.

The use of different methods and their disparate interpretations makes it difficult to create a clear connection with Milky Way studies.  Yet, with sufficient sensitivity and resolution, extragalactic \hi observations should recover the small-scale complexity observed in the Milky Way. New observations of nearby galaxies can provide this connection between galactic and extragalactic approaches. Observations of the Local Group provide the best means for determining this connection as they allow for scales of indiviudal molecular clouds to be resolved, similar to studies within the Milky Way and Magellanic Clouds \citep[e.g.,][]{Wong2009ApJ...696..370W,Fukui2009ApJ...705..144F}. In particular, M33 is an ideal target due to its proximity \citep[840 kpc, ][]{Freedman2001ApJ...553...47F}, moderate disk inclination ($55\degree$), and relatively small angular size.  To that end, we have conducted new observations of M33 with the NSF's Karl G. Jansky Very Large Array (VLA) that focus on high sensitivity coupled with a high spectral resolution.

Recent work on M33 has focused on deep, lower resolution \hi observations seeking the origins of \hi halo gas and high-velocity clouds in the M31 group \citep{Lockman2012AJ....144...52L,Keenan2016MNRAS.456..951K}, or the structure of the dark matter halo \citep{Corbelli2014A&A...572A..23C,Kam2017AJ....154...41K}.  There has been comparatively less attention on the nature of the atomic gas in M33's star-forming disk.  The first complete \hi map with the resolution to discern small-scale structure was presented by \citet{Deul1987A&AS...67..509D}, which unveiled the filamentary structure of M33's \hi morphology.  More recently, \citet{Gratier2010A&A...522A...3G} use archival \hi VLA observations from \citet{Thilker2002ASPC..276..370T} to compare the \hi properties around giant molecular clouds (GMCs).  \citet{Druard2014A&A...567A.118D} further perform a stacking analysis of the \hi and \co~to study the relationship of atomic and molecular gas. \citet{Imara:2011el} use an independent reduction of these same data to examine the \hi environments around 45 GMCs in M33.

In this paper, we present new L-Band (1--2 GHz) observations from the VLA of M33 with a focus on high velocity-resolution observations of the 21-cm \hi line. Taken in the VLA's C-configuration, the beam size at the \hi line is $18-20$\arcsec or a physical size of $\sim80$ pc. The VLA's correlator allows for simultaneous observations of the 21-cm \hi line, four OH transitions, several hydrogen radio recombination lines (RRLs), and the polarized radio continuum. Here we focus on the \hi observations to analyze the \hi profile shapes on 80 pc scales and to determine the structure and kinematics of the \hi disk.  We detect no RRL emission (\S\ref{sub:upper_limits_on_rrl_emission}) and find a single OH maser \citep{koch_maser}. The polarized radio continuum will be presented in a future paper.

In \S\ref{sec:observations} we present the VLA observations and the Green Bank Telescope (GBT) \hi data used to provide short-spacing information \citep{Lockman2012AJ....144...52L}.  Aspects of the \hi imaging, signal masking, and upper limits on the hydrogen radio recombination line (RRL) emission are given in \S\ref{sec:imaging_&_calibration}.  We explore the atomic gas properties of M33 in \S\ref{sec:properties_of_the_atomic_medium} and extra-planar \hi structures in \S\ref{sec:extraplanar_velocity_components}. In \S\ref{sec:discussion}, we critically evaluate the meaning of 21-cm \hi line profiles generated in extragalactic observations.  Appendix \ref{app:imaging_approach} provides a detailed description of the imaging and short-spacing combination.

\section{Observations}
\label{sec:observations}

\subsection{VLA} % (fold)
\label{sub:vla}

Using the VLA in C-configuration, we observed a 13-point mosaic covering the disk of M33 out to a radius of 12 kpc (Project 14B-088). The data were taken in 12 tracks, split equally before and after transit, for a total of 52 hours. We used a hexagonal grid of pointings at the \hi frequency to ensure uniform sensitivity across the mosaic.  To remedy a pointing error that omitted one of the mosaic centres, our final observation used a single $\sim 5$ hour pointing on that location, yielding nearly equal integration times and noise properties across the mosaic.  Due to its close proximity to M33, we use 3C48 as the flux, delay, and gain calibrator. 3C138 is observed as the polarization leakage calibrator for the continuum data.

Using the full capabilities of the VLA correlator, our spectral setup covers the entire 1--2 GHz L-band polarized continuum, the \hi line, four OH lines, and six hydrogen radio recombination lines (RRLs), enabling a variety of science cases to be explored. The setup of the line spectral windows (SPWs) are shown in Table \ref{tab:line_spws}. The \hi spectral window has a high spectral resolution of 977 Hz, corresponding to a velocity of 206 m~s$^{-1}$. This high-spectral resolution is required to detect \hi self-absorption (HISA), based on prior Milky Way observations \citep[e.g.,][]{Gibson2005ApJ...626..195G}.

\begin{table}
\centering
\caption{The VLA SPW setup for spectral lines used in the observations.\label{tab:line_spws}}
\begin{tabular}{llll}
Line           & Rest Freq. (GHz) & Channel Width (km/s) & Channels \\\hline
HI             & 1.42                 & 0.206                & 8192           \\
OH(1612)       & 1.612                & 1.45                 & 512            \\
OH(1665)       & 1.665                & 1.41                 & 512            \\
OH(1667)       & 1.667                & 1.40                 & 512            \\
OH(1720)       & 1.720                & 1.36                 & 512            \\
H(172)$\alpha$ & 1.28                 & 1.26                 & 512            \\
H(166)$\alpha$ & 1.42                 & 1.58                 & 512            \\
H(164)$\alpha$ & 1.48                 & 1.83                 & 512            \\
H(158)$\alpha$ & 1.65                 & 1.64                 & 512            \\
H(153)$\alpha$ & 1.82                 & 1.42                 & 512            \\
H(152)$\alpha$ & 1.85                 & 1.29                 & 512            \\\hline
\end{tabular}
\end{table}

\subsection{GBT} % (fold)
\label{sub:gbt}

We reprocessed the high spectral resolution \hi data presented in \citet[][project AGBT09A\_17]{Lockman2012AJ....144...52L} to provide the short and zero spacing information on \hi emission.  The angular resolution of the GBT in the 21-cm line is $9.\arcmin1$ and the spectral resolution is 0.16 \kms.   Spectra were calibrated and corrected for stray radiation as described in \citet{Boothroyd2011}, and a first-order polynomial was fit to emission-free regions of the spectra to remove any residual instrumental baselines.

The data were originally collected as four separate maps with minimal spatial overlap, so we gridded the data into a cube using a Gaussian kernel, rather than the preferred Gaussian-Bessel kernel.  This choice eliminates edge effects in the data at the centre of the galaxy, but lowers the effective angular resolution to $9.\arcmin8$ (see \S\ref{appsub:combination_tests}), which is a linear resolution of $\sim2.3$ kpc at the distance of M33.   We used the {\sc gbtpipe} package\footnote{v0.1.2, \url{https://github.com/low-sky/gbtpipe}} to build the data cube, which performs spectral preprocessing to eliminate bad scans and a kernel based gridding approach developed for on-the-fly data \citep{Mangum2007}.  The noise in the final cube varies with position ranging from 50 mK to 120 mK in a 0.16 km~s$^{-1}$ channel giving a median column density error of $3.1\times 10^{17}\mbox{ cm}^{-2}$ for a 20 \kms FWHM line.
The shortest baselines in our VLA data provide information on scales of $16.\arcmin5$, providing sufficient overlap in $uv$-space with the GBT beam to calculate comparison statistics (Appendix \ref{app:image_combination}).

% subsection gbt (end)

\section{Imaging \& Calibration} % (fold)
\label{sec:imaging_&_calibration}

We calibrated the data using a modified version of the VLA pipeline (version 1.3.0) with CASA 4.2.2 \footnote{\url{https://science.nrao.edu/facilities/vla/data-processing/pipeline/scripted-pipeline}}. Following observatory recommendations, we used a modified version of the pipeline to better handle line spectral windows (SPWs), namely not using Hanning smoothing or automated RFI flagging on the line SPWs to maintain the spectral resolution and avoid flagging of narrow emission features (e.g., OH maser emission). Using 3C48 as the sole calibrator (excluding polarization) gives a high S/N for all calibration scans, and the automated calibration solutions found by the pipeline are excellent. After the initial pipeline run, we manually flagged the data before running the pipeline once more. Most of the line SPWs required little manual flagging beyond the pipeline solutions, though one RRL SPW could not be recovered due to RFI.

Here we present upper limits for RRL emission and detail the imaging and masking approaches adopted for the \hi data.  A single OH(1665) maser is detected and is presented separately \citep{koch_maser}.  We find no detections in the 1612, 1667, or 1720 MHz OH lines.

\subsection{Radio Recombination Lines} % (fold)
\label{sub:upper_limits_on_rrl_emission}

We detect no RRL emission from the six observed transitions. One RRL SPW, the H(172)$\alpha$ line, was dominated by RFI and was unrecoverable. Prior to imaging, we subtracted a constant continuum background level using the CASA {\sc uvcontsub} task by fitting the velocity ranges found to be emission-free in the \hi imaging (\S\ref{sub:hi_imaging}).  RRL emission is expected to be very faint and based on a prior RRL detection towards NGC 604 \citep[][ see below]{Araya2004ApJS..154..541A}, we do not expect to detect any signal in a single line. To lower the noise level, we average the individual line cubes together\footnote{\url{https://casaguides.nrao.edu/index.php?title=Stacking_Multiple_Spectral_Lines_at_Same_Position}}. To use this process, we generated data cubes for each line at a common spectral resolution ($\sim10$ \kms) and corrected these by the primary beam pattern of the mosaic. Each line was then convolved to a common spatial resolution, set by the lowest resolution line, H(166)$\alpha$. Because the mosaic pointings were set based on the primary beam size at the \hi frequency, the primary beam pattern and sensitivity differs between the RRLs. However, based on the the peak H$\alpha$ flux from the giant \hii region NGC 604, we place limits where the RRL intensity is expected to be brightest.  Because of the pointing issue mentioned in \S\ref{sub:vla}, the region around NGC 604 also has the best primary beam coverage for all RRLs observed. Using these regions, we set a $3\sigma$ upper limit on the RRL intensity at 3.0 mJy within a 60\arcsec\ region that encompasses the optical extent of the \hii region.  This upper limit in agreement with observations by \citet{Araya2004ApJS..154..541A} with Arecibo in C-band, where they detected the H$110\alpha$ line in NGC 604 with a peak line flux of $1.36\pm0.19$ mJy in a 58\arcsec\, beam.

% subsection upper_limits_on_rrl_emission (end)

\subsection{\hib Imaging} % (fold)
\label{sub:hi_imaging}

We subtract a constant continuum component from the \hi data using the CASA task {\sc uvcontsub}.  The continuum level is fit based on identified emission-free channels in both the VLA and GBT data.  The VLA data are inspected in the $uv$-plane to determine emission-free channels.  We then use the GBT data to determine the velocities affected by Milky Way \hi emission. The large bandwidth covered by the \hi line gives an ample number of \hi-free lines, and the fitted background level is well-constrained.

Examining the extent of Milky Way \hi emission in the GBT data is imperative for the red-shifted \hi emission at the southern-most tip of M33, as they become spatially coincident near $v_{\rm LSRK}\sim -71~$\kms. This is indicated by the shaded region in Figure \ref{fig:total_rot_profiles}. This overlap region is clearly shown in the \hi Arecibo data from \citet[][see their Figure 3]{Putman2009ApJ...703.1486P}.  Based on this, we begin imaging the VLA data at a velocity of $-73.0~$\kms to best avoid Milky Way emission.  We estimate that excluding this spectral region removes $\sim50$ Jy \kms of flux from M33,  assuming that the missing emission is similar to the blue-shifted side.  This flux corresponds to $0.5\%$ of the total emission from M33.

The resulting data cube is large with 1178 velocity channels, each of which has a grid size of $2560^2$ pixels. This presents significant computational barriers for imaging; our imaging process is described in Appendix \ref{app:imaging_approach}. Prior to imaging, we determine optimal CLEAN settings by running numerous combinations on a single channel, identifying the imaging parameters that yield the lowest peak residual without the algorithm diverging.  Only natural weighting is tested as we prioritize maximizing sensitivity to extended structure. A CLEAN mask is defined for each channel based on the $3\sigma$ limit in the GBT data, which covers nearly all of the emission in the VLA data.   Some channels with emission near the mask edge used an expanded mask to ensure all emission was included.

We use the multi-scale CLEAN algorithm \citep{Cornwell2008ISTSP...2..793C} for deconvolution with six scales, ranging from a point response to a quarter the largest recoverable scale ($\sim970\arcsec$). Changing the range and the specific scales did not have a significant effect on the resulting image. Based on our single channel tests, we deconvolve each channel until reaching 3.8 mJy~beam$^{-1}$ (7.1 K). This limit is $2.5\sigma$ times the noise level in the final cube.

We then use the GBT data to provide short-spacing information for the deconvolved VLA cube by feathering the VLA and GBT data cubes together with the {\sc uvcombine} package\footnote{\url{https://github.com/radio-astro-tools/uvcombine}}. A detailed explanation of the combination is provided in Appendix \ref{app:image_combination}.  The resulting cube fully recovers the total emission in the GBT data.

The final VLA \hi cubes, with and without short-spacing data, have a $1\sigma$ sensitivity of 2.8 K per 0.2 \kms channel and a beam size of $19\arcsec \times 17\arcsec$.  This corresponds to a hydrogen column density of $1.0\times10^{18}$ cm$^{-2}$, assuming optically thin emission.  The mosaic is cut-off at a primary beam coverage of 0.5 to avoid noise-dominated regions near the edge.

% subsection hi_imaging (end)

\subsection{Signal Masking} % (fold)
\label{sub:signal_masking}

We define a multi-step process for creating a reliable signal mask enabled by the high spectral resolution of our data.  This masking is critical for our analysis to ensure that moment-based estimations are not influenced by noise.  When using the VLA-only data, the masking also removes the influence of negative bowls in the data from missing short-spacing information.

Assuming a typical Gaussian line width of $\sigma=6$ \kms for the broad \hi emission across the disk (see \S\ref{sub:stacking_spectra}), each spectral feature will have a FWHM that spans about 70 channels. For the purpose of characterizing emission features alone (not HISA), we can substantially smooth in the spectral dimension to highlight low surface brightness features.

To determine the mask, we first spectrally smooth the data using a median filter with a width of 31 channels ($\sim6$ km~s$^{-1}$). The noise in each smoothed spectrum is found using the median absolute deviation (MAD), iteratively rejecting points beyond $2\sigma_{\rm MAD}$ until convergence. We then search for valid spectral components by requiring that each component has a maximum intensity above $5\sigma_{\rm MAD}$ and has 30 consecutive channels above $2\sigma_{\rm MAD}$. To ensure low surface-brightness line wings are included, the edges of the valid components are extended until the intensity reaches $1\sigma_{\rm MAD}$. This procedure is performed on all spectra within the cube.

Next, we consider the spatial connectivity. To remove spurious spectral components found in the first step, we use the morphological opening and closing operators with a top-hat kernel equivalent to the FWHM of the beam \citep{mathematical-morphology}. The opening operator will erode the mask edges, removing all features smaller than the beam kernel. The closing operator then dilates the remaining components in the mask to restore their original area. The combination of these two masking steps yields a robust signal mask [$\mathcal{M}(x,y,v)$], consistent in both the spectral and spatial dimensions.

This procedure is essential for high spectral-resolution data. Throughout the map, we find many lines-of-sight with either multiple, resolved components, or significantly skewed profiles. Not recovering the line wings in the masking procedure may hide these asymmetries at low-surface brightness. This also ensures negative bowls in the VLA-only cube are rejected from the signal mask and will not bias measurements of the asymmetry of a spectrum.

\begin{figure}
\includegraphics[width=0.5\textwidth]{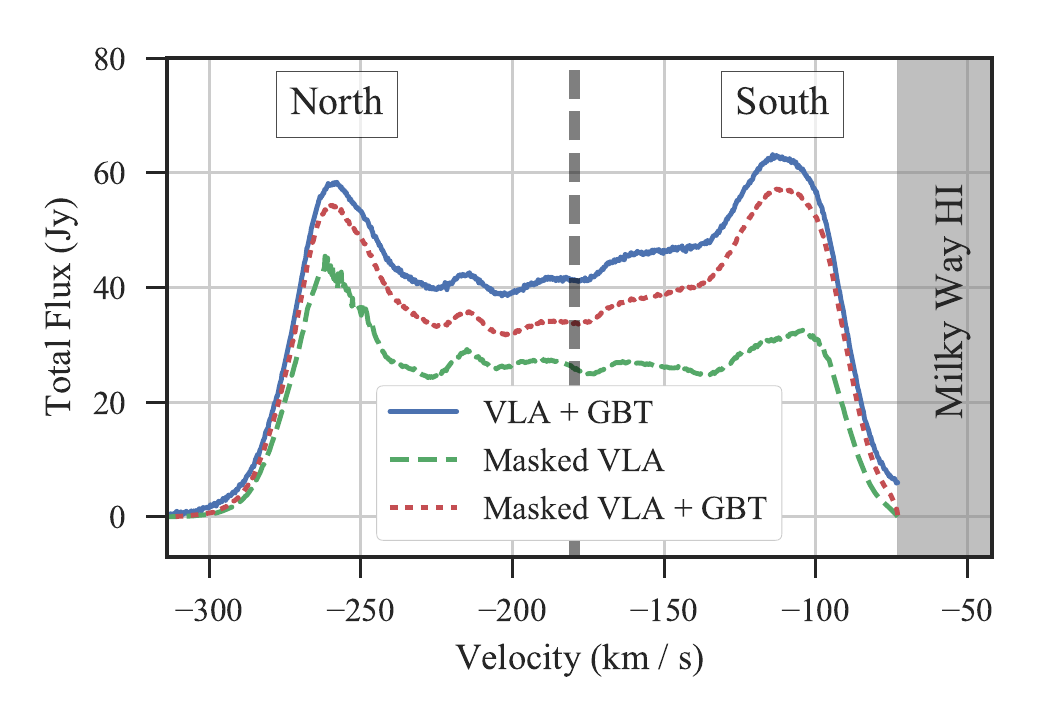}
\caption{\label{fig:total_rot_profiles} Total emission profiles of the VLA-only and the combined VLA+GBT data before and after masking. The curve for the unmasked combined data set is indistinguishable from an umasked GBT-only spectrum. The thick dashed vertical line indicates the systemic velocity of M33 at $-179.2$ \kms. The flux density in the masked VLA map is well recovered by the VLA data in the northern half of M33, but the southern half lacks a significant portion of the emission without short-spacing information.}
\end{figure}

In Figure \ref{fig:total_rot_profiles}, we compare the total GBT emission measured within the VLA mosaic to the emission retained after applying the masking procedure. For the VLA-only data, $60\%$ of the total \hi emission is contained within the mask. The combined VLA+GBT data contains $87\%$ of the total emission in the mask. The missing flux density in the masked version arises from low surface-brightness features that do not satisfy the masking criteria. We discuss these low surface-brightness features in \S\ref{sec:extraplanar_velocity_components}.  The ratio of flux density retained in the mask is roughly constant across all channels for the combined VLA+GBT data, indicating that only low surface brightness \hi is excluded by the masking procedure.

% subsection signal_masking (end)

% section imaging_&_calibration (end)

\section{Properties of the Atomic Medium} % (fold)
\label{sec:properties_of_the_atomic_medium}

In this section, we examine the properties of the \hi emission derived from the new data in the context of previous \hi studies \citep{Corbelli2014A&A...572A..23C,Druard2014A&A...567A.118D,Kam2017AJ....154...41K}.

We calculate an atomic mass of $9\pm2\times10^8$ \msol for the VLA data and $1.3\pm0.3\times10^9$ \msol for the combined data, including a factor of $1.4$ for He and heavier elements.  When including regions outside of the VLA mosaic, the GBT data give a total atomic mass of $1.8\pm0.2\times10^9$ \msol, consistent with other independent \hi mass measurements \citep[e.g., ][]{Putman2009ApJ...703.1486P}.

\begin{figure*}
\includegraphics[width=\textwidth]{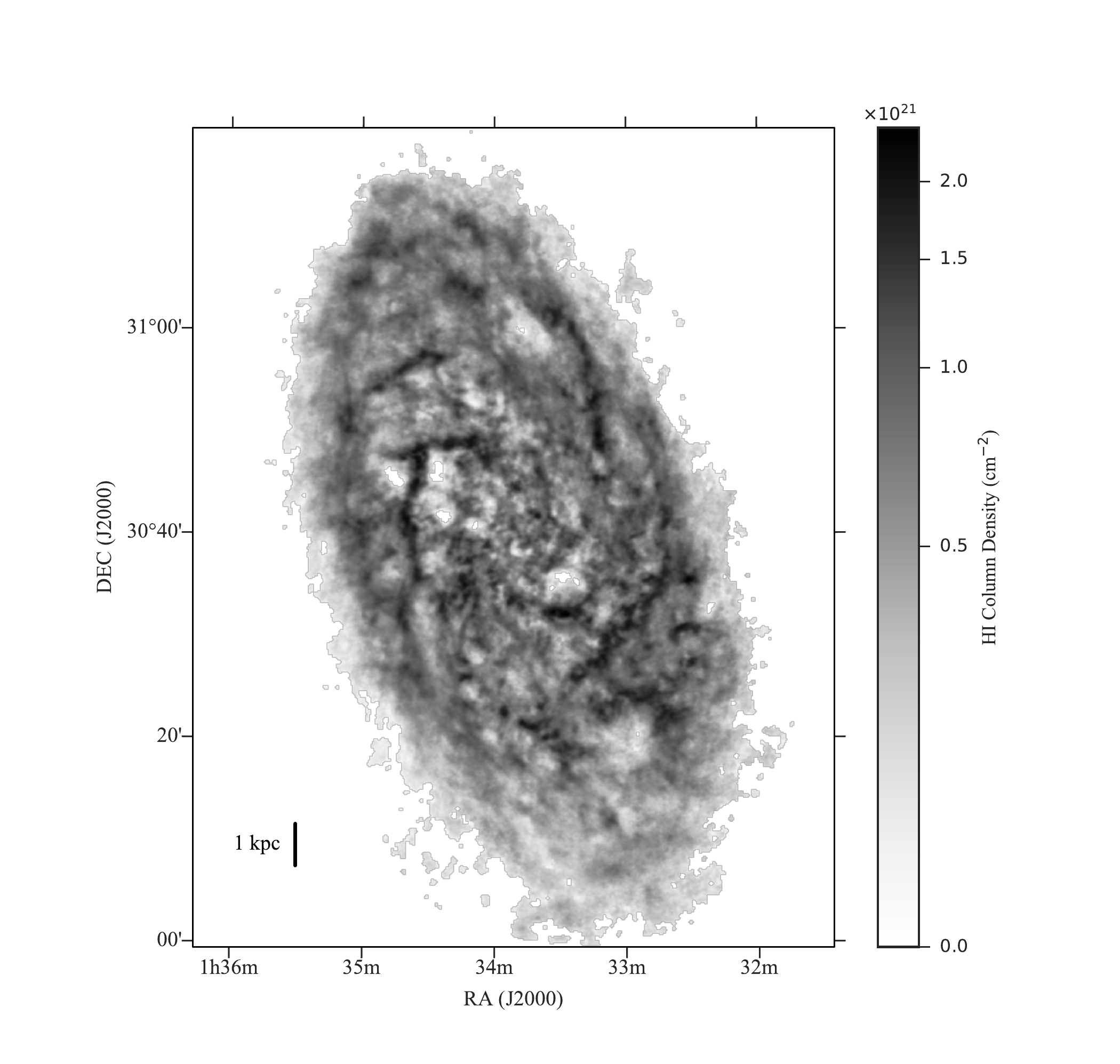}
\caption{\label{fig:zeroth_moment} \hi column density map of the VLA+GBT mosaic, assuming optically thin emission. The map is masked using the technique described in \S\ref{sub:signal_masking}, and shown using an arcsinh stretch.  The mosaic recovers emission well out to a galactic radius of 10 kpc.}
\end{figure*}

\begin{figure*}
\includegraphics[width=\textwidth]{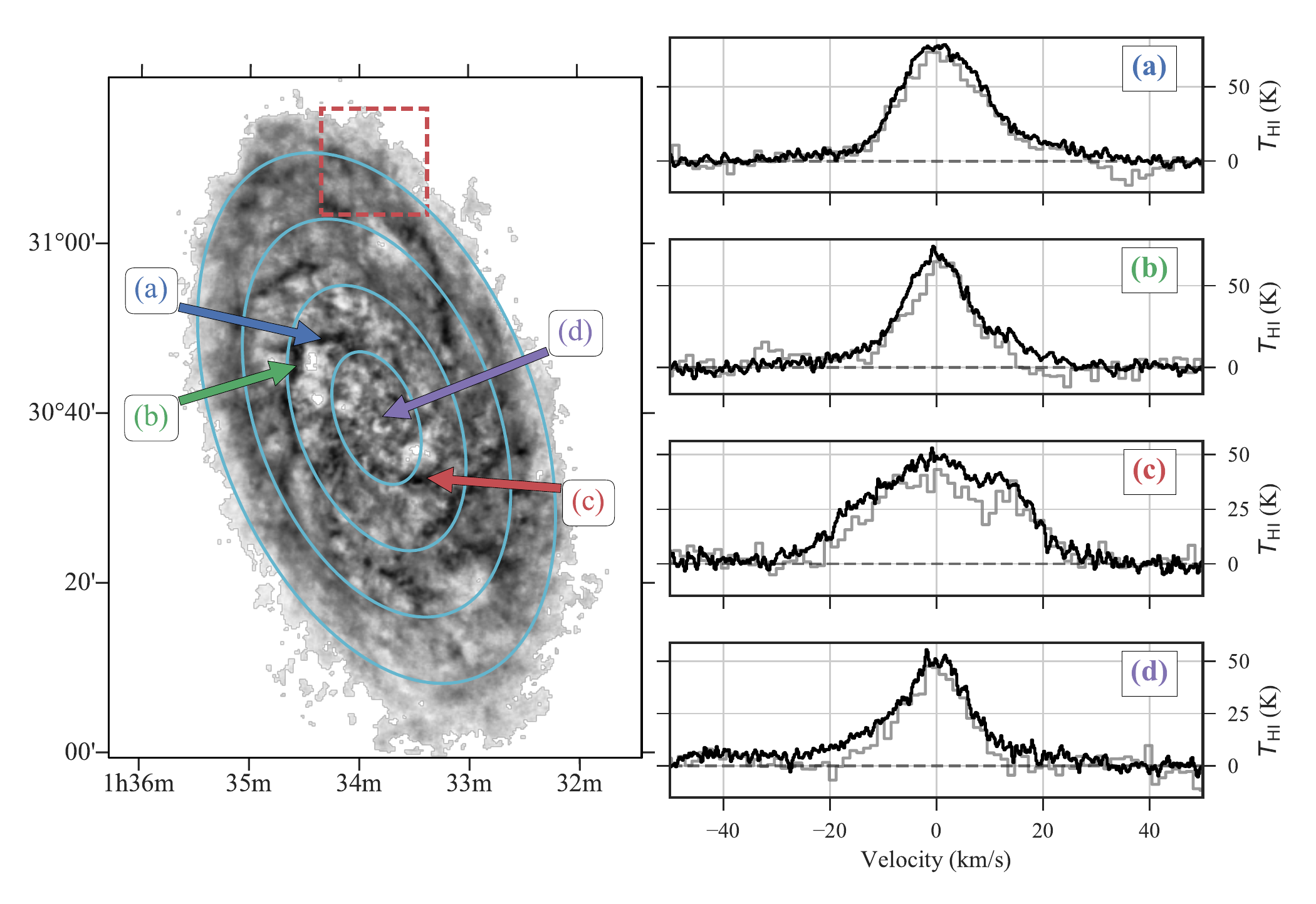}
\caption{\label{fig:hi_spectra} Example \hi spectra compared between the archival VLA (gray) and the new VLA+GBT \hi data (black).  The spectra are centred at the velocity of peak intensity, calculated by smoothing with a $2$ \kms Gaussian kernel.  The location of the spectra are indicated by the arrows on the column density map in the left panel. The new data highlight the complex line shapes near the peaks and demonstrate that signal seen in the extended line wings is real.  The red-dashed box on the left panel shows the position where a high-velocity cloud is found (\S\ref{sub:an_hvc_impacting_the_main_disk_}) at a velocity of -30 \kms from the galaxy's rotation. The cyan contours on the map indicate galactocentric radii of 2, 4, 6 and 8 kpc, respectively, in the plane of the galaxy.}
\end{figure*}

Figure \ref{fig:zeroth_moment} shows the column density map of the VLA data, highlighting M33's flocculent structure. The inner region of the disk ($R_{\rm gal}<2$ kpc; see Figure \ref{fig:hi_spectra} for context) is dominated by small-scale \hi shells and lacks large ($\sim \ $kpc) scale structures.  The ``mid-disk'' ($R_{\rm gal}=2\mbox{--}4$ kpc) is dominated by spiral arms prominent in the optical \citep[e.g., Figure 2 in ][]{Corbelli2014A&A...572A..23C}. The \hi shows similar spiral arm structure in the Northern half, while the Southern arm is dominated by one of the brightest \hi clumps in the galaxy \citep{Rosolowsky2007ApJ...661..830R}.  The \hi in the outer disk ($R_{\rm gal}>4$ kpc) is nearly ubiquitous, with bright spiral-arm segments throughout. The outer edge of our signal-masked \hi map is near the radius where the warp in M33's disk becomes significant \citep{Corbelli2014A&A...572A..23C}.

Figure \ref{fig:hi_spectra} shows four spectra extracted from different positions in the \hi data cube. Each of the spectra shows a non-Gaussian profile with either multiple velocity components, extended line wings, or both. To describe these features, we use moment-based descriptions of the line profiles and velocity-aligned spectral stacking.  Given the lines are clearly non-Gaussian, direct Gaussian fits to the lines do not produce a complete description of the emission.

We note that all uses of the line width in this paper are defined as the Gaussian standard deviation, not the full-width-half-max (FWHM).

\subsection{Rotation Curve and Disk Parameters} % (fold)
\label{sub:rotation_curve}

\begin{figure*}
\includegraphics[width=\textwidth]{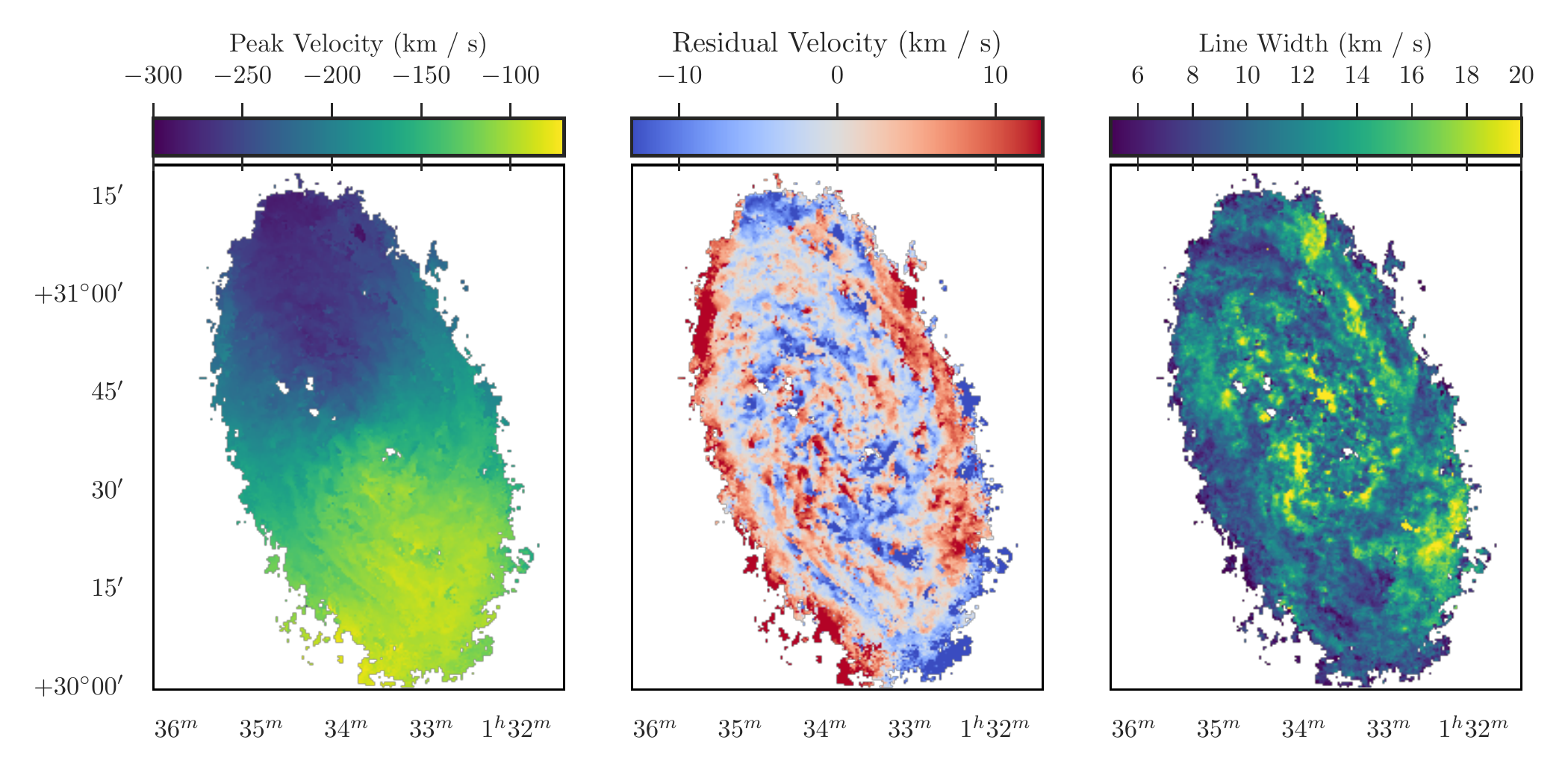}
\caption{Left: \hi \vpeak surface used for fitting the rotation curve. The masked regions from the signal mask (\S\ref{sub:signal_masking}) are shown in white.  Centre: Residual velocities from subtracting the {\sc diskfit} rotation model from the peak velocity surface. Right: The line width map derived from the second moment. \label{fig:peakvel_residual_vel_surface}}
\end{figure*}

Using the masked cube from \S\ref{sub:signal_masking}, we calculate the velocity at the peak intensity, shown in the left panel of Figure \ref{fig:peakvel_residual_vel_surface}, to derive a rotation model.  We define this velocity at peak intensity as $v_{\rm peak} \equiv {\operatorname{argmax}}_v \ I(v)$, which is commonly referred to as the `peak velocity'. To find \vpeak, we smooth the data with a $1$ \kms ($\sim5$ channel) Gaussian kernel to minimize noise before identifying the velocity of the peak.  In Appendix \ref{app:which_velocity_is_the_correct_rotational_velocity_}, we discuss the difference in using \vpeak versus the centroid (first moment) velocity, \vcent, in deriving the rotation curve.  Briefly, \vcent becomes biased by asymmetric line wings (\S\ref{sub:exploring_spectral_complexities}); whereas \vpeak is not biased by the line shape and should provide a more accurate representation of the rotation velocity.

\begin{figure}
\includegraphics[width=0.5\textwidth]{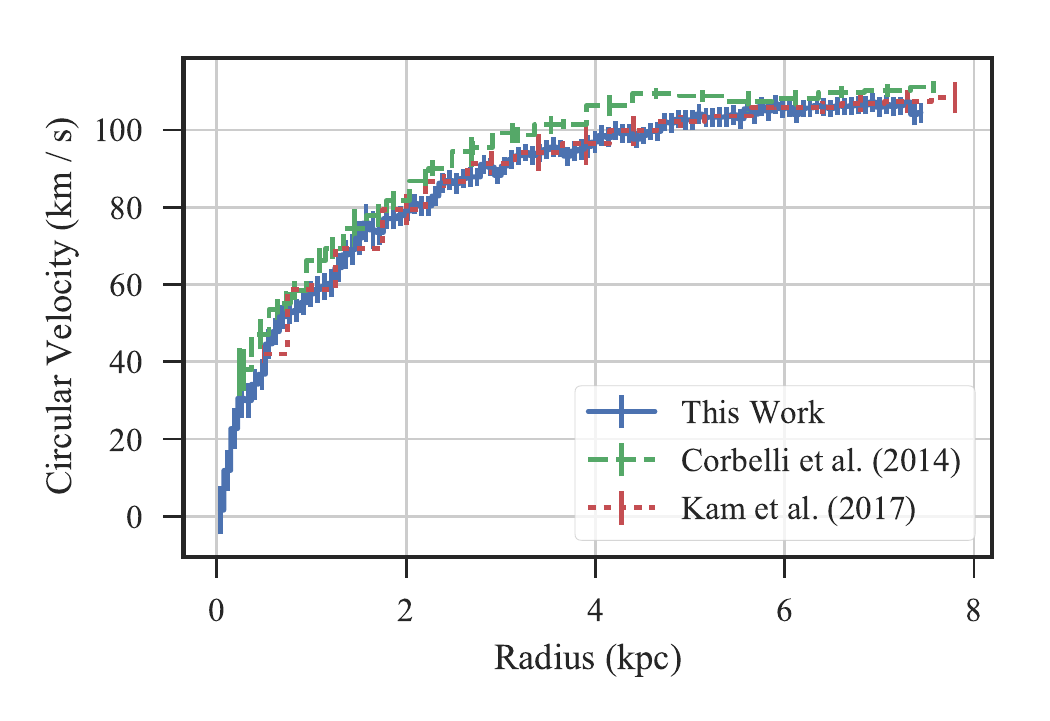}
\caption{\label{fig:rotation_curve} Rotation curve fit by {\sc diskfit} from the VLA+GBT \vpeak surface (blue solid). The rotation curves from \citet[][ green-dashed]{Corbelli2014A&A...572A..23C} and \citet[][ red-dotted]{Kam2017AJ....154...41K} are also shown. Our model is consistent with the \citet{Kam2017AJ....154...41K} model, but has a lower velocity than the \citet{Corbelli2014A&A...572A..23C} curve in the inner 6 kpc.}
\end{figure}

We fit the velocity surface using \diskfit\ \citep{Spekkens2007ApJ...664..204S,Sellwood2015} with a circular velocity model.  \diskfit\ derives a global rotation model by simultaneously fitting the whole velocity surface over a given set of radial bins.  This provides better constraints than minimizing a set of individual rings, but requires that the disk be characterized by a common set of disk parameters (e.g., position angle, inclination).  The rotation curve is shown in Figure \ref{fig:rotation_curve}, and the fit parameters and statistics are given in Table \ref{tab:gal_params}. The rotation velocities in each bin are given in Table \ref{tab:rotvels} in the Appendix.  The parameter uncertainties are calculated from 200 bootstrap iterations in \diskfit\ \citep{Sellwood2010MNRAS.404.1733S}. The $\chi^2$ calculation in \diskfit\ includes an ISM line width parameter for the expected model dispersion, which we set to 8 \kms (see \S\ref{sub:radial_profiles}).  However, since this is a constant factor applied to every position in the fit, it will not affect the resulting disk parameters \citep{2012MNRAS.427.2523K}.  Our rotation model does not include a warp component since a warp only becomes prominent at galactocentric radii beyond $\sim 8$ kpc \citep{Corbelli2014A&A...572A..23C} and we do not fit for a bar component.  We use radial bins with a constant width of 100 pc in the plane of the galaxy, just larger than the VLA beam size, out to a galactocentric radius of 7.5 kpc. Because the radial bins are larger than the beam, we do not include a correction for beam smearing.  Beyond this radius, the uncertainty in the derived moments increases substantially as the integrated intensity decreases. The fitted position of the galactic centre is $7\arcsec$ from the 2MASS position \citep{2MASS}, within the $1\sigma$ parameter uncertainties.

\begin{table}
\caption{\label{tab:gal_params} Galactic disk parameters and fit statistics from {\sc diskfit} for the VLA+GBT peak velocity surface. Errors are the $1\sigma$ intervals based on the {\sc diskfit} bootstrapping. The $n$, $v_{\rm max}$, and $R_{\rm max}$ parameters are the fit parameters to the \citet{Brandt1960ApJ...131..293B} model (Equation \ref{eq:brandt}).}
\centering
\begin{tabular}{lr}
\hline
Points used &  39970 \\
Degrees of Freedom (DOF) for error       &  39663 \\
$\chi^2$/DOF    &      1.44 \\
Iterations  &      4 \\
Centre R.A.      &     $23.4607 \pm 0.0042$\degree \\
Centre Dec.     &     $30.6583 \pm 0.0032$\degree \\
Position Angle          &    $201.12 \pm 0.47$\degree \\
Inclination       &     $55.08 \pm 1.56$\degree\\
$v_{\mathrm{sys}}$ (LSRK)      &   $-179.18 \pm 0.76$ \kms\\\hline
$n$              &     $0.56 \pm 0.04$\\
$v_{\rm max}$     &   $110.0 \pm 1.5$ \kms\\
$r_{\rm max}$     &   $12.0 \pm 1.3$ kpc \\\hline
\end{tabular}
\end{table}

Figure \ref{fig:rotation_curve} shows the circular rotation velocities, \vrot, from our rotation model, along with the recent \hi models from \citet{Corbelli2014A&A...572A..23C} and \citet{Kam2017AJ....154...41K}. There is excellent agreement with the \citet{Kam2017AJ....154...41K} results, though our model has rotation velocities below the \citet{Corbelli2014A&A...572A..23C} rotation curve up to 6 kpc.  The difference may be due to how \citet{Corbelli2014A&A...572A..23C} apply finite disk corrections (see their Appendix B), but since both the data and modelling approaches differ for the three rotation curves, it is difficult to be sure of the cause.

The residual velocity surface from the difference between \vpeak and \vrot is shown in Figure \ref{fig:peakvel_residual_vel_surface}. The residual velocity surface shows only small-scale variations, suggesting that additional velocity components in the rotation model are not needed.  However, the outer edges of the mosaic show what is likely the beginning of the warp. Large deviations ($\sim20$ \kms) both above and below the model values are evident along the edges, similar to the deviations found by \citet{Kam2017AJ....154...41K}.

To remove bin-by-bin variations in the fitted rotation curve we follow the prescription used by \citet{Meidt2008ApJ...688..224M} by fitting a \citet{Brandt1960ApJ...131..293B} rotation curve:
\begin{equation}
  \label{eq:brandt}
  v_{\rm rot} (r) = \frac{v_{\rm max} (r / r_{\rm max})}{\left[ 1/3 + 2/3 (r / r_{\rm max})^n \right]^{3/2n}}.
\end{equation}
The fit parameters for the Brandt model are given in Table \ref{tab:gal_params}.  Using the analytical approximation, we create a smooth version of the rotation surface for use as a model. This smooth rotation model surface is used for creating rotation-subtracted versions of the \hi cube.

% subsection rotation_curve (end)

\subsection{Surface Density Profiles} % (fold)
\label{sub:radial_profiles}

\begin{figure}
\includegraphics[width=0.5\textwidth]{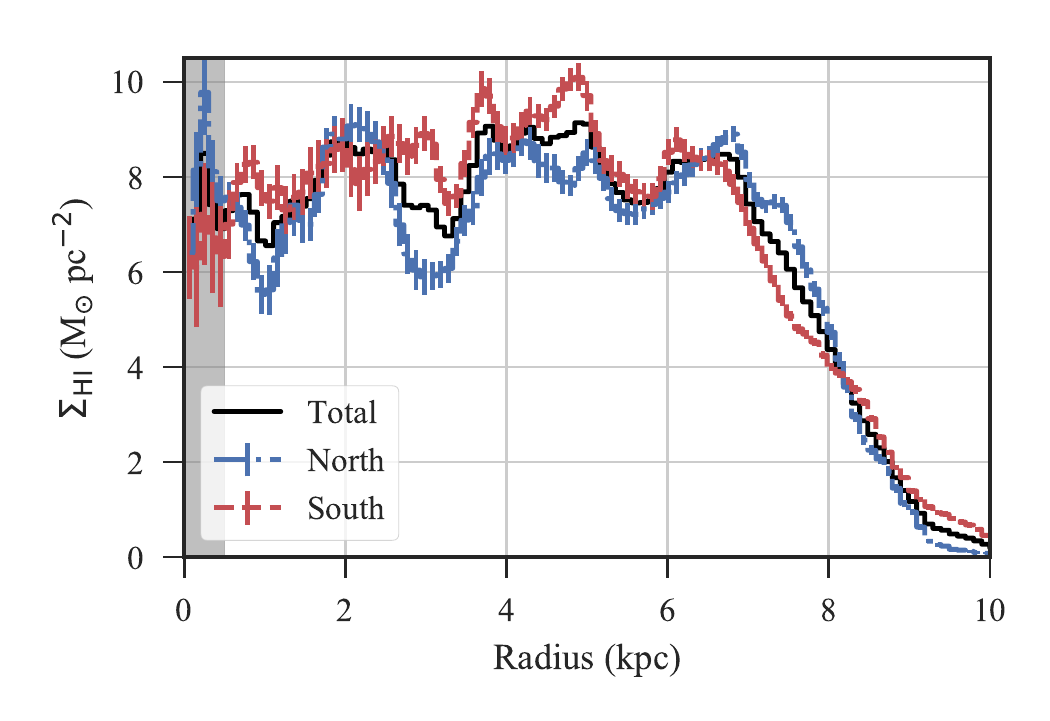}
\caption{\label{fig:surfdens_n_s} The combined GBT and VLA atomic gas surface density profile (black) split into the northern (red dot-dashed) and southern (blue dashed) halves in 100 pc bins and corrected for the disk inclination. The surface density has an average value of $8$ \msolpcsq in the inner 7 kpc, with substantial variation from galactic structure. The error bars are shown for the northern and southern halves; the errors for the entire disk (not shown) are $\sqrt{2}$ smaller. These uncertainties are the standard deviations in each radial bin, corrected by the number of beams (i.e., the independent samples) within the annulus \citep[e.g., ][]{Druard2014A&A...567A.118D}. The shaded region indicates the inner 0.5 kpc where beam smearing and a small number of samples in each bin gives large uncertainties.}
\end{figure}

Using the galactic disk parameters in Table \ref{tab:gal_params}, we create a radial profile of the \hi surface density, corrected for the disk inclination, in 100 pc radial bins out to 10 kpc. Assuming optically thin emission, we use a mass conversion factor of 0.019 \msolpcsq~(\kks)$^{-1}$ to find the \hi surface density, which is the standard \hi column density conversion factor with a factor of 1.4 for the mass of heavier elements.  Figure \ref{fig:surfdens_n_s} shows the radial surface density profile of the combined VLA and GBT data.  The profiles presented here use smaller bin widths than those in \citet{Druard2014A&A...567A.118D}, \citet{Corbelli2014A&A...572A..23C} and \citet{Kam2017AJ....154...41K}, but are consistent with all of these previous works.

The profile has an average value consistent with a surface density of $\sim 8\,$ \msolpcsq out to 7 kpc, where the surface density begins to taper off.  The radius where the tapering begins is comparable to where the stellar surface density equals the gas surface density, based on Figure 10 in \citet{Corbelli2014A&A...572A..23C}. Within this radius, the average stellar surface density exceeds the gas component.

When averaged over galactocentric radii of $<7\,$ kpc, the atomic gas surface density is nearly constant, consistent with the surface density profile from \citet{Kam2017AJ....154...41K}. However, averaging over smaller radii highlights the deviations driven by the large-scale disk structure.  When averaging with the VLA-only data, the average value is $\sim 6\,$ \msolpcsq for the inner 7 kpc, roughly consistent with the fraction of emission recovered without the GBT data.

The profile shows two significant peaks located around 2 and 4.5 kpc; these occur at radii that enclose the prominent spiral arm structure. Within $\sim2$ kpc, the disk lacks spiral structure and is dominated by wind-driven shells \citep{Rosolowsky2007ApJ...661..830R}, making it difficult to connect these large-scale averages to the morphology of the emission. The outer disk also lacks significant coherent large-scale structure, with the emission dominated by multiple spiral arm fragments.

To further examine these variations in the surface density profile, we create radial profiles split into the northern and southern halves. The aforementioned peaks are far more prominent in the northern half than the southern. This is even more dramatic for the dip at 3 kpc, which is entirely driven by variations in the northern half. These differences are driven by the asymmetric main spiral arms: the northern arm, in \hi emission, is more prominent and distinct than the southern arm.

The surface densities between the two halves are approximately equal between a radius of 1.5 to 2 kpc and beyond 5.5 kpc.  The region from 1.5 to 2 kpc contains the beginning of the main spiral arms.  The \hi arms begin at $\sim1.8$ kpc, are offset by nearly $180\degree$, and are nearly symmetrical, despite tracing different extents at larger radii.  The other matching region at 5.5 kpc is beyond the extent of the main spiral arms where the disk becomes dominated by flocculent structure.  These matching regions over all angles suggest that this middle region of the disk is dominated by an arm driving mechanism and distinct from the inner and outer regions in the disk. However, the spatial scale of these variations may be different due to the large area averaged over at larger radii.

The outer region of the disk shows less variation between the halves compared to the inner and middle disk.  While this region contains variations in the surface density due to the spiral arm fragments, the area averaged over is also larger and will tend to smooth out variations more than the regions at smaller radii.

% subsection radial_profiles (end)

\subsection{Exploring Spectral Complexities} % (fold)
\label{sub:exploring_spectral_complexities}

With the high spectral resolution in our data, there are a sufficient number of channels to use higher-order spectral moments for describing line profile shapes.  Figure \ref{fig:hi_spectra} demonstrates that simple spectral models (one or two Gaussians) are inappropriate to represent the data at this resolution.  We quantify these features by calculating maps of the skewness and kurtosis of the velocity profiles. These higher-order moments require a larger number of channels than the line width (second moment) to reduce uncertainties, which has limited their use in previous extragalactic \hi studies with coarser resolution. Instead, previous studies have relied on adding equivalent parameters to Gauss-Hermite polynomials \citep[e.g.,][]{Young2003ApJ...592..111Y,2011AJ....141..193O}. However, this analytic form is limited in how well it can explain extended line wings, making the moments-based description better suited for our data.

The skewness is calculated from the third moment of each spectrum, normalized by the line width cubed:
\begin{equation}
S(x,y) = \frac{\sum_{v\in \mathcal{M}(x,y,v)}[v - M_1(x,y)]^3 I(x,y,v)~\delta v}{M_0 \sigma_v^3},
\end{equation}
where $M_0$ is the integrated line intensity (zeroth moment), and $M_1$ is the line centroid (first moment). The line width here is defined from the second moment: $\sigma_v^2 = M_2(x,y)$. A positive skewness indicates a spectrum with a red-shifted tail, while a spectrum with negative skewness has a blue-shifted tail.

Kurtosis is similarly calculated from the fourth moment. We report the {\it kurtosis excess} by subtracting the value of 3 expected for a normal distribution:
\begin{equation}
K(x,y) = \frac{\sum_{v\in \mathcal{M}(x,y,v)}[v - M_1(x,y)]^4 I(x,y,v)~\delta v}{M_0 \sigma_v^4} - 3.
\end{equation}
A spectrum with a positive kurtosis will have strong tails and have a peak sharper than a Gaussian.

\begin{figure*}
\includegraphics[width=\textwidth]{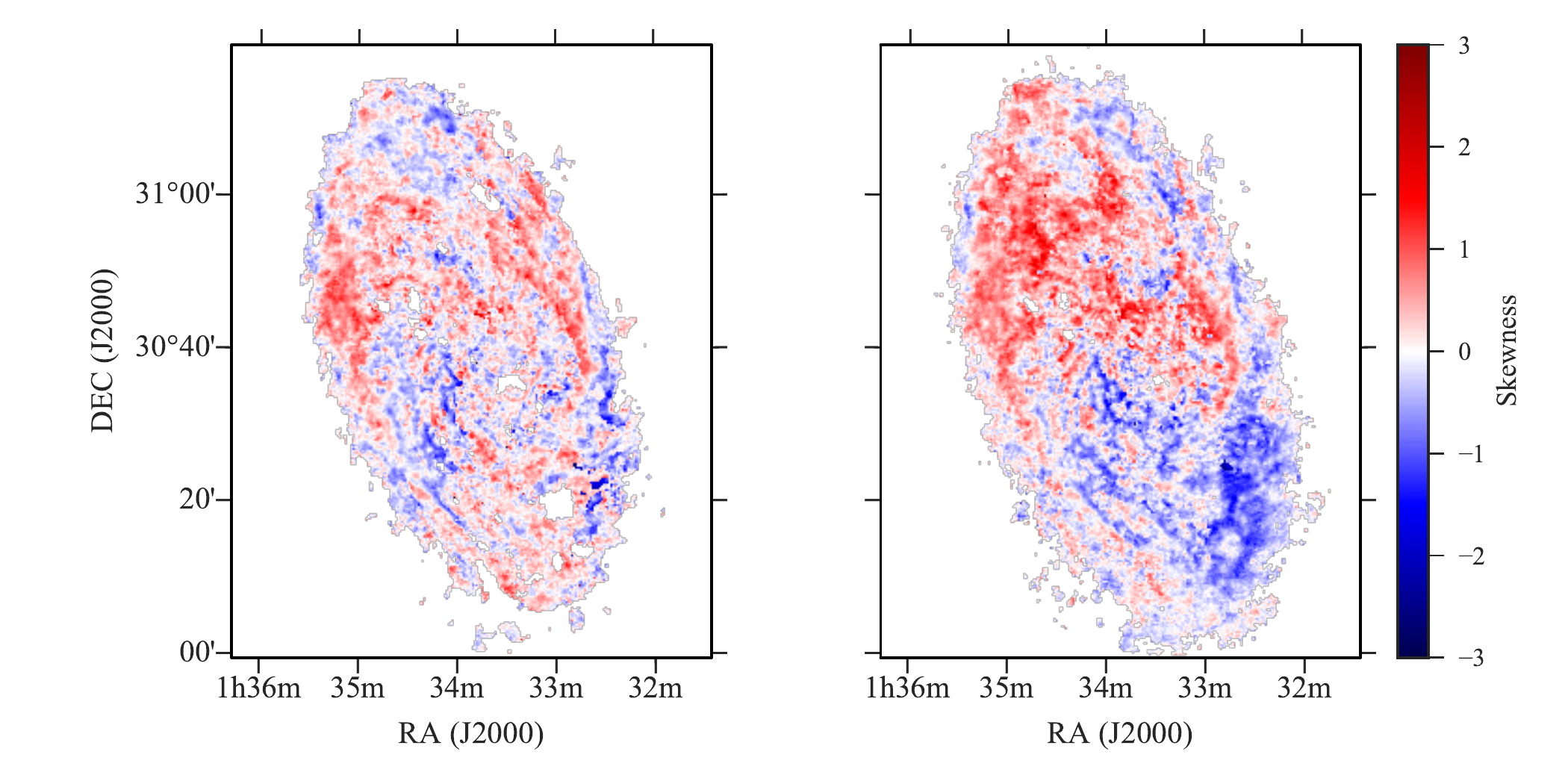}
\caption{\label{fig:skew_maps} The skewness maps of the VLA (left) and combined data (right). Positive skewness indicates a red-shifted tail, while a negative value is a line profile with a blue-shifted tail.  The low surface-brightness component, primarily added to the line wings, from the GBT data drives the large-scale skewness properties.}
\end{figure*}

\begin{figure}
\includegraphics[width=0.5\textwidth]{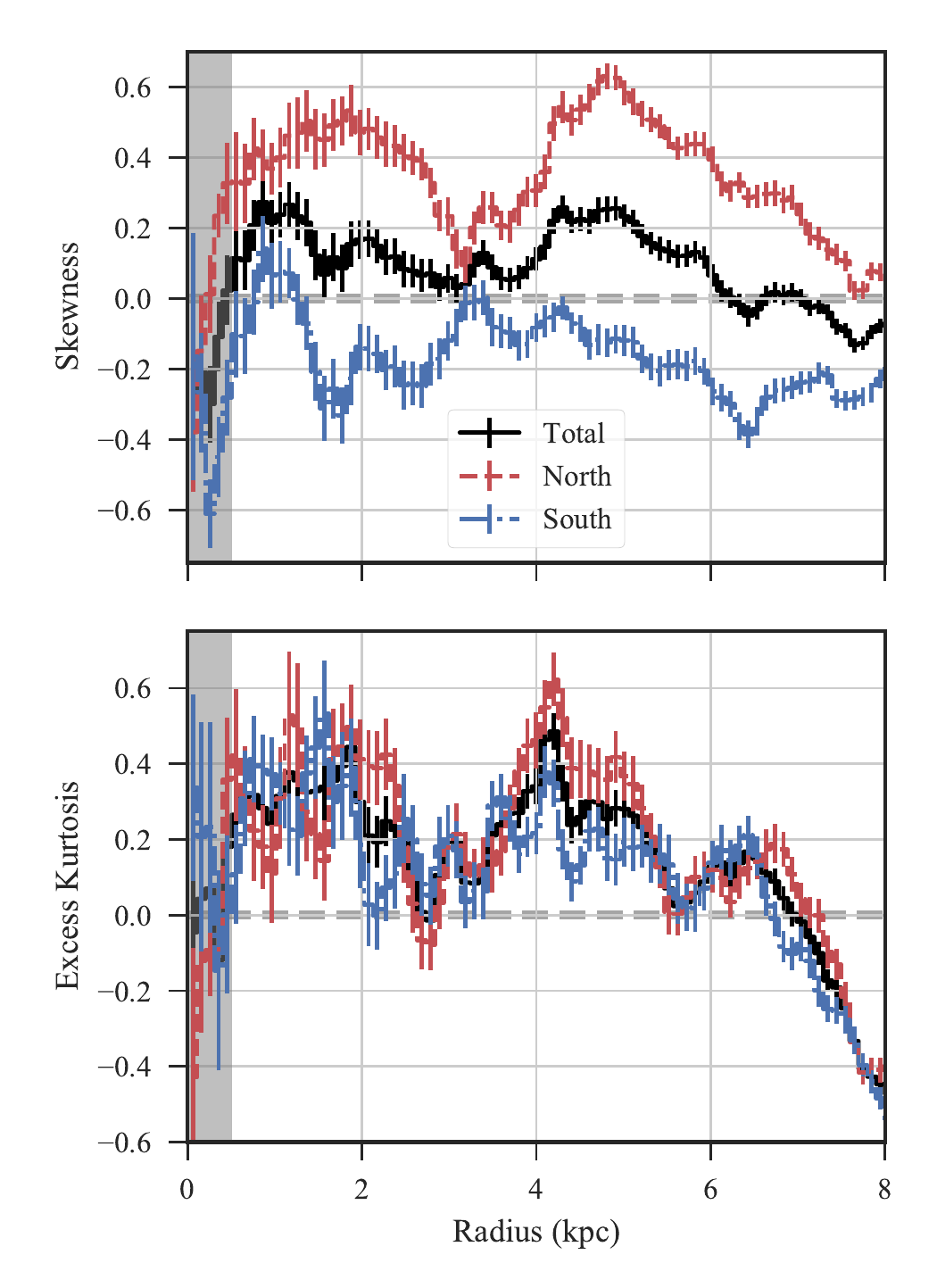}
\caption{\label{fig:skew_kurt_profile} Radial profiles of the skewness (top) and kurtosis (bottom) of the VLA+GBT data averaged over the entire disk (black), and the northern (blue) and southern (red) halves.  The shaded region indicates the inner 0.5 kpc where beam smearing and a small number of samples in each bin gives large uncertainties.  The skewness profile demonstrates the differing skewness properties between the halves of the galaxy.  The excess kurtosis indicates that the average \hi spectrum has heavier tails relative to a Gaussian, though the bias in the second moment requires caveats for interpreting the kurtosis (\S\ref{subsub:correlations_with_skewness_and_kurtosis}).}
\end{figure}

The skewness maps are shown in Figure \ref{fig:skew_maps} for the VLA-only and combined data.  The addition of the total power component from the GBT data has a significant effect on the skewness.  The VLA-only map shows small-scale structure across much of the disk, indicating variations of the spectral shape down to the resolution of the data.  This map also shows some larger-scale variations, particularly for the outer disk in the northern half.  With the inclusion of the GBT data, these large-scale variations become far more prominent.  The skewness structure is split into a postively-skewed northern half and a negatively-skewed southern half.  This separation is highlighted in Figure \ref{fig:skew_kurt_profile}, where the radial profiles of skewness are shown for the northern and southern halves.  The spectral shapes are skewed to the red-shifted side in the northern half and to the blue-shifted side in the southern half.  This is consistent with the position-velocity slices in \S\ref{sec:extraplanar_velocity_components} that show a lagging rotational \hi component across the disk.

\begin{figure*}
\includegraphics[width=\textwidth]{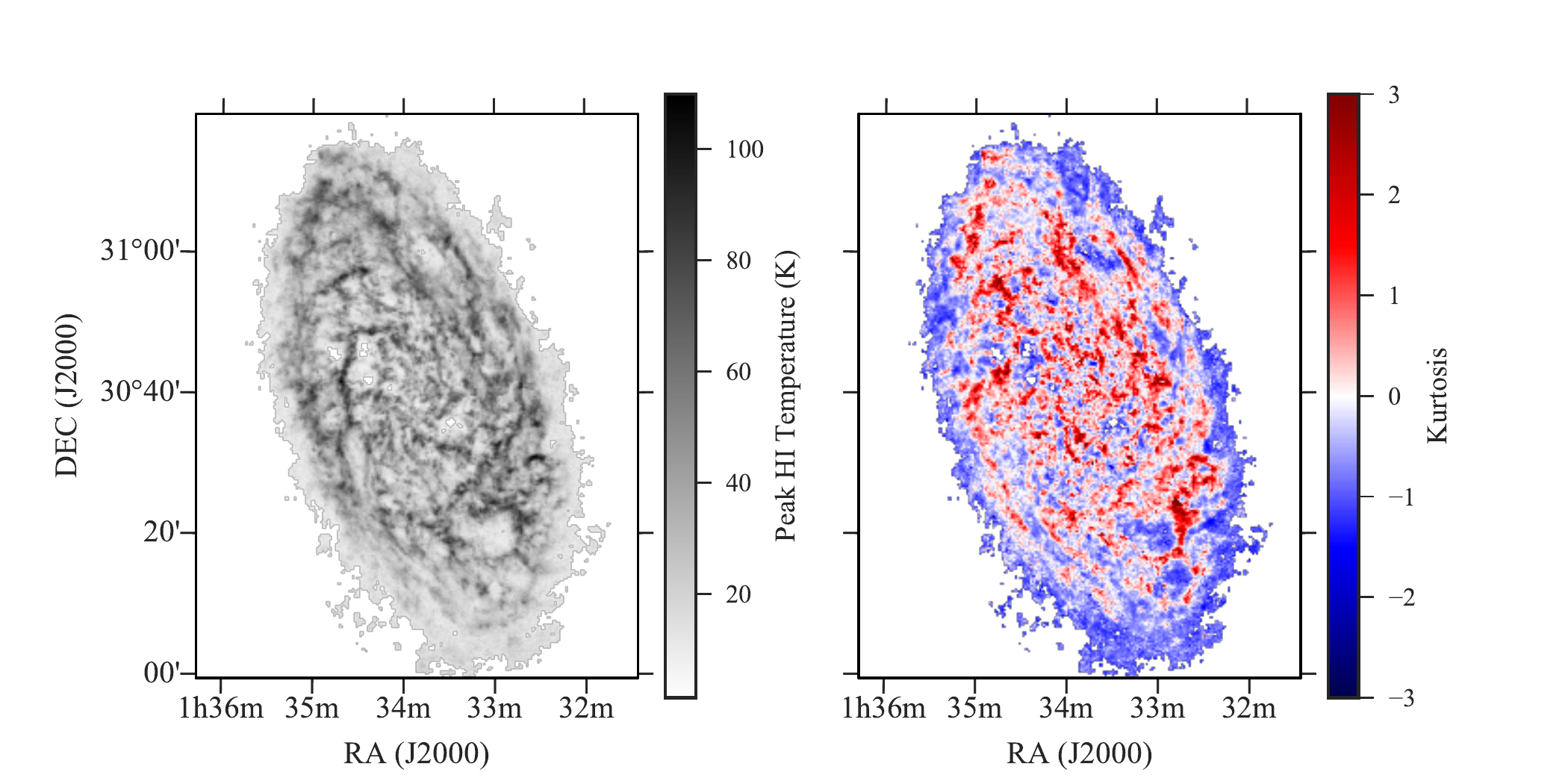}
\caption{\label{fig:tpeak_kurt_maps} The peak \hi temperature (left) and the kurtosis (right) maps of the combined data.  A positive kurtosis indicates a shape with strong tails and more peaked relative to a Gaussian.  There is a strong correlation between the two maps that results from the influence of the line wings on the kurtosis (\S\ref{subsub:correlations_with_skewness_and_kurtosis}).}
\end{figure*}

The kurtosis maps are less interesting due to the influence of the line wings, and this drives a strong correlation to the peak temperature map.  Figure \ref{fig:tpeak_kurt_maps} shows the close resemblance between the peak temperature and kurtosis maps for the combined data.  The excess line wings bias the kurtosis, making it a measure of the spectral shape relative to the line wings rather than the bright \hi component.  A higher peak temperature then leads to a larger positive kurtosis.  We discuss this relation further in \S\ref{subsub:correlations_with_skewness_and_kurtosis}.  Since kurtosis is velocity-weighted by the fourth power, the assumption that line shapes are well-described by single-peaked near-Gaussian breaks down and it provides less-useful empirical relations compared to the skewness.

The skewness profiles in Figure \ref{fig:skew_kurt_profile} are nearly mirror images of each other in the northern and southern halves.  By averaging over the entire disk, this distinction between the halves is lost, and the average values are near to a skewness of zero over most radii.  This hides interesting variations seen separately in the halves, particularly near 3 kpc, where the average skewness in both halves is close to zero.  In the surface density profiles (Figure \ref{fig:surfdens_n_s}), this radius corresponds to a dip in surface density, though it is more prominent in the northern half.  This region contains a large portion of the northern spiral arm, including some of the more active star-forming regions (e.g., NGC 604). Apart from this, it is unclear why only this region has a skewness close to zero and equal between the halves.

Though this asymmetry dominates both skewness maps, there are ``limbs'' in the outer disk that do not follow the same trend. The western side of the northern half in Figure \ref{fig:skew_maps} has a predominantly blue-shifted component along most of the map edge, while the eastern side of the southern half is mostly red-shifted. These features are nearly symmetric and are roughly pointed towards the position angle of the warped disk's semi-major axis \citep[$\sim165\degree$;][]{Corbelli2014A&A...572A..23C,Kam2017AJ....154...41K}. The skewness ``limbs'' may indicate where the warped disk becomes more prominent than the main disk. Indeed, the northern limb has larger line widths relative to the surrounding structure in Figure \ref{fig:peakvel_residual_vel_surface}.  These enhanced line widths are also consistent with the line width map from \citet{Kam2017AJ....154...41K}, which includes the structure at larger radii.  Note that this does not contradict the lack of a warped disk component in the rotation model (Appendix \ref{app:rotation_velocities}) since the peak velocities will not be sensitive to extended tails.

The kurtosis profiles in Figure \ref{fig:skew_kurt_profile} do not show variation between the halves and are consistent within the uncertainties with the whole disk average at most radii.  The positive kurtosis values indicate that the typical \hi spectrum in the inner 6 kpc has strong tails that do not vary significantly in shape.  The negative kurtosis beyond 7 kpc occurs where the brightest \hi emission tapers off in the surface density profile (Figure \ref{fig:surfdens_n_s}). A negative kurtosis value in these regions indicates a lack of narrow and bright \hi components.

We note that the profile values in the inner $\sim0.5\,$ kpc are more uncertain due to the low number of spectra in these bins and may be the cause of the negative skewness slope out to $1\,$ kpc.  If the spectra were preferentially red-shifted at the centre, the PV-slices in \S\ref{sub:anomalous_velocity_component} would show this asymmetry.

\subsubsection{Correlations with Skewness and Kurtosis} % (fold)
\label{subsub:correlations_with_skewness_and_kurtosis}

In Table \ref{tab:skewkurt_corrs}, we show the Spearman correlation coefficients calculated by comparing either the kurtosis or skewness with other \hi properties.  Most lines-of-sight have \hi detected along them, giving a population of $\sim1.2$ million spectra to compare properties\footnote{We do not correct for small-scale correlations over the beam size since the map extent is significantly larger. This will not affect correlations on large-scales.}.  Since the distribution of these parameter values need not be Gaussian, we perform a bootstrap to calculate the uncertainty of the correlation and the p-value. In each bootstrap iteration, we randomly choose 30\% of the sample and re-calculate the correlations; the standard deviation of the resulting distribution from all iterations gives the uncertainty.  The p-value calculation uses a randomly shuffled version of one of the parameters to calculate the correlation with the other parameter. Both steps are run for 1000 iterations, and all correlation values are found to be significant with small uncertainties.

\begin{table}
\caption{\label{tab:skewkurt_corrs} Correlation between skewness and kurtosis with other \hi properties. The uncertainties on each correlation are $\pm0.002$, using bootstrap resampling of one-third of the data to recompute the correlation 1000 times. We also perform a permutation test with 1000 iterations. All correlations have a p-value of $<0.001$. The mask width is the number of pixels in the spectral dimension contained within the signal mask.}
\centering
\begin{tabular}{lrr}
{}               & Skewness & Kurtosis      \\\hline
Mask Width       & $ 0.019$ & $ 0.408$ \\
Zeroth Moment    & $ 0.065$ & $ 0.682$ \\
Peak Temperature & $ 0.088$ & $ 0.800$ \\
Centroid         & $-0.463$ & $-0.155$ \\
Peak Velocity    & $-0.516$ & $-0.162$ \\
Line Width       & $-0.033$ & $-0.050$ \\\hline
\end{tabular}

\end{table}

Skewness is strongly correlated with \vcent and \vpeak for the combined VLA and GBT data.  This is well-explained by the north-south asymmetry in skewness evident in Figures \ref{fig:skew_maps} and \ref{fig:skew_kurt_profile}.

The kurtosis shows strong correlation with the peak temperature, zeroth moment, and the spectral width of the signal mask. We test whether these correlations are driven only by the shape of the signal mask by recalculating the correlations using points with similar mask widths (e.g., the correlation between kurtosis and peak temperature only for points with a mask width of 300 to 320 pixels). We find that the correlations with peak temperature and the zeroth moment persist when controlling for the mask width, indicating the correlations are not solely dependent on the signal mask.

The correlations with kurtosis are instead the result of the line width bias due to extended line wings. As discussed in Appendix \ref{app:second_moment}, the presence of line wings in the profile leads to an overestimate of the line width. This results from having a larger number of channels associated with the line wings than the bright component (e.g., spectrum d in Figure \ref{fig:hi_spectra}), and the kurtosis suffers from the same issue. The measured kurtosis is then with respect to the shape of the line wings, rather than the shape of the bright component.  The positive kurtosis reflects the strong line wings present in most spectra relative to a Gaussian shape.  Since peak temperature and zeroth moment are measures of how ``peaked'' that bright component is, they are positively correlated with the kurtosis.  This highlights where the naive moments approach to examining complex line shapes breaks down and that the results should be treated with caution.

Neither skewness nor kurtosis shows a strong correlation with the line width. This is encouraging since both depend on the line width, with increasing power, and a strong correlation may imply the variations in skewness and kurtosis are only driven by changes in the line width. Our findings above also show that correlations with kurtosis are not driven by the mask shape. These results show that our measurement of significant non-zero values in the higher order moments is robust and that single-Gaussian descriptions of the \hi line becomes inadequate at 80 pc resolution.  However, since moment-based descriptions are subject to biases in the centroid, we further explore \hi line profile shapes using stacked profiles below.

% subsubsection correlations_with_skewness_and_kurtosis (end)

% subsection exploring_spectral_complexities (end)

\subsection{Stacking Spectra} % (fold)
\label{sub:stacking_spectra}

Stacking spectra is a common procedure to enhance the signal-to-noise in regions of faint emission.  By aligning spectra to a common central velocity, combining the spectra will coherently add emission components, while the noise will add incoherently. This technique has often been used for studying typical \hi line properties in nearby galaxies.

We create stacked profiles using three different definitions of the line centre: the rotational velocity (\vrot; \S\ref{sub:rotation_curve}), the centroid velocity (\vcent), and the peak velocity (\vpeak). From the resulting stacked profiles, we can assess how the reference velocity affects the stacked profile shape. A cursory test with Gauss-Hermite polynomials, as used in several previous works \citep[e.g., ][]{Petric2007AJ....134.1952P,Ianjasm2012AJ....144...96I,Stilp2013ApJ...765..136S}, gave poor results due to spectra with multiple bright \hi components and those with extended line wings.  The typical S/N in the data used in these studies prohibited directly estimating \vpeak, which is well-defined in our higher S/N data \citep[see also ][]{Braun1997ApJ...484..637B}.

The different definitions of the line centre will highlight different ISM properties. Previous studies have typically stacked spectra based on \vpeak, which will minimize the stacked profile width and recover the typical spectrum properties at the resolution of the data \citep{Caldu2016AJ....151...34C}.  This is ideal for attempting to recover the thermal and turbulent properties of the medium.  Stacking with \vrot yields stacked profiles with widths sensitive to large-scale turbulence and non-circular motions.  For example, variations from the circular rotation velocity are clear from the residual velocity map in Figure \ref{fig:peakvel_residual_vel_surface}.  Using \vcent as the common velocity gives, for an optically-thin tracer, the mass-weighted velocity average. As discussed in \S\ref{sub:exploring_spectral_complexities}, this can make \vcent biased if the line profiles are asymmetrical.  This bias in \vcent makes a physical interpretation less clear.  Nonetheless, we explore these three stacking methods since they are commonly used in spectral-line studies.

We shift the spectra using Fourier transforms before co-adding them to create stacked profiles.  Fourier shifting allows shifting by fractions of the channel width without interpolating,  minimizing artificial broadening due to a finite-sampled grid.  We then add the aligned spectra to create the stacked profile.  The uncertainty in each channel of the stacked profile is the sum-in-quadrature of the uncertainty from each PPV-pixel.  Since the noise is approximately constant in our data, this reduces to the noise of one PPV-pixel multiplied by the square root of the number of beam elements in the stacked region.  We include all of the spectra in the stacked profiles where \vcent and \vpeak can be calculated after applying the signal mask (i.e., the signal mask contains at least one component along the line-of-sight).  This excludes only the small masked regions in Figure \ref{fig:peakvel_residual_vel_surface}.

To model the stacked spectra, we adopt the half-width-half-max (HWHM) scaling method from \citet[][]{Stilp2013ApJ...765..136S}.  We favour this approach rather than other models due to the large variation in the shapes of the stacked profiles we find using different definitions of the line centre.  In particular, we found that the two-Gaussian model used in other works \citep[e.g.,][]{Young1996ApJ...462..203Y,Ianjasm2012AJ....144...96I} cannot account for how peaked the \vpeak stacked profiles are and the ratios between the two components are highly-sensitive to covariances between parameters.  We provide a more thorough comparison in Appendix \ref{app:modelling_super_profiles}.

The HWHM method assumes that the stacked profile within the FWHM of the peak intensity can be modeled as a Gaussian line profile.  Since these stacked profiles are created over large areas, the S/N is extremely high and the FWHM is a reliable measure of the width.  Additional properties describing the stacked profile shape are defined based on the residual between the stacked profile and the assumed Gaussian profile for the peak.  We use the definitions for four of these parameters from \citet{Stilp2013ApJ...765..136S}.  The effective Gaussian width, $\sigma_{\rm HWHM}$, is estimated by scaling to a Gaussian shape within the FWHM of the stacked profile.  This assumed Gaussian shape is centred at the velocity of the maximum intensity, $v_{\rm centre}$.  The line wings are defined as the excess relative to this Gaussian peak beyond the FWHM points.  The fraction of excess flux is given as $f_{\rm wing}$:
\begin{equation}
  \label{eq:f_wings}
  f_{\rm wings} = \frac{\sum\limits_{|v|>{\rm HWHM}} \left[S(v) - G(v)\right]}{\sum\limits_{v} S(v)},
\end{equation}
where $S(v)$ is the stacked profile, $G(v)$ is the Gaussian model within the FWHM and ${\rm HWHM} = {\rm FWHM} / 2$. The sum in the denominator is over the entire profile. The line wing excess can be used to define an effective width of the wings equivalent to the second moment:
\begin{equation}
  \label{eq:sigma_wings}
  \sigma_{\rm wing}^2 = \frac{\sum\limits_{|v|>{\rm HWHM}} \left[S(v) - G(v)\right]v^2}{\sum\limits_{|v|>{\rm HWHM}} \left[S(v) - G(v)\right]}.
\end{equation}
Note that the definition of the second moment relies on a Gaussian line profile. Since the residual $S(v) - G(v)$ is not Gaussian, this may not have a clear connection to a Gaussian line width. We include it here to compare with the values presented in \citet{Stilp2013ApJ...765..136S}.

We define two additional parameters to describe the profile shapes. While higher-order moments present one avenue for characterizing deviations from a Gaussian shape, we favour developing empirical measures that do not weight the statistic by deviations from the line centroid.  These metrics provide a more uniform description of the line profile over its entire extent. First, we define the asymmetry of a profile as the difference between the total flux at velocities smaller than the reference velocity and the total flux at velocities greater than the reference velocity. The difference is then normalized by the total flux over all velocities:
\begin{equation}
  \label{eq:aymm_hwhm}
  a = \frac{\sum\limits_{v > v_{\rm centre}} S(v) - \sum\limits_{v < v_{\rm centre}} S(v)}{\sum S(v)}.
\end{equation}
The sum in the denominator is over the whole profile.  \citet{Stilp2013ApJ...765..136S} define a different asymmetry parameter that, while similar, returns an absolute value.  We choose this alternate definition since it retains information on the side of the profile containing more flux, which is determined by the asymmetry in the line wings.  In this manner, $a$ is analogous to the skewness, without weighting by the velocity offset and line width.

The final parameter, $\kappa$, provides the fractional difference between the central peak and the Gaussian model within the FWHM, similar to $f_{\rm wings}$:
\begin{equation}
  \label{eq:kappa_hwhm}
  \kappa = \frac{\sum\limits_{|v|<{\rm FWHM}} \left[ S(v) - G(v) \right]}{\sum\limits_{|v|<{\rm FWHM}} G(v)}.
\end{equation}
This is another parameterization for the kurtosis: a negative value indicates the profile is more peaked than the Gaussian, while a positive value is flatter.  We note, however, that the kurtosis used in \S\ref{sub:exploring_spectral_complexities} will primarily be sensitive to the line wing structure, whereas $\kappa$ describes the shape of the peak since it is only measured within the HWHM relative to a Gaussian of equivalent width.

\begin{figure}
\includegraphics[width=0.5\textwidth]{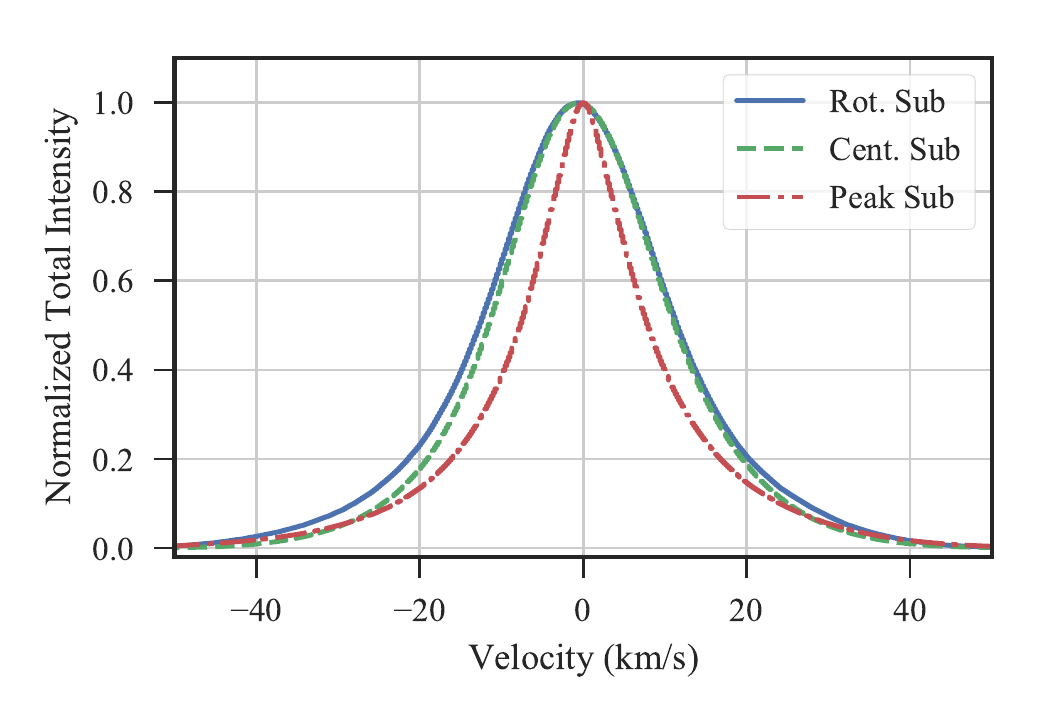}
\caption{\label{fig:hwhm_profiles} Stacked spectra over the entire disk using \vrot (blue solid), \vcent (green dashed), and \vpeak (red dot-dashed) as the definitions of the line centre.  The peak intensities are set to unity to emphasize the line shapes and the velocity axis is centred at the systemic velocity.  Stacking based on \vrot and \vcent give similar profiles with near Gaussian peaks within the FWHM.  However, using \vpeak as the reference velocity yields a sharper peak shape and more prominent line wings relative to a Gaussian.}
\end{figure}

Figure \ref{fig:hwhm_profiles} shows the stacked spectra over the entire disk with different reference velocities, and Table \ref{tab:stacked_fits} gives the fitted model parameters.  The parameter uncertainties are calculated in one of two ways. For $\sigma$ and $v_{\rm centre}$, we assume that the uncertainty is set by the channel size; thus the uncertainty is half the channel width ($\sim 0.1$ \kms). For the rest of the parameters, we perform 100 bootstrap iterations where 1) the data are re-sampled by adding Gaussian noise within the flux uncertainty for each channel, and 2) allowing the estimated peak velocity and Gaussian width to vary by their uncertainties.  This propagates the uncertainty in the flux and the assumed Gaussian peak to the other parameters.  Due to the high S/N in the stacked profiles, most of the uncertainty comes from the latter source.

\begin{table*}
\caption{\label{tab:stacked_fits} HWHM fit values to the stacked profiles over the entire disk. These profiles are shown in Figure \ref{fig:hwhm_profiles}. The uncertainties are propagated based on the uncertainty in each channel of the profiles, and assuming an uncertainty of half the spectral resolution ($\sim0.1\,$ \kms) for the peak velocity and FWHM line width.  Note that $\sigma_{\rm wing}$ may not be treated as an equivalent Gaussian width due to the non-Gaussian shape of the residuals; see Equation \ref{eq:sigma_wings}.}
\begin{tabular}{lrrr}
{} & \multicolumn{3}{c}{Method of Stacking Spectra} \\
{}                          & Rotation Vel. (\vrot)              & Centroid Vel. (\vcent)              & Peak Vel. (\vpeak)      \\\hline
\\[-1em]
$\sigma_{\rm HWHM}$ (\kms)             & $10.3\pm0.1$               & $9.7\pm0.1$                & $6.6\pm0.1$    \\
\\[-1em]
$v_{\rm centre}$ (\kms)       & $-0.5\pm0.1$               & $-0.5\pm0.1$               & $0.0\pm0.1$    \\
\\[-1em]
$f_{\rm wing}$              & $0.11\pm0.01$              & $0.09\pm0.01$              & $0.26\pm0.01$  \\
\\[-1em]
$\sigma_{\rm wing}$ (\kms)  & $28.9_{-0.4}^{+0.5}$       & $26.3_{-0.4}^{+1.5}$       & $23.4_{-0.2}^{+0.3}$   \\
\\[-1em]
$a$                         & $-0.009_{-0.010}^{+0.011}$  & $0.027_{-0.003}^{+0.004}$ & $0.021_{-0.014}^{+0.005}$  \\
\\[-1em]
$\kappa$                    & $-0.013_{-0.004}^{+0.004}$ & $-0.011_{-0.004}^{+0.004}$ & $-0.059_{-0.004}^{+0.005}$ \\\hline
\end{tabular}
\end{table*}

Figure \ref{fig:hwhm_profiles} shows how the use of different reference velocities significantly changes the stacked profile shapes. Stacking based on \vrot and \vcent gives the widest profiles and both have similar shapes. The shapes of their peaks are close to Gaussian ($\kappa\sim0$; Table \ref{tab:stacked_fits}) and the widths are $\sim10$ \kms in both. The widths of these profiles are dominated by the residual velocity in the \vrot model (Figure \ref{fig:peakvel_residual_vel_surface}) and the bias in \vcent due to extended line wings evident in the skewness maps (Figure \ref{fig:skew_maps}).
Stacking using \vcent or \vpeak gives profiles with non-zero $a$ parameters (Equation \ref{eq:aymm_hwhm}).  This may be highlighting an asymmetry in the low surface-brightness component of the line wings not evident in the position-velocity (PV) slices in \S\ref{sec:extraplanar_velocity_components}.

\begin{figure}
\includegraphics[width=0.5\textwidth]{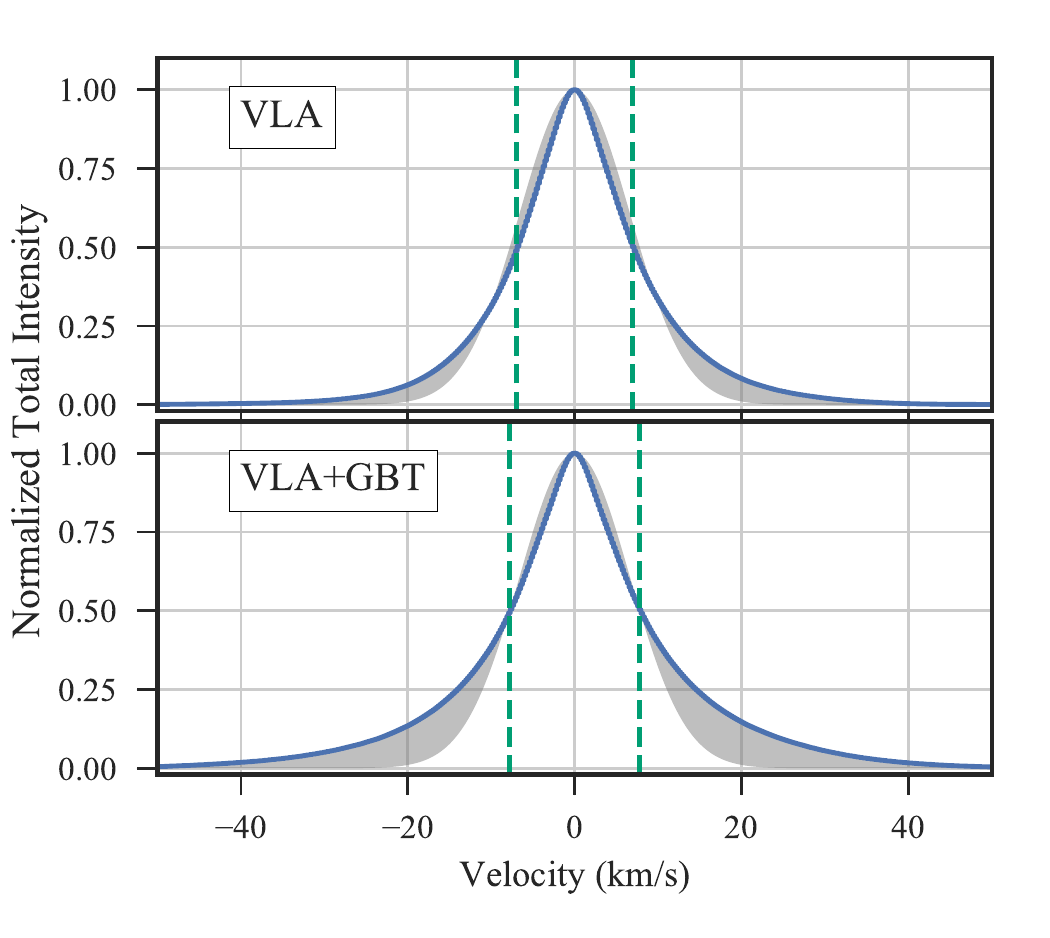}
\caption{\label{fig:peakvel_profiles} Stacked profiles of the VLA (top) and VLA+GBT data (bottom) when centered using \vpeak. The dashed green lines are the FWHM line widths and the shaded gray regions show the difference between the equivalent Gaussian profile and the data.  The GBT data mostly adds emission to the line wings and does not affect the shape within the FWHM.}
\end{figure}

The stacked profile from using \vpeak is clearly non-Gaussian in its central shape ($\kappa=-0.059$).  The profile is both narrower and more peaked relative to the other profiles and is similar to the shape found in other studies \citep[e.g.,][]{Petric2007AJ....134.1952P}. The narrow and peaked profile contains a larger fraction of flux in the line wings relative to a Gaussian ($f_{\rm wings}=0.26$).  The difference in this stacked profile compared to those from \vcent and \vrot is the source of broadening in the width, which we discuss further in \S\ref{sub:mix_models}. If the broadening source has a Gaussian distribution of velocities around the peak, like the the residuals from \vrot, the central-limit theorem gives that the shape will be close to Gaussian within the FWHM. When using \vpeak, the dominant source of broadening is the uncertainty on \vpeak ($\sim0.1$ \kms), which is small compared to typical \hi line widths.

The width and shape of the \vpeak profile is relatively unchanged without the total power component. In Figure \ref{fig:peakvel_profiles}, we show the \vpeak stacked profiles with and without the GBT data. Including the GBT data affects the line wings, which are far more prominent in the combined VLA+GBT data set. These differences highlight that most of the line wing emission is driven by a low surface brightness, extended component.  This combination of a steep peak within the FWHM and heavy tails makes these shapes difficult to model. They are clearly non-Gaussian, but neither of these profiles has tails strong enough to be well-modeled by a Lorentzian.

The assumed Gaussian shape for sharply peaked profiles violates some of the assumptions for the HWHM model.  The fractional difference in the peak shape, defined with $\kappa$, indicates that the assumed Gaussian shape overestimates the emission in the peak by 6\% for the \vpeak stacked profile.  Though the two-Gaussian model also cannot account for the peak shape (Appendix \ref{app:modelling_super_profiles}), this highlights the difficulty in defining a simple model for the stacked profile shapes. We discuss the factors that lead to this difficulty in \S\ref{sub:mix_models}.

\subsubsection{Radial Variations in Stacked Profiles}
\label{subsub:radial_variations_in_stacked_profiles}

We next examine how the stacked profile parameters vary with galactocentric radius, focussing on the \vpeak stacked profiles.  We create stacked profiles in galactocentric rings with $100$ pc widths, as used for the surface density profiles (\S\ref{sub:radial_profiles}). Given the opposing line wing directions found in the skewness map (\S\ref{sub:exploring_spectral_complexities}), we also examine stacked profiles over the northern and southern halves separately.

\begin{figure}
\includegraphics[width=0.5\textwidth]{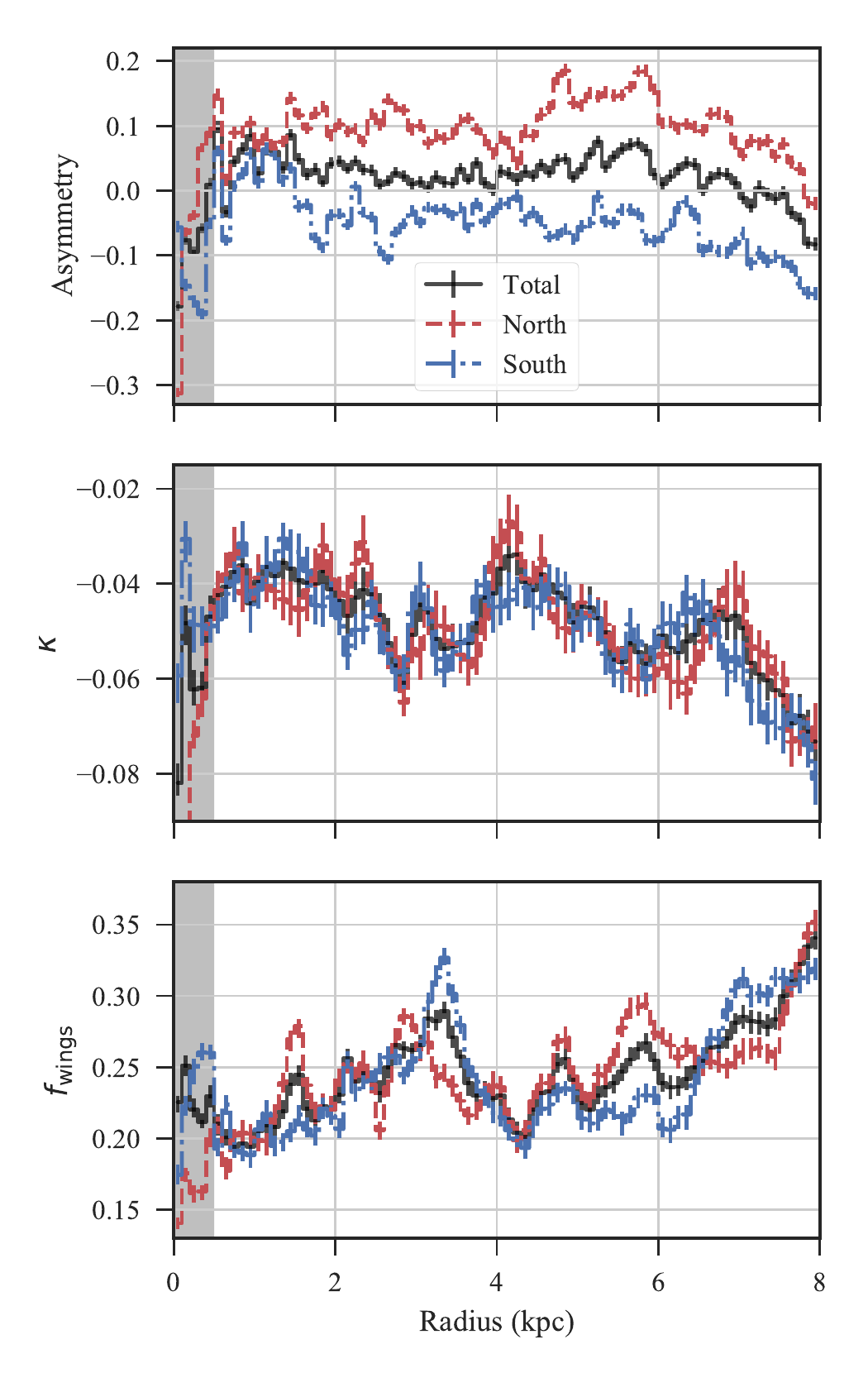}
\caption{\label{fig:hwhm_properties} Top: Asymmetry, $a$, of \vpeak stacked profiles in 100 pc radial bins. The parameter values are shown for stacked profiles over the whole disk (black), and the northern (blue dashed) and southern (red dot-dashed) halves. Middle: The peak shape $\kappa$. Bottom: The line wing fraction $f_{\rm wings}$. The shaded region in both plots highlights the inner $0.5$ kpc, where the lack of data points make the estimates more uncertain.  The uncertainties in this region are underestimated since they do not account for the additional velocity uncertainty that results from spatial smearing where the rotation curve is steep \citep{Stilp2013ApJ...765..136S}.  The top two panels are qualitatively similar to the skewness and kurtosis profiles from Figure \ref{fig:skew_kurt_profile}: the asymmetry varies between the halves of the galaxy, but the peak shape is consistent between both.}
\end{figure}

Figure \ref{fig:hwhm_properties} shows the three profile parameters describing asymmetries ($a$) and shape ($\kappa$ and $f_{\rm wings}$) for the \vpeak stacked profiles; we explore the line width variations in \S\ref{sub:comparing_estimates_of_velocity_dispersion}.  The asymmetry parameter $a$ shows a clear difference between the two halves, matching the variations identified in the skewness analysis (\S\ref{sub:exploring_spectral_complexities}). The southern half of the galaxy has an excess of emission at negative velocities while the northern half has an excess at positive velocities. This is qualitatively the same as is shown in the skewness radial profile in Figure \ref{fig:skew_kurt_profile}, though we find less variation with radius in $a$.  The decreased radial variation in $a$ results from the $\sigma^3$ weighting in the skewness, which makes it sensitive to small line width variations, and from the difference in the order of operations, since stacked profiles are already an averaged quantity.

The shape within the FWHM of the stacked profiles, parameterized with $\kappa$, is largely consistent between the two halves.  However, $\kappa$ varies by $\sim33\%$ with radius in the inner 6 kpc.  There are two dips at 2.5 and 3.5 kpc where $\kappa$ increases, indicating a larger discrepancy from a Gaussian.  There are similar features in the kurtosis profile in Figure \ref{fig:skew_kurt_profile}, however, those variations suggest that the typical spectrum becomes {\it closer} to a Gaussian profile.  This apparent discrepancy may result from an increase in the fraction of multi-component spectra in these regions. These radii contain portions of the optical spiral arms, particularly the region around NGC 604 in the northern arm and the bright clump in the southern arm (see spectrum c in Figure \ref{fig:hi_spectra}).  The kurtosis is driven by the shape with respect to the line wings (\S\ref{subsub:correlations_with_skewness_and_kurtosis}) and multiple components will give a smaller excess kurtosis since the bright component is effectively wider.  We see the opposite effect in $\kappa$ because the spectra are added coherently and multiple components will tend to enhance the line wings.

The fraction of flux in the line wings is on average $f_{\rm wings} \sim 0.22$ in the inner 6 kpc, but there are substantial ($\sim50\%$) variations.  There are peaks at radii of 2.5 and 3.5 kpc, consistent with variations in $\kappa$ being driven more by lines-of-sight having multiple spectral components.  However, peaks in $f_{\rm wings}$ result from an increase in one half of the galaxy, whereas the variations in $\kappa$ are consistent between both halves.  The variations in  $f_{\rm wings}$ cannot then be entirely driven by an increase in multi-component spectra. Another process, possibly outflows from star formation in the spiral arms, may lead to the increase in the line wing fraction.

None of the parameters show a monotonic trend with galactocentric radius in the inner 6 kpc, suggesting that the variations are due to local processes (i.e., outflows, local galactic structure).

A trend is clear, however, in all of the parameters beyond 6 kpc.  The line profiles, over both halves, become more asymmetric ($a < 0$), have sharper and narrower peaks ($\kappa$ and $\sigma$ decrease) and have increased line wing fractions.  A similar trend is seen for the kurtosis profile (Figure \ref{fig:skew_kurt_profile}), though the kurtosis {\it decreases} for the reasons discussed above. At $\sim6.5$ kpc, the surface density profile strongly decreases to the edge of the map (Figure \ref{fig:surfdens_n_s}).  We attribute these trends to the lack of a bright \hi component when averaged on large-scales. The change in asymmetry is due to the beginning of M33's warped disk, which starts to dominate the velocity surface at $\sim 7$ kpc, consistent with increases in the line width at the edge of the map (Figure \ref{fig:peakvel_residual_vel_surface}).  The variations we find beginning at 6 kpc suggests the warped disk has a significant effect on the spectral shapes before it dominates the large-scale kinematics.

\subsubsection{Asymmetry between the Northern and Southern Halves}
\label{subsub:asymmetry_between_the_northern_and_southern_halves}

The skewness maps in Figure \ref{fig:skew_maps} indicate a large-scale asymmetry in profile shapes between the northern and southern halves of the galaxy.  By comparing the line wing excess in each half of the galaxy, we can estimate what fraction of $f_{\rm wings}$ is symmetric and asymmetric in velocity.

\begin{figure}
\includegraphics[width=0.5\textwidth]{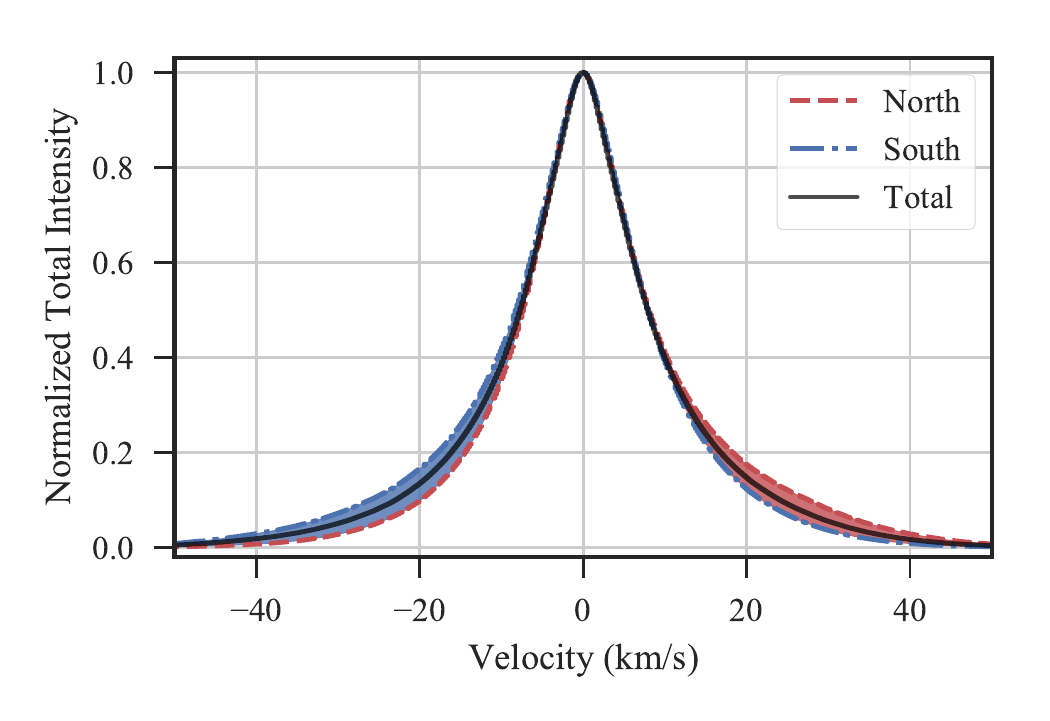}
\caption{\label{fig:hwhm_ns_asymm} The \vpeak stacked profiles over the whole galaxy (solid black), the Northern half (blue dashed), and the Southern half (red dot-dashed).  The line wings asymmetry between the halves are highlighted with the shaded regions, where the colours correspond to the velocity the profiles are skewed towards.}
\end{figure}

The \vpeak stacked profiles in each half and over the entire disk are compared in Figure \ref{fig:hwhm_ns_asymm}.  The peaks of all three profiles are essentially identical in shape and width, but the line wings differ in each half.  We can re-write the line wing fraction (Equation \ref{eq:f_wings}) as:
\begin{equation}
  \label{eq:line_wing_n_s}
  f_{\rm wing} = \frac{\sum\limits_{|v|>{\rm HWHM}} \left[S_\mathrm{N}(v) + S_\mathrm{S}(v) - G(v)\right]}{\sum\limits_{v} S_\mathrm{N}(v) + S_\mathrm{S}(v)},
\end{equation}
where $S_\mathrm{N}(v)$ and $S_\mathrm{S}(v)$ are the stacked profiles of the northern and southern halves, respectively.  The asymmetric line wing component is the total excess on opposite sides of the peak, located at $v=0$ \kms in this case. Based on the total profiles, the southern half will have an excess towards negative velocities, while the northern half will have an excess at positive velocities, both pointing toward the systemic velocity. Thus we define the asymmetric line wing component as:
\begin{equation}
  \label{eq:line_wing_asymm}
  f_{\rm asymm} = \frac{\sum\limits_{v< - {\rm HWHM}} \left[S_\mathrm{S}(v) - S_\mathrm{N}(v)\right] + \sum\limits_{v > {\rm HWHM}} \left[S_\mathrm{N}(v) - S_\mathrm{S}(v)\right]}{\sum\limits_{v} S_\mathrm{N}(v) + S_\mathrm{S}(v)}.
\end{equation}
Note that the asymmetric component is not dependent on the assumed Gaussian peak $G(v)$, though the limits on the sums are set by the HWHM. The symmetric line wing component is the difference between the total and asymmetric components:
\begin{equation}
  \label{eq:line_wing_symm}
  f_{\rm symm} = f_{\rm wing} - f_{\rm asymm}.
\end{equation}
Because the asymmetric component is not defined as the absolute value of the difference, it is possible to have a negative value if the line wings excesses are opposite of the definition assumed based on Figure \ref{fig:hwhm_ns_asymm}.

Table \ref{tab:f_wing_symm_asymm} shows the asymmetric and symmetric components of the total profiles from Figure \ref{fig:hwhm_ns_asymm}. Relative to the total line wing fraction, we find that the $\sim70\%$ arises from a symmetric component while $30\%$ stems from an asymmetric one.  In both halves, the asymmetric component points toward the systemic velocity of M33 (since $f_{\rm asymm} > 0$) and could trace a rotationally-lagging disk component. We discuss this further in \S\ref{sec:extraplanar_velocity_components}.

\begin{table}
\centering
\caption{The line wing fractions from the peak velocity stacked profiles split into the Northern and Southern halves. The total line wing fraction is the same as given in Table \ref{tab:stacked_fits}.\label{tab:f_wing_symm_asymm}}
\begin{tabular}{lr}
                         & Line Wing Fraction \\\hline
$f_{\rm wing}$           & $0.26\pm0.01$ \\
$f_{\rm symm}$           & $0.18\pm0.01$ \\
$f_{\rm asymm}$          & $0.08\pm0.01$ \\\hline
\end{tabular}
\end{table}

\begin{figure}
\includegraphics[width=0.5\textwidth]{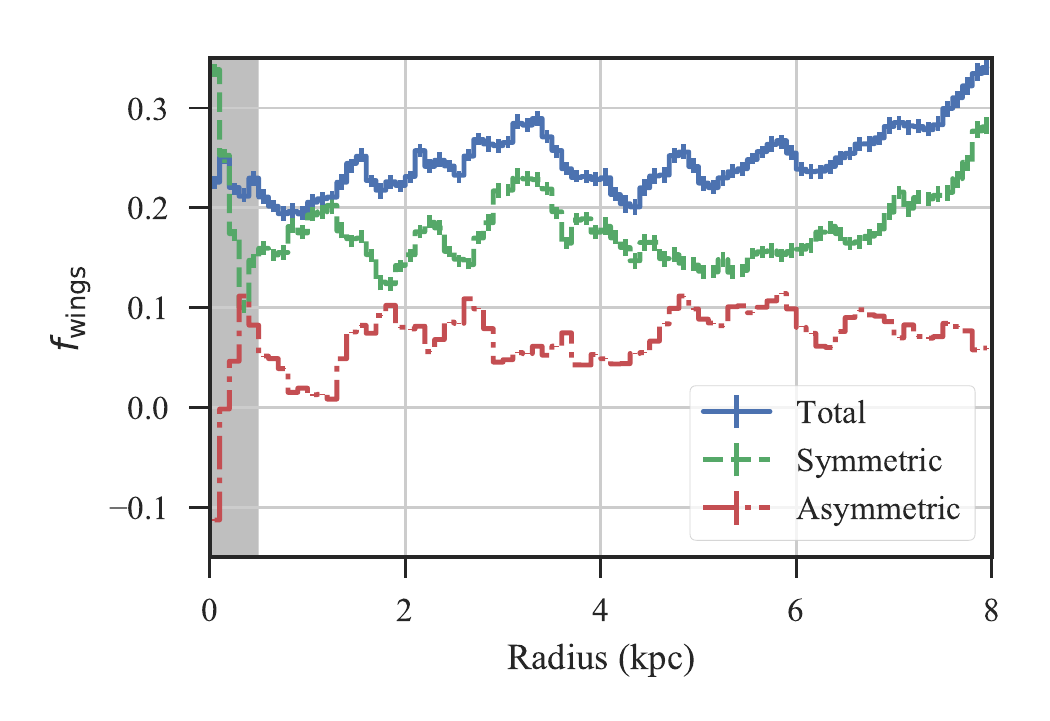}
\caption{\label{fig:hwhm_wing_ns} The total (blue solid), symmetric (green dashed), and asymmetric (red dot-dashed) line wing fractions determined from Equations \ref{eq:line_wing_n_s} -- \ref{eq:line_wing_symm} for \vpeak stacked profiles in 100 pc radial bins.  The inner 0.5 kpc is shown in gray, where the small area and beam smearing cause deviations in the profile properties. The asymmetric component accounts for $\sim30\%$ of the line wing fraction and is roughly constant throughout the disk.  The variations in the total line wing fraction are mostly driven by variation in the symmetric component.}
\end{figure}

Figure \ref{fig:hwhm_wing_ns} shows the symmetric and asymmetric line wing fractions calculated from profiles in 100 pc radial bins.  The asymmetric component is largely constant throughout the entire disk.  The symmetric line wing component contains most of the variations previously seen in Figure \ref{fig:hwhm_properties} for the total line wing fraction. In particular, the increase in $f_{\rm wings}$ beyond 6 kpc is entirely due to the symmetric component.

% subsubsection asymmetry_between_the_northern_and_southern_halves (end)

\subsubsection{Stacking Profile with Peak \hi Temperature} % (fold)
\label{subsub:stacking_profile_with_peak_hi_temperature}

We also explore the properties of stacked profiles binned by their peak \hi temperature. We created 20 stacked profiles in bins with a width of $5\%$ of the peak \hi temperature distribution.  This provides an equal number of spectra ($48900$) in each stacked profile. As in the previous section, we consider the profiles stacked with respect to \vrot, \vcent and \vpeak.

\begin{figure}
\includegraphics[width=0.5\textwidth]{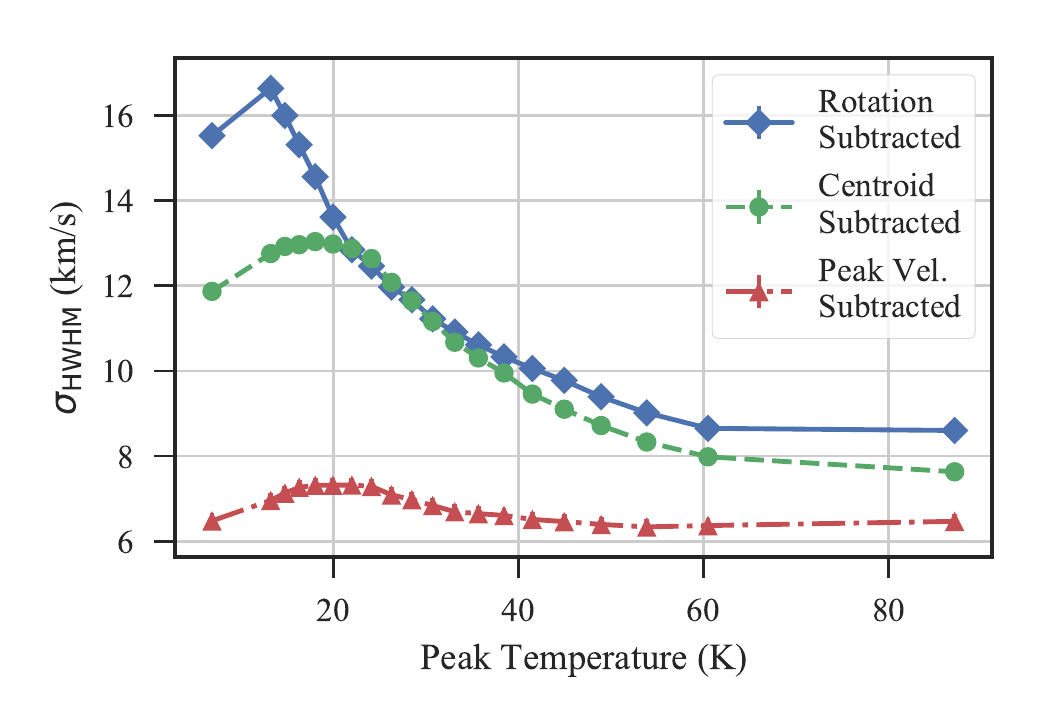}
\caption{\label{fig:peaktemp_lwidths} The \hi line width of profiles stacked based on their peak \hi temperature. Each stacked profile covers a range of $5\%$ in the peak temperature distribution --- the symbols in the plot indicate the middle of each bin.}
\end{figure}

Figure \ref{fig:peaktemp_lwidths} shows the line widths for each stacking method in $5$-percentile bins.  Similar to radially-binning the data (\S\ref{sub:comparing_estimates_of_velocity_dispersion}), we find an offset between the widths when using different definitions of the line centre, with the peak velocity stacking providing the smallest widths.  Each stacking method has a line width that decreases with increasing peak temperature, though the change in the widths of the \vpeak stacked profiles is only $\sim1$ \kms.

Despite the small changes we find in the \vpeak stacked profiles over the range in peak temperature, the shape parameters change substantially.  The line wing fraction $f_{\rm wing}$ monotonically decreases with increasing peak temperature, ranging from $0.17\mbox{--}0.32$.  The shape within the FWHM of the profile, described by $\kappa$, also increases with the peak temperature, with values ranging from $-0.15$ to $-0.02$. Note that this is the opposite behaviour from the radially-binned profile parameters (Figure \ref{fig:hwhm_properties}).  The combination of the trends in these two parameters and the near-constant line width demonstrates a shift in the profile shape that is near-Gaussian at higher peak temperatures and significantly non-Gaussian to low peak temperatures.

The profile shapes at low peak temperatures ($<20$ K) show two distinct components: a narrow sharply-peaked component on top of a broad near-Gaussian component.  Since, by area, most of the lines-of-sight with peak temperatures of $<20$ K are located at $R_{\rm gal}>7$ kpc, where the average surface density begins decreasing (Figure \ref{fig:surfdens_n_s}),  we interpret the increased importance of the broad component as arising from an increase in the proportion of emission from M33's warped disk component. This is consistent with our analysis of the skewness maps (\S\ref{sub:exploring_spectral_complexities}), where the reversal in the sign of the skewness near the edges of the map in Figure \ref{fig:skew_maps} are roughly aligned with the position angle of the warped disk component \citep{Corbelli2014A&A...572A..23C,Kam2017AJ....154...41K}.  The increase in $f_{\rm wings}$ from the radially-binned stacked profiles in Figure \ref{fig:hwhm_properties} is also consistent with most spectra with low peak temperature ($<20$ K) arising at $R_{\rm gal}>7$ kpc.

We also find no clear evidence for \hi spectra with flattened peaks in the highest peak temperature bins as would result from optically thick \hi emission \citep{Braun2009ApJ...695..937B,Braun2012ApJ...749...87B}.  If the stacked spectra had flattened tops, $\kappa$ should decrease with increasing peak temperature, opposite of the consistent increase we find.

% subsection stacking_profile_with_peak_hi_temperature (end)

% subsection stacking_spectra (end)

\subsection{Comparing Estimates of Line Width} % (fold)
\label{sub:comparing_estimates_of_velocity_dispersion}

\begin{figure}
\includegraphics[width=0.5\textwidth]{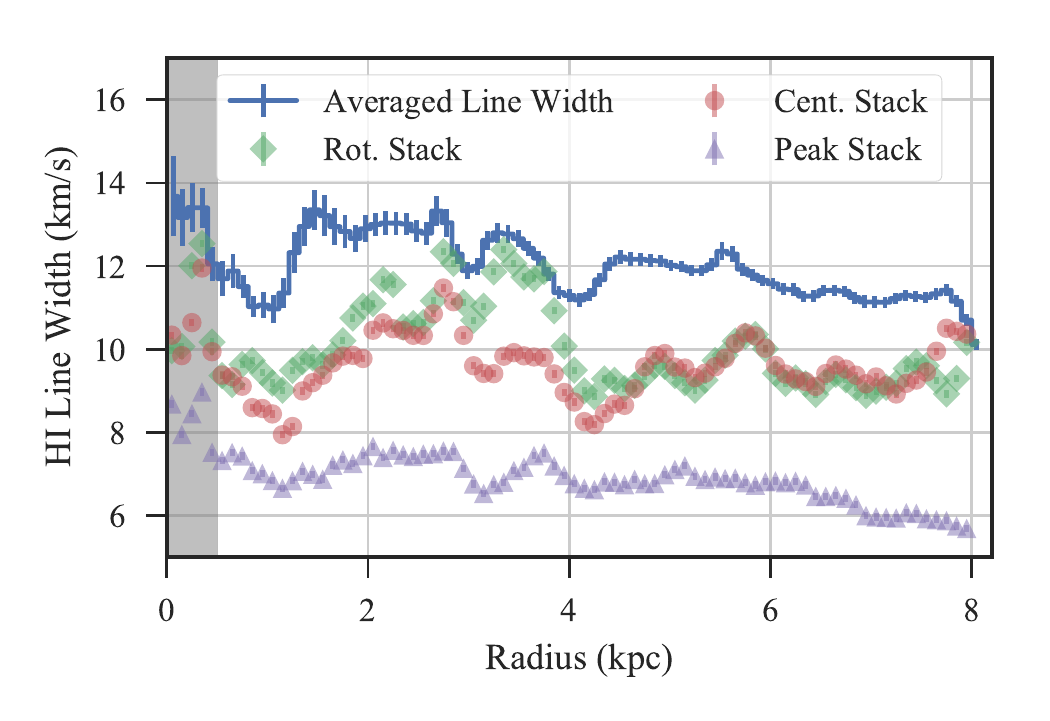}
\caption{\label{fig:veldisp_profile} \hi line width profiles in 100 pc bins for the azimuthally-averaged line width map (solid line; Figure \ref{fig:peakvel_residual_vel_surface}) and the stacked profiles from \vrot (green diamonds), \vcent (red circles), and \vpeak (purple triangles). The shaded gray region highlights the inner 0.5 kpc where the lack of samples and beam smearing cause large uncertainties in the line widths.  The different methods give a large range in the \hi line width. However, none of the methods shows a strong radial decrease.}
\end{figure}

Figure \ref{fig:veldisp_profile} shows the line width as a function of galactocentric radius as derived from the azimuthally-averaged second-moment line width map (Figure \ref{fig:peakvel_residual_vel_surface}) and the stacked spectra aligned with \vrot, \vcent and \vpeak.  The enhanced line widths in the inner $0.5$ kpc result from the small areas averaged over and from beam smearing where the rotation curve is steep.

Stacking by \vpeak gives the narrowest profiles, consistent with the whole-disk stacked spectra in Figure \ref{fig:hwhm_profiles}.  The typical width is $\sim7$ \kms in the inner 6 kpc with a shallow gradient; beyond 6 kpc, the line widths decrease to $\sim6$ \kms.  Our line widths are consistent with the stacked profile widths found by \citet{Druard2014A&A...567A.118D} using the archival \hi VLA observations presented in \citet{Gratier2010A&A...522A...3G}.

The \vrot and \vcent stacked profiles also have similar widths ($\sim10$ \kms) to the whole-disk stacked profiles.  The widths are consistent with each other throughout most of the disk, suggesting that the line wing bias in \vcent in the northern and southern halves of the disk are similar to the dispersion of velocity residuals from the \vrot model.  At radii from 2 to 4 kpc, where the spiral arms dominate the galactic structure, the  \vrot stacked profiles have larger widths than the \vcent stacked profiles.  This indicates enhanced motion in the spiral arms due to multi-component spectra or coherent flows across the spiral arms, which are not included in the rotation model. Apart from the spiral arm region, there is no trend between \vrot or \vcent stacked line widths with radius.

We find that the azimuthally-averaged line widths from the second moment have a typical value of $\sim12$ \kms and are larger than all of the stacked profile widths. Since stacking based on \vrot yields the typical dispersion to non-circular motion and the \vcent is biased by asymmetric line wings, it is worrying that the second moment estimates are substantially larger.  Previous studies have shown that the second moment can overestimate the line width relative to stacked profiles \citep{Ianjasm2012AJ....144...96I} and Gaussian fits to individual spectra \citep{Mogotsi2016AJ....151...15M}, particularly in the inner disks of nearby galaxies.  The discrepancy we find shows that, when extended line wings are detected throughout the disk, the second moment line widths overestimate the line width.  In Appendix \ref{app:second_moment}, we find that the second moment line widths increase by $50\%$ between the VLA and VLA+GBT data sets.

% subsection comparing_estimates_of_velocity_dispersion (end)

% section properties_of_the_atomic_medium (end)

\section{Extra-Planar Velocity Components} % (fold)
\label{sec:extraplanar_velocity_components}

Many nearby galaxies have velocity components offset from galactic rotation, either in the form of discrete \hi clouds or extended line wings related to the main disk structure \citep{Sancisi2008A&ARv..15..189S}.  Both features relate to neutral gas accretion onto the main disks and may play an important role in galaxy evolution.  We explore the preferential directions of line wings, as shown in the skewness maps (Figure \ref{fig:skew_maps}), in \S\ref{sub:anomalous_velocity_component}, and discrete \hi sources offset in velocity from the main disk in \S\ref{sub:discrete_hi_clouds}.

\subsection{Anomalous Velocity Component} % (fold)
\label{sub:anomalous_velocity_component}

\begin{figure*}
\includegraphics[width=\textwidth]{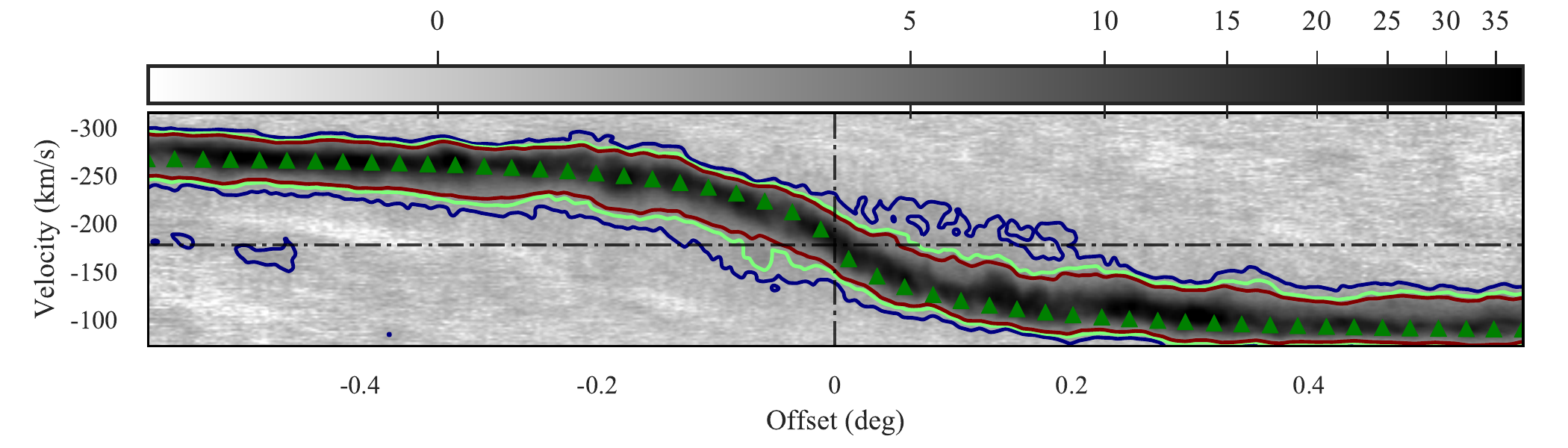}
\caption{\label{fig:pvslice_wide} A position-velocity slice spatially-averaged over a 200\arcsec region centred along the major axis. The $x$-axis is the offset angle measured from the rotation centre of the disk (Table \ref{tab:gal_params}). The gray-scale, indicated by the colorbar, is in K, while the contours are at 1, 2, and 3 K, respectively. The green triangles are the velocities from the rotation model (Appendix \ref{app:rotation_velocities}). The horizontal dot-dashed line is at the systemic velocity, and the vertical dot-dashed line indicates the centre of M33. The region near the centre shows significant emission off the galactic rotation curve. The lowest contour highlights two features near the systemic velocity that are independent with emission from the main disk down to the noise level. Across the slice, the contours show faint emission that is preferentially skewed towards the systemic velocity (Table \ref{tab:gal_params}).}
\end{figure*}

Figure \ref{fig:pvslice_wide} shows a position-velocity (PV) slice along the major axis of the disk from the North to South. Similar to \citet{Kam2017AJ....154...41K}, we find several features either unassociated with the M33's main disk or lagging relative to the rotation velocity.  The bright emission from the main disk, shown in gray-scale, is well-matched to the rotation model (green triangles). The contours highlight the extent of the low-surface brightness features. In particular, the 1 K contour, shown in blue, shows the asymmetry in the line wings directed towards the systemic velocity, consistent with what we find from individual spectra (\S\ref{sub:exploring_spectral_complexities}).  The extents of the blue contours in Figure \ref{fig:pvslice_wide} suggests that the line wing structure is similar on large-scales within the disk, implying that the line wings do not entirely arise from localized outflows from stellar feedback.

The blue contours in Figure \ref{fig:pvslice_wide} highlight asymmetrical features that point towards the systemic velocity. This indicates a rotationally-lagging disk component that is consistent with gas on high-latitude orbits in a thick disk.  The extensive excesses near the centre of the disk may be related to the halo structure. \citet{Kam2017AJ....154...41K} find extensive ``beard''-like structures --- velocity elongations in the PV-slice --- in this region, which are more evident at lower surface brightnesses.

The excess intensity at lagging velocities extends across the disk, as is also indicated by the skewness of the line wings (Figure \ref{fig:skew_maps}). Similar large-scale skewed profiles are found by \citet{Ianjasm2012AJ....144...96I} for a sample of nearby galaxies in the THINGS survey (see their Figure 17).  Their measure of skewness for individual profiles comes from Gauss-Hermite fitting, where higher-order skewness and kurtosis terms can be included directly in the fit.  They find a clear bias in the skewness between the approaching and receding halves for a number of galaxies in their sample.

Similar to the asymmetrical ``limbs'' of opposing skewness along the edge of the map in Figure \ref{fig:skew_maps}, \citet{Ianjasm2012AJ....144...96I} find several other systems with similar features towards the outer regions of galactic disks.  In \S\ref{sub:exploring_spectral_complexities}, we suggest that these limbs are roughly aligned with the position angle of M33's warped disk component \citep[e.g.,][]{Corbelli2014A&A...572A..23C} and such features may arise from similar warped or lagging components in other nearby galaxies. For example, the direction of the asymmetrical regions found by \citet{Ianjasm2012AJ....144...96I} in NGC 2403 is consistent with the position angle of its lagging rotational disk \citep{Fraternali2001ApJ...562L..47F,Fraternali2002AJ....123.3124F}.

Lagging disk components have been found in a number of nearby galaxies, including NGC 2403 \citep{Fraternali2002AJ....123.3124F}, NGC 4559 \citep{Barbieri2005A&A...439..947B,Vargas2017ApJ...839..118V}, NGC 891 \citep{Barnabe2006A&A...446...61B}, and NGC 925 \citep{Sancisi2008A&ARv..15..189S,Heald2011A&A...526A.118H}. As mentioned above, NGC 2403 provides an interesting comparison to M33 because of its flocculent structure.  \citet{Fraternali2002AJ....123.3124F} suggest the lagging component has an origin internal to NGC 2403, rather than being primordial, due to the coherent structure and connection to the main disk. NGC 2403 also appears to have been undisturbed by dynamical interactions \citep[e.g., ][]{Williams2013ApJ...765..120W}.  \citet{Fraternali2002AJ....123.3124F} suggests that the lagging component results from a galactic fountain mechanism, whereby hot ionized gas ejected into the halo from the fountain cools into atomic clouds that accumulate and fall back to the disk.  Observations of ionized gas tracers confirm the presence of the lagging component \citep{Fraternali-Xray2002ApJ...578..109F,Fraternali2004A&A...424..485F}, consistent with the galactic fountain model. For a sample of nearby galaxies, \citet{Sancisi2008A&ARv..15..189S} also find that the extra-planar gas likely results from a galactic fountain, with a small component accreted from the intergalactic medium. \citet{Barnabe2006A&A...446...61B} find that the lagging component in NGC 891 is well-modeled by a baroclinic pressure model dependent on both the density and temperature of the medium.  The model treats the medium as a mixture of a hot homogeneous component surrounding discrete cold \hi clouds. From this, they reproduce the observed velocity gradient of the lagging component in NGC 891.

A lagging disk component from the galactic fountain mechanism may be related to cold \hi ``low-velocity clouds'' (LVCs), which have been observed in the Milky Way. \citet{Stanimi2006ApJ...653.1210S} find a population of small ($\sim$ few pc) LVCs with velocities of $<30$ \kms relative to the Milky Way rotation velocity \citep[see also][]{Lockman2002ApJS..140..331L,Stil2006AJ....132.1158S}.  Adopting their typical column densities of clouds ($>2\times 10^{19}\mbox{ cm}^{-2}$), we compare this to the average column density of material at $v\sim 30\mbox{ km s}^{-1}$ off M33 projected rotation speed ($<2\times 10^{18}\mbox{ cm}^{18}$).  If the off velocity gas were organized into such clouds, it would have a reasonable filling factor of $f<0.1$ and be broadly distributed across the disk.  Our observations of this gas are consistent with an unresolved population of small clouds at high altitudes above the disk.  The material could be raining back down as per a galactic fountain model, or the material could be directly accreting from the hot medium.

Only $1/3$ of the line wing emission ($\sim 8\%$ of the total atomic gas mass) is found in the asymmetric wing components, which would be identified as the ``lagging'' gas.  The remaining $2/3$ of the line wing emission in the symmetric component may require other explanations, especially since the wings are significantly wider than the thermal width of the warm neutral gas. One possible explanation is neutral gas accretion from the halo.  Recently, \citet{Zheng2017AApJ...834..179Z,Zheng2017ApJ...840...65Z} favour a model where $M\sim 10^{8}~M_{\odot}$ of ionized gas in M33's halo is accreting onto the disk at $\sim100$ \kms.  They suggest that these features result from the infall of material produced by a galactic fountain or past interactions between M33 and M31.  A key observation in this analysis is the split between the symmetric and asymmetric components of the line wings (\S\ref{subsub:asymmetry_between_the_northern_and_southern_halves}).  Our symmetric line wings are consistent with this accretion flow.

% subsection anomalous_velocity_component (end)

\subsection{Discrete \hibold Clouds} % (fold)
\label{sub:discrete_hi_clouds}

 Substantial \hi structure has been found within a few degrees of M33 \citep{Grossi2008A&A...487..161G,Putman2009ApJ...703.1486P,Keenan2016MNRAS.456..951K}, including discrete features in an \hi bridge that stretches towards M31 \citep{Lockman2012AJ....144...52L}.  These structures are thought to comprise a population of high-velocity clouds (HVCs) or halo gas surrounding M33, similar to the HVC population in the Milky Way \citep{Sancisi2008A&ARv..15..189S}.  Studies of these structures require excellent surface-brightness sensitivity and have mostly been limited to single-dish surveys at low resolution. \citet{Kam2017AJ....154...41K} provides the first resolved study of these features nearest to the main disk, including elongated velocity structures connected to M33's disk and discrete \hi clouds offset in velocity.  The data used by \citet{Kam2017AJ....154...41K} has a resolution of 120\arcsec, much better than previous deep \hi studies.

\begin{figure*}
\includegraphics[width=\textwidth]{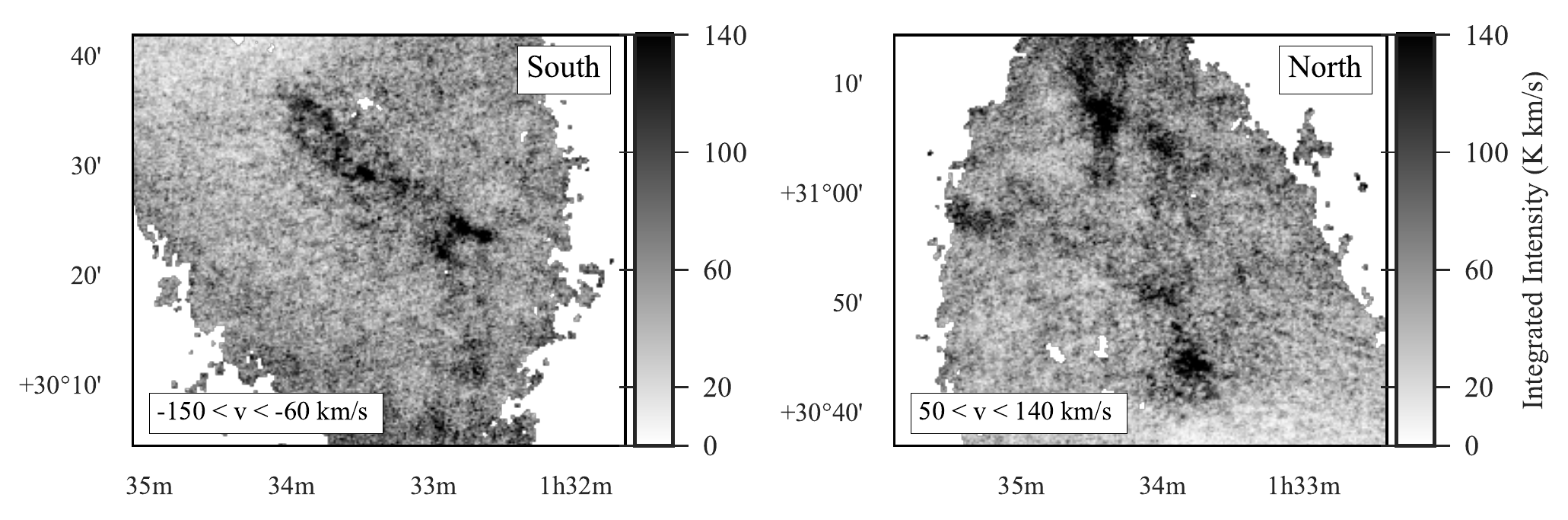}
\caption{Integrated intensity maps of components displaced in velocity from M33's main disk in the southern half (left) and the northern half (right).  The maps are integrated over the rotation-subtracted data cube; a velocity of $v=0$ \kms is set at the systemice velocity of M33.  The emission from the main disk is nearly all confined to velocities of $-60$ to $+50$ \kms (see \S\ref{sub:stacking_spectra}). The left shows the emission structure blue-shifted from M33 (southern half), while the right shows the red-shifted components (northern half). \label{fig:offrot_mom0}}
\end{figure*}

We search for similar discrete \hi structures within the spatial limits of our map. Though most of the \hi emission in the aforementioned surveys is outside this area, we find significant extra-planar structure within our data.  Figure \ref{fig:offrot_mom0} shows the extra-planar emission with integrated intensity maps in the northern and southern halves after subtracting the circular rotation model from the data cube.  While most of the previously known \hi clouds from single-dish studies fall outside of the area mapped our survey area, we find two structures that are consistent with the catalogue from \citet{Keenan2016MNRAS.456..951K}.

We find a total atomic mass of $1.3\pm0.5\times10^7$ \msol in the velocity-integrated regions shown in Figure \ref{fig:offrot_mom0}, which accounts for about 20\% of the total HVC and disk-halo mass in M33 \citep{Lockman2017ASSL..430...49L}, and $\sim1\%$ of the atomic mass of the main and lagging disks.  This mass is a factor of $\sim3$ less than the estimated HVC mass in the Milky Way \citep{Putman2012ARA&A..50..491P}.

The left panel of Figure \ref{fig:offrot_mom0} shows the extent of the blue-shifted components and the right panel shows the red-shifted components. The atomic mass over these regions is $5\pm2\times10^6$ \msol and $8\pm3\times10^6$ \msol for the blue- and red-shifted components, respectively.  In both cases, the extra-planar \hi emission has significant small-scale structure, with features that were unresolved in previous studies of extra-planar gas in M33 \citep[][]{Westmeier2005A&A...436..101W,Grossi2008A&A...487..161G,Putman2009ApJ...703.1486P,Keenan2016MNRAS.456..951K}.  At 80 pc resolution, the structures are filamentary with `blob'-like concentrations.  This is particularly prominent for the blue-shifted components, which appear highly elongated.

The red-shifted integrated intensity shows four structures, one of which is partially cut-off by the edge of the map. The structure nearest to the centre of M33 is a portion of the lagging rotation component highlighted in Figure \ref{fig:pvslice_wide}, shown by the deviation in the green contour at an offset of $-0.05\degree$. The last two structures near the top of the disk (1$^{\mathrm{h}}$34$^{\mathrm{m}}$00$^{\mathrm{s}}$ +31\degree05\arcmin) have been catalogued by \citet{Grossi2008A&A...487..161G} as AA1 and \citet{Keenan2016MNRAS.456..951K}  as AGESM33-1. These structures appear blended at the resolutions of both studies, but are shown here to have multiple faint filamentary structures surrounding the brighter centre. \citet{Keenan2016MNRAS.456..951K} report a mass of $1.18\times10^6$ \msol for the blended structure, about $1/8$ of the total mass we find over this velocity range.

A second structure identified as a HVC by \citet{Keenan2016MNRAS.456..951K}, AGESM33-22, is within the spatial extent of our data but is too faint to be detected in our data at 80 pc resolution. The HVC, located at a velocity of $-338\pm15\,$\kms, has a column density of $7.2\times10^{17}$ cm$^{-2}$ at a resolution of $3.5\arcmin$, well below the sensitivity of our full resolution data.

The structure of the blue-shifted component is dominated by a single filamentary feature that stretches across most of the disk. Thus its mass is that of the blue-shifted region: $5\pm2\times10^6$ \msol.  The portion nearest to M33's centre is comprised of two connected filamentary structures, forming a loop, that then extends further south.  This feature has not been discussed in previous studies, though it is noted in Figure 7 of \citet[][]{Sancisi2008A&ARv..15..189S} using the \citet{Thilker2002ASPC..276..370T} archival VLA data.  It can also be seen in Figure 3 of \citet{Putman2009ApJ...703.1486P} in their Arecibo data at velocities of $-159$ to $-123$ \kms. The Arecibo data indicate that the two components of the filament are indeed connected.  The filament then has a projected length of $\sim8$ kpc.  Similarly, the velocity of the northern tip of the filamentary structure approaches the velocity of the lagging rotational component and the extended structure between $0\mbox{--}0.2\degree$ in Figure \ref{fig:pvslice_wide}.  Filamentary \hi structures with similar lengths have been found in NGC 2403 and NGC 891 \citep{Sancisi2008A&ARv..15..189S}. \citet{Fraternali2001ApJ...562L..47F,Fraternali2002AJ....123.3124F} find a mass of $1\times10^7$ \msol for the filament in NGC 2403, about twice the total mass we find in the blue-shifted off-axis emission in M33.  Structures like this \hi filament in M33 are strong evidence for accretion onto the main disk \citep{Sancisi2008A&ARv..15..189S}.

\subsection{An \hibold Cloud Impacting the Main Disk?} % (fold)
\label{sub:an_hvc_impacting_the_main_disk_}

\begin{figure}
\includegraphics[width=0.5\textwidth]{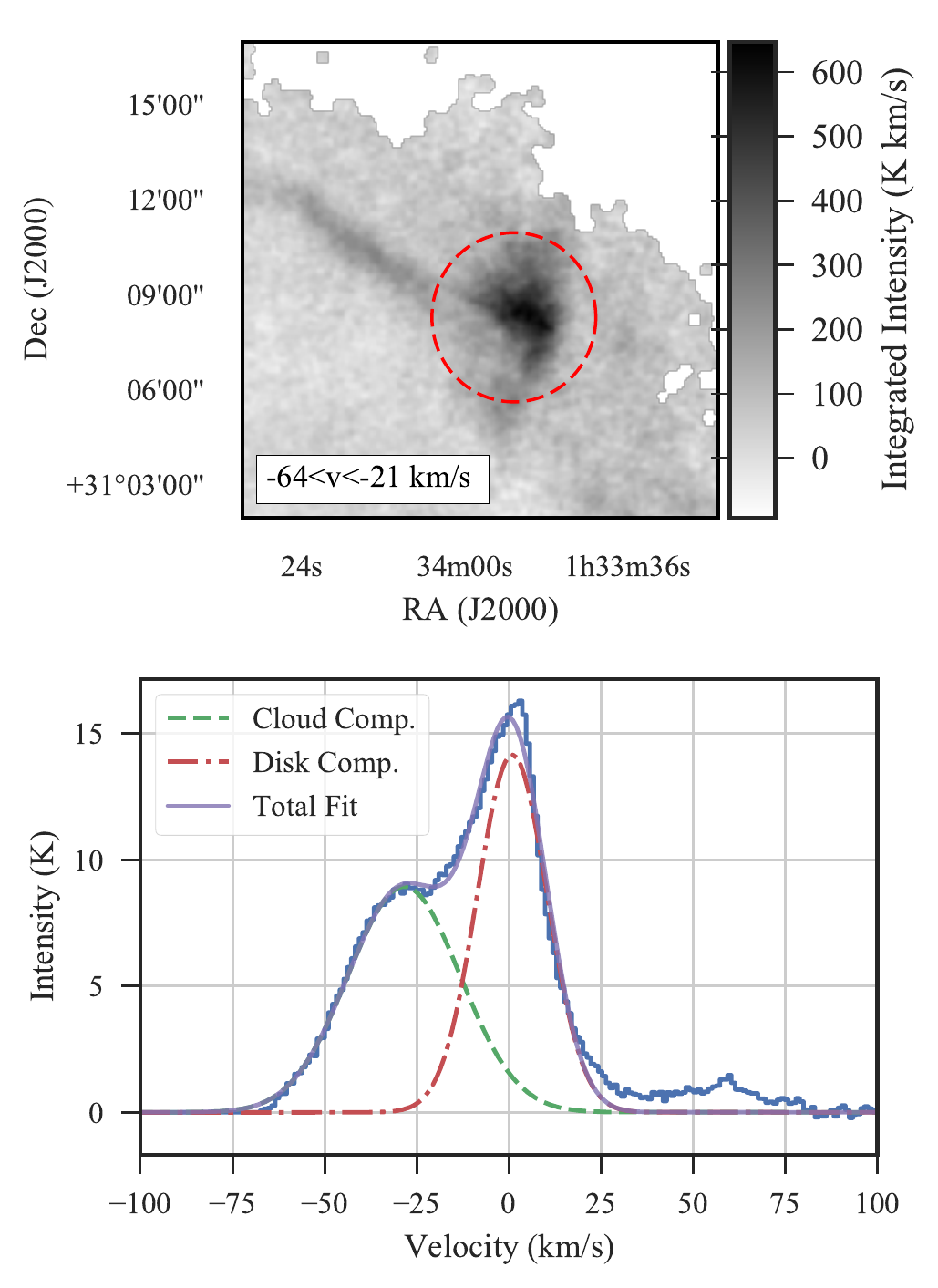}
\caption{\label{fig:nplume_props} Top: An integrated intensity map highlighting the structure of the \hi cloud. The red dashed box in Figure \ref{fig:hi_spectra} shows this position relative to the rest of the galaxy. The velocity range is chosen to minimize emission from the main disk. Bottom: The average spectrum within the red-dashed circle in the top panel.  The velocity axis is defined with respect to the \vrot model from \S\ref{sub:rotation_curve}. We fit two Gaussian components to the spectrum, shown as the green-dashed and red dot-dashed components.}
\end{figure}

An interesting \hi cloud near the northeast edge of the map overlaps in velocity with the main disk, possibly indicating an interaction between an infalling cloud and the disk.  We show the spatial structure of this cloud in the top panel of Figure \ref{fig:nplume_props}.  The morphology of the cloud is similar to the larger-scale clouds in \S\ref{sub:discrete_hi_clouds}, with a central blob-like structure with connecting filamentary structure.  The prominent filamentary feature in the figure reaches toward the northern edge of the map and overlaps in velocity with the main blob.  The blob is about $600$ pc in diameter.

The bottom panel of Figure \ref{fig:nplume_props} shows the average spectrum over the cloud, where the velocity axis is defined with respect to the circular rotation velocity at $v=0$.  There are clearly two spectral components and possibly a third faint component around $60$ \kms. We model the two bright components with Gaussians.  The widths of the fitted components are $15.2\pm0.5$ \kms for the \hi cloud and $9.5\pm0.2$ \kms for the disk.

\citet{Putman2009ApJ...703.1486P} show that this cloud overlaps in velocity with a portion of M33's warped disk, labeled as the ``northern warp'' in their Figure 3, outside of our survey area.  What we detect as a separate \hi cloud may be part of a larger structure related to the warped disk.

Treating the \hi cloud as a discrete structure, we use the cloud's spectral component from Figure \ref{fig:nplume_props} with the total emission over the cloud to derive a mass of $6.5\pm0.5\times10^6$ \msol. This is similar to the total mass found in the blue- and red-shifted \hi clouds in \S\ref{sub:discrete_hi_clouds}.  With this cloud included, the total mass of off-rotation components becomes $\sim2\times10^7$ \msol and there are likely other fainter off-rotation \hi clouds blended with the lagging rotational disk that will also increase this total mass.  A full search of these features will require a more thorough spectral decomposition of the data.

% subsection an_hvc_impacting_the_main_disk_ (end)

% subsection discrete_hi_clouds (end)

% section anomalous_velocity_components (end)

\section{Discussion}
\label{sec:discussion}

Measuring the kinematic properties of the atomic ISM is vital for evaluating and comparing to theoretical models.  We compare two common techniques for estimating kinematic properties in extragalactic observations: the second moment of the line shape and spectral-stacking.  Our results demonstrate that these common techniques are inherently limited and can yield vastly different results.  For example, the method used to estimate the line width can yield values that differ by a factor of two (\S\ref{sub:comparing_estimates_of_velocity_dispersion}).  In this discussion, we compare our results to those found in nearby galaxies and present a simple Gaussian Mixture Model for describing stacked profile shapes. This model recovers the qualitative profile shape found in previous studies, which we use to compare with previous model interpretations of stacked \hi profiles.

\subsection{Previous Work and Interpretations}
\label{sub:disc_prevwork}

Previous work on extragalactic \hi spectra focuses on the \hi line width, measured using the second moment \citep{Tamburro2009AJ....137.4424T}, fitting Gaussian models to individual spectra \citep{Petric2007AJ....134.1952P,Warren2012ApJ...757...84W,Mogotsi2016AJ....151...15M}, or stacking spectra over large spatial regions \citep{Young1996ApJ...462..203Y,Braun1997ApJ...484..637B,Ianjasm2012AJ....144...96I,Stilp2013ApJ...765..136S}.  The first method is attractive since second moments are simple to compute, though it assumes the profile is close to a Gaussian.  \citet{Tamburro2009AJ....137.4424T} use line widths from the second moment for the THINGS survey and find typical values of $10\pm2$ \kms across the sample.  Since these line widths are larger than the thermal line width expected for the WNM ($\sim8$ \kms), the profiles are interpreted as being broadened due to turbulent motions.

Modeling individual spectra by fitting Gaussian components has the advantage of measuring variations down to the resolution of the data, and being able to account for multiple components.  \citet{Young2003ApJ...592..111Y} and \citet{Warren2012ApJ...757...84W} find that most \hi spectra can be modelled by up to two Gaussian components.  Both studies find a range of line widths in both components, suggesting emission components arise from both CNM and WNM phases.  The narrow components have widths of $<6$ \kms and are attributed to the CNM, with the line width broadened substantially from the thermal width by turbulence.  However, \citet{Warren2012ApJ...757...84W} do not find a clear connection between the location of narrow components and star-forming regions.  Using a similar data set as \citet{Tamburro2009AJ....137.4424T}, \citet{Mogotsi2016AJ....151...15M} fit a single Gaussian model to all \hi spectra where \co\, is detected. They find line widths of $10\pm2$ \kms, consistent with estimates from the second moment.

Alternatively, non-Gaussian line profiles have been modeled using Gauss-Hermite polynomials, which can account for variations in the skewness and kurtosis.  While this method often provides a good analytical description, the connections between the model parameters and the underlying physical properties of the medium are unclear \citep{Young2003ApJ...592..111Y}. \citet{Young1996ApJ...462..203Y}, \citet{Young2003ApJ...592..111Y}, \citet{Ianjasm2012AJ....144...96I} and \citet{Stilp2013ApJ...765..136S} use this model to estimate the velocity at the peak intensity in their data, which are then shifted to a common centre and stacked.

The \hi stacking method has the advantage of vastly increasing the S/N in the data, allowing low surface brightness emission to be studied at the expense of removing spatial variations.  Nearly all studies find the same basic stacked profile shape: a near-Gaussian peak with enhanced line wings.  This shape is not well-described by physically-motivated profile shapes. The enhanced line wings lead to poor single Gaussian fits, yet they are not prominent enough to match a Lorentzian profile.  This has led to different modeling approaches with differing interpretations of the results.  \citet{Young1996ApJ...462..203Y} model the stacked profiles as a wide and narrow Gaussian component. These two components have widths of $9$ \kms and $3$ \kms, respectively.  Since these are similar, though slightly larger, than the expected thermal line widths for WNM and CNM phases, they attribute the model as separating out the two different phases, with some broadening from turbulence.  A larger sample of 34 galaxies from the THINGS survey was analyzed by \citet{Ianjasm2012AJ....144...96I} using the same two-Gaussian model; the sample overlaps with the independent analyses by \citet{Tamburro2009AJ....137.4424T} and \citet{Mogotsi2016AJ....151...15M}.  They find similar results to \citet{Young1996ApJ...462..203Y}, with narrow components arising from the CNM with Gaussian widths of 3.4 to 8.6 \kms, and wide WNM components with widths from 10.1 to 24.3 \kms.  These ranges imply that both the CNM and WNM must be highly turbulent since the line widths are much larger than the expected thermal line widths of these phases.  We note that, for comparison with our results, the data from the THINGS survey has a coarser spectral resolution compared to our data, with channel sizes from $1.3$ to $5.2$ \kms.

\citet{Petric2007AJ....134.1952P} propose an alternate interpretation of the enhanced line wings in stacked profiles.  They argue that the flux ratio of the narrow component, if exclusively tracing the CNM, to the wide component should steeply decline with galactocentric radius, matching the observed star formation activity.  Instead, they find that the ratio stays roughly constant to large radii.  They attribute the central component to a mixture of atomic gas, with enhanced line wings that describe bulk motions, possibly from stellar feedback or infalling halo gas. \citet{Stilp2013ApJ...765..136S} test this interpretation by modelling the stacked profiles as a single Gaussian peak with line wings measured by their fractional increase relative to the Gaussian.  For a sample of 24 dwarf galaxies, they find line widths ranging from $\sim6 - 10$ \kms and $f_{\rm wings} \sim 0.05$ to $0.2$.

An alternative model to the previous studies is introduced by \citet{Braun2009ApJ...695..937B} and \citet{Braun2012ApJ...749...87B} where spectra with flattened tops are attributed to a cool \hi layer sandwiched between two warmer layers \citep{Braun1997ApJ...484..637B}.  It is difficult to make comparisons to this model without accounting for multi-component spectra when calculating the line width.  We defer further discussion of this model to future work.

Most methods have also been used to study radial trends in the \hi properties. \citet{Tamburro2009AJ....137.4424T} examine radial variations in the \hi line width, finding that the line width tends to decrease rapidly in $R_{\rm gal}$ to $R_{25}$\footnote{Defined as the semi-major axis of the $25$ mag arcsec$^{-2}$ isophote in B-band.}, similar to the radial decline in the star formation rate surface density.  The slope beyond this radius in some galaxies flattens significantly.  \citet{Stilp2013bApJ...773...88S} and \citet{Ianjam2015AJ....150...47I} examine radial trends by creating stacked profiles in radial bins.  They find the same qualitative result despite using different models: the line width tends to decrease with radius.  This trend is found in most nearby galaxies, though with three exceptions: NGC 2403, NGC 2976, and NGC 628 \citep{Ianjam2015AJ....150...47I,Mogotsi2016AJ....151...15M}. These systems have a shallow radial decrease in the line width throughout the disk. The first two galaxies are flocculent spirals similar to M33. NGC 628 is classified as a grand spiral galaxy, though its mass and star formation rate is within a factor of 2 compared to M33 \citep{Walter2008AJ....136.2563W}.

These studies cover the \hi properties of most nearby galaxies, spanning a range of galaxy properties and types.  \citet{Young1996ApJ...462..203Y}, \citet{Young2003ApJ...592..111Y}, and \citet{Stilp2013ApJ...765..136S} restrict their sample to dwarf galaxies, where the low-metallicity properties and lack of large-scale galactic structure may lead to differences in the ISM properties compared to the canonical view of the ISM.  \citet{Tamburro2009AJ....137.4424T}, \citet{Ianjasm2012AJ....144...96I}, and \citet{Mogotsi2016AJ....151...15M} include a mix of galaxy types in their sample. \citet{Ianjasm2012AJ....144...96I} find no difference in the \hi line widths of stacked profiles between dwarf and spiral galaxies.  Furthermore, \citet{Caldu2013AJ....146..150C} analyze the radial line width trends in the \hi and \co\ for a similar set of nearby galaxies and find no common radial variation.  These results are consistent with all stacked profiles in these studies having the same qualitative shape \citep{Petric2007AJ....134.1952P}, suggesting that the physical mechanism governing typical line profiles is similar regardless of galaxy type.
% However, these studies also find a large range in line width, highlighting galaxy-to-galaxy variations that are not due to the galaxy type.

\citet{Druard2014A&A...567A.118D} present an analysis of \hi stacked profiles in M33, using the archival VLA \hi data presented in \citet{Gratier2010A&A...522A...3G}.  In this work, they emphasize comparisons of the \hi to stacked profiles of \cotwoone.  In 1 kpc radial bins, they find an average \hi stacked profile width of $\sim6.5$ \kms by fitting a single Gaussian to the stacked profiles. The line width decreases slowly with radius --- a decrease of just $\sim2$ \kms between the bins at $R_{\rm gal}=0\mbox{--}1$ kpc and the $7\mbox{--}8$ kpc bin --- similar to the three galaxies noted above.

\subsection{Key Results in M33}
\label{sub:disc_m33results}

Our results are broadly consistent with previous studies. In particular, we find a line width of $6.7\pm0.2$ \kms from the \vpeak stacked profile, consistent with the M33 study from \citet{Druard2014A&A...567A.118D}.  The line widths of the stacked profiles over radial bins, ranging largely between $5.5$ and $7.5$ \kms, are also consistent with \citet{Druard2014A&A...567A.118D}.  This is encouraging since: (1) the stacked profiles of the archival \hi data should have a similarly high S/N as our data, with the only major difference being the lower spectral resolution of the archival data ($1.2$ \kms), and (2) a different methodology is used in parameterizing the line shapes.  The stacked profiles between the two studies also have a consistent shape, with a kurtosis excess in the central peak and prominent line wings.  The kurtosis excess is similar to that found by \citet{Braun1997ApJ...484..637B} and more extreme than other recent studies \citep[e.g.,][]{Ianjasm2012AJ....144...96I,Stilp2013ApJ...765..136S}.

The line widths from the \vpeak stacked profiles are near the lower limit of the range of line widths found by \citet{Ianjasm2012AJ....144...96I} and \citet{Stilp2013ApJ...765..136S} for nearby galaxies (see Appendix \ref{app:modelling_super_profiles}).  However, since these profiles have non-Gaussian shapes, we note that the equivalent Gaussian width should be treated as an upper limit on the line width.  While the HWHM-scaling method from \citet{Stilp2013ApJ...765..136S} is appealing because it does not require fitting an analytical model, it still has limitations on how well the profile shape can be described.  We discuss the source of these difficulties in \S\ref{sub:mix_models}.

There are some discrepancies between the range of values we find for the other HWHM model parameters compared to the set of dwarf galaxies in \citet{Stilp2013ApJ...765..136S} and \citet{Stilp2013bApJ...773...88S}:

\begin{enumerate}
    \item We find consistently larger $f_{\rm wing}$ values, ranging from 0.25 to 0.35 in radial bins, than the 0.1 to 0.25 range in the \citet{Stilp2013bApJ...773...88S} sample (Figure \ref{fig:hwhm_properties}).  This range is, however, consistent if we only consider the symmetric line wing fraction (Figure \ref{fig:hwhm_wing_ns}).  Thus the symmetric line wing component, possibly arising from inflowing or outflowing material (\S\ref{sec:extraplanar_velocity_components}), is similar with those measured in nearby dwarfs and the difference is due to the presence of a lagging rotational disk in M33.

    \item \citet{Stilp2013bApJ...773...88S} find a large range of asymmetries when stacking spectra in radial bins\footnote{Note that the asymmetry parameter in their study returns the fraction of asymmetry, bounded between 0 and 1: \\ $a_{\rm full} = \frac{\sum_v \sqrt{[S(v) - S(-v)]^2}}{\sum S(v)}$}.  We only find large asymmetries in M33 when stacking over radial bins in the northern or southern half, respectively (Figure \ref{fig:hwhm_properties}).  The shape within the HWHM of the stacked profiles in \citet{Stilp2013bApJ...773...88S} appear close to symmetrical (see their Figure 4), and most of the asymmetry occurs in the line wings.
\end{enumerate}

The estimated line widths from the second moment have an average of $\sim12$ \kms, consistent with the upper limit of the range found by \citet{Tamburro2009AJ....137.4424T}.  These widths are $50\%$ larger than largest widths of the stacked profiles (Figure \ref{fig:veldisp_profile}). In Appendix \ref{app:second_moment}, we demonstrate that the second moment is sensitive to the line shape, and can drastically alter the second moment line width when extended line wings are present. Thus, we discourage the use of the second moment for estimating line widths. Similar differences between other line width methods and the second moment are noted in several studies \citep{Ianjasm2012AJ....144...96I,Stilp2013ApJ...765..136S,Mogotsi2016AJ....151...15M}.

\subsection{Mixture Models for Stacked Profiles}
\label{sub:mix_models}

As we explain in \S\ref{sub:disc_prevwork}, the shape of stacked profiles has led to different modelling approaches and different interpretations of their meanings.  To illustrate why stacked profiles have shapes that are difficult to model using simple analytic forms, we invoke a Gaussian mixture model as a framework for understanding the line wings and widths.  We consider the case where a stacked profile $S(v)$ is an average over an ensemble of Gaussian line profiles with purely thermal line widths:

\begin{equation}
S(v) = \sum_i w_i \exp\left[-\frac{(v-v_i)^2}{2\sigma_{\mathrm{T},i}^2}\right]
\end{equation}
where $w_i$ is the weight of each spectrum, $v_i$ is the central velocity and $\sigma_\mathrm{T}$ is the thermal line width for temperature $T$, i.e., $\sigma_\mathrm{T} = \sqrt{kT/\mu m_{\mathrm{H}}}$.  The weights are normalized such that $\sum_i w_i = 1$.  We can describe the shape of the stacked profile $S(v)$ based on the moments of the distribution.  We use the derivation from \citet{Wang2015_mixmod} to describe the first four moments. The centroid (first moment) of the stacked profile is $v_0 = \sum_i w_i v_i$, which depends only on the distribution of the line centre $v_i$.  The variance (second moment) of the stacked spectrum is then
\begin{equation}
    \label{eq:mix_sigma}
    \sigma^2 = \sum_i w_i (\sigma_{\mathrm{T},i}^2 + v_i^2) - v_0^2
\end{equation}
and the skewness (third moment) and kurtosis (fourth moment) are respectively
\begin{align}
\label{eq:mix_skew}
\mathrm{skew}&= \frac{1}{\sigma^3} \sum_i w_i (v_i - v_0) \left[3\sigma_{\mathrm{T},i}^2 + (v_i - v_0)^2 \right] \\
\label{eq:mix_kurt}
\mathrm{kurt}&= \frac{1}{\sigma^4} \sum_i w_i \left[3\sigma_{\mathrm{T},i}^4 + 6(v_i-v_0)^2 \sigma_{\mathrm{T},i}^2 + (v_i-v_0)^4\right] - 3.
\end{align}
The critical result of these moments is that the stacked profile will depend on the distributions of the line centre and width of the individual spectra that were averaged over.  For example, the shape of the stacked profile could be reproduced by adopting a reference probability density function for $w_i = f(T,v_i)$ encoding the amount of material found at a given temperature and velocity offset from the reference velocity.

We investigate how the common features in stacked profiles found in different studies can be attributed to the dependence on the distributions of the line centre and width of individual spectra that comprise the stacked profile.

\begin{enumerate}
    \item Velocity offsets contribute to the line width. --- Equation \ref{eq:mix_sigma} has an equal dependence on the line width of the spectra and the distribution of their central velocities.  The differences between the \vrot, \vcent, and \vpeak stacked profile widths is due to this dependence (Figure \ref{fig:veldisp_profile}). Stacking based on \vpeak minimizes the variation of the central velocity to within the uncertainty of \vpeak ($0.1$ \kms). It represents the optimal recovery of the mixture of the line widths.  The \vcent and \vrot stacked profiles are instead dominated by the distribution of the central velocities, not the distribution of the line widths. The purpose of the different approaches is discussed further in \S\ref{sub:disc_interp_stacking}.

    \item Skewness depends on the distribution of line centres. --- From Equation \ref{eq:mix_skew}, an asymmetric stacked profile can only occur if the distribution of line centres is also asymmetric since the sign of the skewness is determined by the $(v_i - v_0)$ term. Stacked profiles using \vrot or \vcent are both close to symmetric about $v=0$, but \vcent can be biased by the asymmetric line wings.  To get asymmetric stacked profiles from the former two line centres, an additional component (i.e., the lagging disk) must be present in the line profiles.

    \item Stacked profiles of the atomic medium almost always have a line wing excess. --- The kurtosis from Equation \ref{eq:mix_kurt} has a strong dependence on the distribution of line centres and the line widths.  Since the terms all have even powers, non-Gaussian features in either distribution will always increase the kurtosis, causing the mixture model to have an excess in the tails. For the atomic medium, the distribution of temperatures, which only affect $\sigma_{T, i}$ will be non-Gaussian since it will contain a mixture of CNM and WNM temperatures, with some fraction in the unstable intermediate regime. Thus, without requiring a separate physical process to produce line wings, stacked profiles of the atomic medium will always have an excess in the line wings.  The only case where the kurtosis will decrease is if the majority of the spectra have multiple line components. This case is not explicitly handled in this toy model, but would tend to broaden the profile and flatten the peak, resulting in a kurtosis deficit.
\end{enumerate}

Without requiring additional physical processes, this mixture model can qualitatively reproduce properties of observed stacked profiles based only on a set of Gaussian components.  This accounts for the near universal stacked line shape noted in \citet{Petric2007AJ....134.1952P} and \citet{Stilp2013ApJ...765..136S} when stacking based on the peak velocity.

By adopting simple distributions for the line centres and widths, we test whether the mixture model of thermal-Gaussian components can produce simulated stacked profiles with the quantitative properties of observed stacked profiles.  We find that reasonable distributions of \hi temperatures creates narrower stacked profiles with smaller line wing fractions than the observed stacked profiles.  We simulate a stacked profile from the mixture model by sampling $10^4$ spectra from a uniform temperature distribution ranging from $10^2$--$10^4$ K.  We also sample the central velocities from a Gaussian distribution with width $\sigma=1$ \kms to compare with the \vpeak stacking in our data.  The resulting stacked profile is similar in shape to the \vpeak stacked profiles, with a sharp peak and enhanced line wings. Using the same model used for the stacked profiles, the mixture model profile has a width of $\sigma=4.8$ \kms, $\kappa=-0.03$ and $f_{\rm wings}=0.09$.  When using a wider central velocity distribution to compare with the \vrot stacked profiles, we find that the mixture model also has a narrower width and smaller $f_{\rm wings}$.

We create another simulated stacked profile, but alter the temperature distribution to range from $10^3$--$10^4$ K.  This increases the width of the profile to $\sigma=5.2$ \kms, but makes the shape closer to a Gaussian than the observed profiles, with $\kappa=-0.01$ and $f_{\rm wings}=0.06$.  This suggests that a cold atomic component in the stacked profiles is responsible for the high kurtosis excess in observed stacked profiles.

The mixture model demonstrates that the qualitative stacked profile shape can be reproduced from a mixture of Gaussian components and a range of plausible thermal temperatures.  This highlights the difficulty in using simple analytical models for describing stacked profiles.  We find that observed stacked profiles are wider and have more prominent line wings than the simplified mixture model presented here (Figure \ref{fig:hi_spectra}). These differences must arise from a combination of the CNM/intermediate/WNM fractions, the lagging rotational disk, multi-component spectra, and turbulence. We conclude that stacked profiles reflect this full set of physical effects in M33, rather than being solely the distribution of CNM/WNM temperatures. In particular, there must be significant contributions from a multi-modal distribution of line centres (e.g., from multiple components) to explain the profiles. Because our data can resolve multiple velocity components both spatially and spectrally, we will revisit this analysis in future work through the multi-component fitting to our data, mirroring the approaches used in Galactic studies \citep[e.g.,][]{Lindner2015AJ....149..138L,Henshaw:2016el}.

\subsection{Interpreting Line Wings}
\label{sub:disc_wings}

The various models used for \hi stacked profiles are driven by different interpretations of the line wings. The two-Gaussian models used by \citet{Ianjasm2012AJ....144...96I} argues that the line wings trace a turbulent WNM component, while \citet{Stilp2013ApJ...765..136S} assign the wings as the product of feedback.  Our findings suggest that line widths result from a mixture of different physical processes, making it difficult to uniquely identify the source of the line wing excess without a detailed knowledge of the processes that can broaden the line profiles. The mixture model presented in \S\ref{sub:mix_models} shows that the excess line wings naturally result from combining a set of Gaussian components. Since both turbulent and thermal motion produce Gaussian profiles, the effect of each component cannot be determined from the line wing fraction without additional information.

Both in individual spectra and stacked profiles (over the approaching or receding halves of the galaxy), we find that line wings are not symmetric and are consistently skewed towards the systemic velocity (Figure \ref{fig:skew_kurt_profile}).  Using the \vpeak stacked profiles in the northern and southern halves, we find that about $1/3$ of the line wing fraction results from this asymmetric component (Figure \ref{fig:hwhm_wing_ns}).  Based on our comparison of off-rotation and extra-planar structure in \S\ref{sec:extraplanar_velocity_components}, we identify this component as arising from the lagging rotational disk.  Figure \ref{fig:hwhm_wing_ns} shows that the lagging rotational component has roughly an equal contribution to the line wing fraction throughout the disk.

The symmetric component of the line wing fraction, however, shows substantial variation with galactocentric radius, though there is no consistent trend. Instead, the variations appear to result from local processes within the disk.  For example, there is an increase in the mid-disk ($R_{\rm gal}=2\mbox{--}4$ kpc) where the spiral arms dominate, possibly driven by enhanced recent star formation in the region and bulk non-circular motions driven by the arm mechanism.

The radial trend is inconsistent with the symmetric line wing component being driven solely by stellar feedback.  Line wings resulting from stellar feedback should correlate well with tracers of the star formation rate, which drops exponentially with radius in M33 \citep{Heyer2004ApJ...602..723H,Boquien2015A&A...578A...8B}, as does the \htwo surface density \citep{Druard2014A&A...567A.118D}.  \citet{Stilp2013ApJ...765..136S} find a correlation between the SFR and line wing fraction for stacked profiles over the full galaxy of their sample of dwarf galaxies. However, the correlation is tenuous when the stacking is performed in radial bins \citep{Stilp2013bApJ...773...88S}.  Many of the galaxies in their sample show relatively constant $f_{\rm wings}$, or otherwise peak far from the centre of the disk, as we find in M33.  The lack of a radial trend may also be inconsistent with the turbulent WNM interpretation, since turbulent motion is expected to decrease with radius.  We discuss this in context with the line width of the stacked profiles in \S\ref{sub:disc_constlinewidth}.

The increase in $f_{\rm wings}$ beyond 6 kpc, driven entirely by the symmetric component, emphasizes that multiple processes must contribute to the line wings. The increase in this region is consistent with the behaviour in skewness and kurtosis analyses in \S\ref{sub:exploring_spectral_complexities}, and we attribute the variations in \S\ref{subsub:radial_variations_in_stacked_profiles} to M33's warped disk component. \citet{Corbelli2014A&A...572A..23C} and \citet{Kam2017AJ....154...41K} both find evidence for the kinematic effects of the warp at similar radii.

\subsection{A Shallow Radial Gradient in the Line Width}
\label{sub:disc_constlinewidth}

As mentioned in \S\ref{sub:disc_prevwork}, the central regions of many nearby galaxies have a steep radial gradient in the velocity dispersion, with notable exceptions including the flocculent spirals NGC 2403 and NGC 2976. From Figure \ref{fig:veldisp_profile}, we find that all of the line width measurements have a shallow radial trend.  The line widths from the \vpeak stacked profiles decrease by just $\sim1.5$ \kms between a radius of 2 and 8 kpc.  This variation is consistent with the radial decrease found in M33 by \citet{Druard2014A&A...567A.118D} using 1 kpc bins.

The measured line widths from the \vpeak stacked profiles are narrower than the expected thermal line width of $\sim8$ \kms for the WNM \citep[for a temperature of $\sim8500$ K;][]{Wolfire1995ApJ...443..152W}.  And as found using the Gaussian Mixture model, the excess kurtosis universally found in stacked profiles requires a contribution of narrow Gaussian components with equivalent thermal widths of $T<1000$ K.  The line widths of the stacked profiles should then be some combination of varying thermal line widths, ranging from CNM to WNM temperatures, with turbulent broadening, which plausibly varies as well.

The line width traces the distribution of thermal and kinetic energy in the atomic medium.  Cooling and dissipative mechanisms will remove this kinetic energy, so maintaining these line widths will require energy injection.  Previous work has used the energy balance in the ISM of other galaxies to determine sources for the energy injection \citep[e.g.][]{Tamburro2009AJ....137.4424T, Stilp2013bApJ...773...88S}.  All 11 galaxies examined in \citet{Tamburro2009AJ....137.4424T} have a steep line width gradient, which tend to flatten towards the outer disk.  They find that feedback via supernova is not sufficient to maintain the observed line widths in the outer regions, but magneto-rotation instabilities (MRI) plausibly could.

In particular, M33 has an increasing (not flat) rotation curve (Figure \ref{fig:rotation_curve}), which contrasts with models used in the previous work.  Similarly the star formation rate declines exponentially with radius \citep{Heyer2004ApJ...602..723H,Corbelli2014A&A...572A..23C}, providing insight into the rate of energy injection by core collapse supernova.  We estimate the volumetric energy dissipation rate by assuming motions are turbulent with an outer scale set by the disk scale height of the atomic ISM in the galaxy \citep[$H$, taken to be 100 pc;][]{Tamburro2009AJ....137.4424T} and dissipate on crossing times ($\tau_c = H \sigma^{-1}$) for the system:
\begin{equation}
\dot{u}_{\mathrm{diss}} = \frac{3}{2}\frac{\Sigma \sigma^2}{H \tau_c} = \frac{3}{2}\frac{\Sigma \sigma^3}{H^2},
\end{equation}
where $\Sigma$ is the surface density of the atomic ISM. The rate of MRI energy injection has been estimated as
\begin{equation}
\dot{u}_{\mathrm{MRI}} = 0.6 \frac{B^2}{8\pi} R \frac{d\Omega}{dR}
\end{equation}
following \citet{MacLow2004RvMP...76..125M}.  Here $B$ is the characteristic magnetic field strength, taken to be a constant $B=8~\mu\mathrm{G}$ with radius based on modeling of the non-thermal radio continuum by \citet{Taba2008A&A...490.1005T}. We use our rotation curve model to evaluate orbital frequency $\Omega$ at galactocentric radius $R$. We take the energy injection from core-collapse supernova from \citet{MacLow2004RvMP...76..125M}:
\begin{equation}
\dot{u}_{\mathrm{SN}} = \epsilon_{\mathrm{SN}} E_{\mathrm{SN}} f_{\mathrm{SN}} \frac{\dot{\Sigma}_{\star}}{\langle m \rangle H}.
\end{equation}
Here $\epsilon_{\mathrm{SN}}\sim 0.1$ is the efficiency with which supernovae inject energy into the ISM, $E_{\mathrm{SN}} = 10^{51}\mbox{ erg}$ is the energy of a supernova, $\dot{\Sigma}_{\star}$ is the star formation rate surface density, $\langle m \rangle=10^2~M_{\odot}$ is the average mass of stars that forms per core collapse supernova \citep{Tamburro2009AJ....137.4424T} and $H$ is the scale height of star formation, taken to be 100 pc again.  We use the star formation rate maps from \citet{Boquien2015A&A...578A...8B}, specifically the combined maps from the Galex FUV and Spitzer/MIPS 24 $\mu$m data. We generate a radially averaged star formation rate profile as the median of data in annular bins with width of 50 pc.

\begin{figure}
    \centering
    \includegraphics[width=\columnwidth]{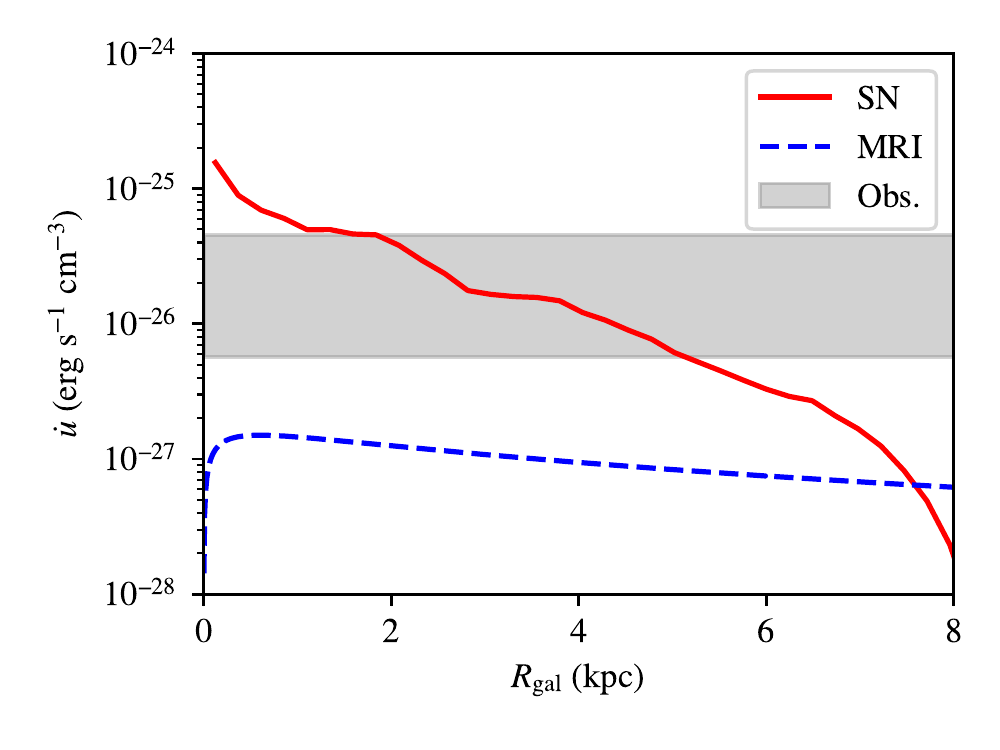}
    \caption{Radial profiles of energy injection for core-collapse supernova (SN; solid red line) and magnetorotational instability (MRI; dashed blue line) and estimates of the turbulent dissipation rates in the atomic ISM. Neither mechanism, by itself, matches the range and constant profile of the observed dissipation rates.}
    \label{fig:energy}
\end{figure}

In Figure \ref{fig:energy}, we plot the radial profiles of the turbulent energy dissipation adopting a range of line widths from 6 to 12 km~s$^{-1}$ and a constant surface density of $8~M_{\odot}\mbox{ pc}^{-2}$, which describe our data well.  Neither injection mechanism can explain the line widths in isolation.  Because of the rising rotation curve and comparatively low shear, MRI driving rates are significantly below the dissipation rate estimates, but the MRI mechanism shows a flat profile consistent with the observed shallow trends in line width and surface density.  The star formation rates falls off exponentially and cannot explain the flat profile of the line widths and surface density but it does broadly agree with the range of required energy injection rates. Appealing to Type Ia supernovae as an additional source of turbulent motion will not change the profile substantially since the old stellar disk in M33 also declines exponentially with a scale length of 1.4~kpc \citep{Regan1994ApJ...434..536R}.

This model adopts a constant atomic gas scale height throughout the observed region, but the scale height likely increases with galactocentric radius.  While the radial variation of the scale height of gas in M33 has not been measured, other disk galaxies show that the atomic gas scale height increases linearly by a factor of two over the optical radius of the galaxy, $R_{25}$ \citep[e.g.,][]{Yim2014AJ....148..127Y}. For M33, $R_{25} = 9$ kpc \citep{RC31991rc3..book.....D}, implying that $\dot{u}_{\mathrm{diss}}$ would decrease by a factor of $<4$ over the plotted region.  The supernova driving would also show a steeper decline, dropping by an additional factor of two.  However, the thickness of the star-forming disk is relatively constant \citep{Yim2014AJ....148..127Y} so the scale height used in $\dot{u}_{\mathrm{SN}}$ may remain constant with radius.  Even with this assumption, a flaring \hi disk does not clearly match the factor of $10^2$ decline in supernova driving in this simple model.  Using higher resolution \hi data, it will become possible to map spatial variations in the scale height through changes in the turbulent properties of the gas \citep{Padoan2001ApJ...555L..33P} and investigate these effects in more detail.

Without a more detailed study, we thus arrive at a similar conclusion as \citet{Stilp2013bApJ...773...88S}, who examine a number of energy sources for driving turbulence in their sample of dwarf galaxies and find that a single source cannot maintain the observed line widths.

\subsection{Interpretation of Different Stacking Models}
\label{sub:disc_interp_stacking}

The properties of stacked profiles depend on how the spectra are aligned.  The appropriate choice of which definition of the line centre to use --- the rotation velocity (\vrot), centroid velocity (\vcent), or velocity at peak intensity (\vpeak) --- depends on the goal of the analysis.

Stacking based on \vrot maximizes contributions from non-circular motions in the disk.  In M33, the width of the velocity residuals is significant when compared to the typical WNM thermal width.  The line width of this stacked profile will measures the average kinetic energy in the ISM in the galactic potential.

The properties of the atomic ISM components at the spatial resolution of the data are best recovered by stacking based on \vpeak since this minimizes the variance in the line centre distribution for the bright components.  Line wings are also the most significant in this method but trace distributions of temperature in the ISM as well as multiple velocity components along the line of sight (\S\ref{sub:mix_models}).   Assuming the statistical uncertainty on \vpeak is small, the resulting stacked profile will recover the average line shape at the resolution of the data.  We explicitly test this in a forthcoming paper (Koch et al., in prep) and our preliminary results show that individual \hi components have a median line width of $\sim6$ \kms based on a Gaussian fit, which matches the $6$--$7$ \kms found for the \vpeak stacked profiles (Table \ref{tab:stacked_fits}). However, in our upcoming work, we find that individual components have widths ranging from $4$--$10$ \kms.  While the \vpeak profiles appear to recover the typical spectrum's properties, it necessarily removes spatial variations in the widths.

The line width of the \vcent stacked profile can be interpreted as tracing the kinematics of the internal motions of the neutral medium, since, for an optically thin tracer, the centroid is the mass-weighted average velocity.  The dispersion around this velocity represents the energy of the gas-to-gas relative motions, as opposed to the gas-to-galactic potential motions traced by the rotation-stacked profile. Because \vcent is biased by the line shape, however, caution must be used if the line shapes have a preferential skewness as this will artificially broaden the velocity distribution.  Creating a map of the skewness (Figure \ref{fig:skew_maps}) or checking the asymmetry of stacked spectra over limited regions of the galaxy (Figure \ref{fig:hwhm_properties}) can be used to diagnose how substantial the bias will be.

Thus, the line width in each of the stacked profile types result in a different but physically meaningful measure of typical atomic ISM properties.  Differences in the interpretations of these line widths present in the literature is partly attributable to real variations, but also arise from the method used for modeling. Most of the literature aims to characterize components of the ISM, however we caution that these interpretations are necessarily uncertain, even with good velocity and spatial resolution.  We demonstrate this uncertainty in the source of stacked profile properties with the Gaussian Mixture Model in \S\ref{sub:mix_models}.  Reasonable distributions of ISM properties --- for example the temperature --- can produce the qualitative properties of observed stacked profiles.  However, a realistic mixture model must also include distributions for other physical processes, such as turbulent broadening, outflows from stellar feedback, bulk non-circular motion and lagging rotational motion, amongst others.

For the M33 data, the best recourse is to perform a multi-component Gaussian decomposition on a per line-of-sight basis. Such methods can realistically only work for good-quality observations of nearby galaxies.  These detailed analyses are able to separate the effects of different processes by retaining spatial relationships.  We recommend that studies seeking to separate physical processes either model individual spectra \citep[e.g., ][]{Warren2012ApJ...757...84W} or average spectra over the size scale of the targetted processes. For more distant targets where the scale of most processes will be unresolved, or for poor S/N data, stacking methods are still useful. Though the stacked profile will combine many physical processes together, it can be used as a powerful measure of similarity between different galaxies. For example, the consistent line widths found by \citet{Ianjasm2012AJ....144...96I} between dwarf and spiral galaxies suggests that some atomic ISM properties are unchanged by galaxy type.

% section discussion (end)

\section{Summary}
\label{sec:summary}

We present new L-band observations of M33 taken by the VLA in C-configuration.  The new data set yields a spatial resolution of $\sim18$\arcsec, tracing $\sim80$ pc scales, and significantly improves the sensitivity and spectral resolution compared to previous observations with an rms brightness temperature of 2.8 K in a 0.2 \kms channel.

\begin{enumerate}

\item The new \hi VLA data recover 72\% of the total \hi emission in M33 compared to estimates from single-dish data within the VLA survey region.  After combining the VLA data with GBT observations, the total emission matches the emission from the GBT data alone. This gives a total atomic mass of $1.3\pm0.2\times10^9$ \msol.  The combined data cube is fully-sampled down to $80$ pc scales.

\item We fit a circular rotation model to constrain the kinematic parameters on $100$ pc scales. The rotation curve is well represented by a \citet{Brandt1960ApJ...131..293B} model with $v_{\mathrm{max}} = 110.0\pm 1.5$ \kms at $r_{\mathrm{max}}=12.0\pm 1.3\mbox{ kpc}$. In general, we find good agreement with previously published rotation curves for the inner 8 kpc of the disk.

\item For galactocentric radii $R<7$~kpc, the azimuthally averaged atomic gas profile has a nearly-constant average surface density of $\Sigma_{\mathrm{HI}}=8~M_{\odot}\mbox{ pc}^{-2}$, though the observations highlight local variations of 25\% with radius. These variations are seen in both limbs (north and south) of the galaxy.

\item The \hi line profiles are consistently non-Gaussian.  We parameterize this by calculating higher-order moments --- skewness and kurtosis --- for the line profiles.  We find that \hi profiles are asymmetrically skewed towards the systemic velocity, resulting in the northern and southern halves having oppositely-signed skewness.  By examining PV-slices, we find evidence for a lagging rotational component in M33, consistent with \citet{Kam2017AJ....154...41K}.  We find a typical excess in kurtosis of $\sim0.2$, indicating that the typical spectrum has an excess in the line wings relative to a Gaussian.

\item  We stack spectra over the entire galaxy, and in $100$ pc radial bins. By modeling the profiles as Gaussian peaks with enhanced line wings, we find that stacking based on the velocity at the peak intensity (\vpeak) of spectra gives the smallest widths of $\sim7$ \kms.  This is consistent with arising from a combination of cool and warm atomic gas with some turbulent broadening.  Stacking based on the centroid (\vcent) and rotation (\vrot) velocities give large linewidths of $\sim10$ \kms, which provides an estimate of internal motions in the atomic medium.  However, the centroid is biased by the asymmetric line wings, which will tend to broaden the stacked profile.  Line width estimates from the second moment are larger than all of the stacked profile widths and are sensitive to extended line wings.

\item All estimates of the line width show a shallow decrease with radius, dropping $\sim2$ \kms over the inner 8 kpc, though with significant (20\%) fluctuations.  This is atypical for most nearby galaxies, which have steeper slopes at small galactocentric radii.  We find that simple estimates of the volumetric energy dissipation rates from core-collapse supernova and magneto-rotational instability cannot explain the radial trend in the line width.  The rising rotation curve and low shear in M33 leads to MRI energy injection rates that are significantly lower than the estimated range from the observed line widths.  More careful measurements of the scale height of atomic gas in M33 are needed to make these conclusions robust.

\item  The fraction of excess emission in the stacked profile line widths ranges from 9\% to 26\% depending on the choice of line centre.  There is no clear radial trend in the inner 6 kpc, but local variations of $50\%$ are significant.  We split the line wing fraction into asymmetric and symmetric parts based on the stacked profiles in the northern and southern halves. We find the asymmetric part, from the lagging disk, accounts for $1/3$ of the line wing excess and is nearly constant with radius.  The radial variations in the line wings are almost entirely driven by the symmetric part.  Beyond 6 kpc, the symmetric part steadily increases, which we attribute to the start of M33's warped disk.

\item We present a Gaussian Mixture Model to explain stacked profile shapes.  Using only a set of Gaussian components with thermal line widths, we can qualitatively reproduce the stacked profile shape found in many previous studies.  In particular, the highly peaked centre of the profile can only be reproduced by including line widths expected for the CNM.  However, the model line widths are smaller than the observed range, suggesting that a combination of the CNM/intermediate/WNM fractions, turbulence, a lagging rotational disk, and multi-component spectra act to broaden the observed stacked profiles.  This large possible combination of physical processes cannot be extracted from the stacked profile without additional information.

\item We identify discrete high-velocity gas structures on the blue- and red-shifted sides of the disk. These structures have a total mass of $1.3\pm0.5\times10^7$ \solmass in this high-latitude component, about $1\%$ of the total atomic mass in the system.  Some of these structures were previously identified with single-dish studies. Here, we can resolve these structures and their extent across the main disk.  All high-velocity structures appear as a clump surrounded by fainter filamentary structure.  We find two features of particular interest: a long filament in the southern half of the disk with a project length of $8$ kpc, and a cloud $600$ pc in diameter overlapping in velocity with the main disk, indicating a possible interaction point.  These structures highlight the complexities of M33's halo, and possibly a connection with the warped disk component.

\item We find no detections of RRL emission, based on stacking six RRL lines.  We set a $3\sigma$ upper limit of 3.0 mJy in a $60\arcsec$ region towards NGC 604.

\end{enumerate}

In conclusion, the high spectral resolution observations of M33 show the galaxy harbours a kinematically rich atomic medium.  With complete spatial sampling down to 80 pc scales, our analysis shows that several common analysis paths used in extragalactic observations have limitations in describing these spectra.  We highlight that overcoming these limitations requires decomposing the spectra into their individual components to gain a full understanding of the atomic ISM kinematics.  This provides new opportunities for exploring the data using tools from the Milky Way community.

The spectral-line VLA data and derived analysis products are available at \url{https://doi.org/10.11570/18.0003}, including derived moment maps and azimuthally-averaged properties.  Scripts to reproduce the reduction, imaging, and analysis are available at \url{https://github.com/e-koch/VLA_Lband}\footnote{v1.1; \url{https://doi.org/10.5281/zenodo.1300005}}.

\section*{Acknowledgments}

We are grateful for helpful discussions with Doug Johnstone in developing this paper, as well as Adam Ginsburg and Hauyu Baobab Liu for discussions on imaging.  We also thank the anonymous referee for their feedback and comments that improved the paper.  EWK is supported by a Postgraduate Scholarship from the Natural Sciences and Engineering Research Council of Canada (NSERC). EWK and EWR are supported by a Discovery Grant from NSERC (RGPIN-2012-355247; RGPIN-2017-03987).  This research was enabled in part by support provided by WestGrid (\url{www.westgrid.ca}), Compute Canada (\url{www.computecanada.ca}), and CANFAR (\url{www.canfar.net}). The National Radio Astronomy Observatory and the Green Bank Observatory are facilities of the National Science Foundation operated under cooperative agreement by Associated Universities, Inc.

\textbf{Code Bibliography: }
CASA \citep[version 4.4 \& 4.7;][]{casa_mcmullin2007} --- astropy \citep{astropy} --- radio-astro-tools (spectral-cube, radio-beam, uvcombine; \url{radio-astro-tools.github.io}) --- matplotlib \citep{mpl} ---  seaborn \citep{seaborn} --- \diskfit\ \citep{Spekkens2007ApJ...664..204S,Sellwood2015} --- image\_registration (\url{http://image-registration.readthedocs.io})\\

\bibliographystyle{mn2e}
\bibliography{ref}

\appendix

\section{Imaging Approach} % (fold)
\label{app:imaging_approach}

At a spectral resolution of $0.2$ \kms and a spatial grid size of $2560^2$ pixels needed to cover the entire mosaic, imaging and deconvolution requires significant computational time and power. The size of the resulting cube is $1178\times2560\times2560$ pixels, giving a size of $\sim29$ GB.  Rather than imaging the cube as a whole, we split the data into individual velocity channels, image and deconvolve each channel separately, then recombine the imaged channels into a final data cube. When working on a cluster, this approach allows channels to be simultaneously imaged, providing a significant speed-up in the time required to image the entire cube. Furthermore, since the size of the channel measurement set (MS) is much smaller than the original MS, slow I/O operations are relatively minimized. Using CASA 4.4, we found that imaging a single channel from the complete MS was $\sim 5\times$ slower than when imaging from a split channel MS.

This approach to imaging large data-cubes has a significant bottleneck during the splitting stage: the time required to split all 1178 channels from the complete MS was nearly two weeks on the Jasper cluster\footnote{\url{https://www.westgrid.ca/resources_services}}, which uses a lustre-based file system. In hindsight, we note that this is not an optimized choice of operation; a binary-split algorithm that progressively splits an MS would likely achieve a significant speed-up in the operation, particularly since this can be naturally parallelized. Furthermore, the storage requirements are more than double the size of the original MS. The new CASA task {\sc tclean} allows for the measurement set to be opened in a read-only mode, largely mitigating the need to split off individual channels. This approach will be used in the future.

\subsection{Image Combination} % (fold)
\label{app:image_combination}

We explore the effects of combining the VLA data with Arecibo \citep{Putman2009ApJ...703.1486P} and Green Bank Telescope \citep[GBT;][]{Lockman2012AJ....144...52L} observations to provide short-spacing information. Both are well suited to be combined with our VLA data as they have similar spectral resolution and at least a factor of two spatial overlap in the $uv$-plane. We note that older GBT data were combined with the archival VLA observations presented in \citet{Braun2012ApJ...749...87B}, but these data have a spectral resolution of $1.42$ \kms and are not well-suited for combination with the $0.2$ \kms resolution VLA data presented here.

The Arecibo data from \citet{Putman2009ApJ...703.1486P} has a spatial resolution of $3.\arcmin4$ and a spectral resolution $0.4$ \kms. This provides a significant spatial overlap with the VLA data, but also requires up-sampling in the spectral domain, thereby increasing the channel-to-channel correlation.  We first spatially-register the data using the cross-correlation method in {\sc image\_registration}\footnote{\url{http://image-registration.readthedocs.io}} and find that no correction is needed. We use the {\sc feather\_simple}\footnote{This matches CASA's feather implementation.} task from the {\sc uvcombine} package to combine the data. This method of combining the data assumes the single-dish beam is well-approximated by an isotropic Gaussian kernel. However, the Arecibo data contains significant side-lobe structure, which leads to enhanced negative bowling in the feathered image on scales of $\sim1$ \arcmin. Further, the power-spectrum of the Arecibo data, clipped to match the region covered by the VLA mosaic, shows a drop in power on the scale of the entire image, indicating a large-scale ripple. This ripple is far less significant in the larger region around M33 used by \citet{Putman2009ApJ...703.1486P}. For these reasons, we did not pursue further image combination with the Arecibo data.

The GBT data from \citet{Lockman2012AJ....144...52L} has a nominal spatial resolution of $9.\arcmin1$ and a spectral resolution $0.16$ \kms. Unlike the Arecibo data, no spectral up-sampling is required to match the VLA data.  These data were taken in four $2\degree \times 2\degree$ regions centred on M33; aspects of the GBT gridding are presented in \S\ref{sub:gbt}. The effective resolution of the data is $9.8\arcmin$ (see below), which gives sufficient $uv$-overlap for combining with the VLA data. The GBT beam has minimal side-lobe structure after calibration and the combined images with the VLA data does not show the enhanced bowling we encountered with the Arecibo data.  Before feathering, we test if the data are spatially-registered and find that the GBT data require a $3\arcsec$ shift in declination. Visually, this shift appears to provide a better combination with the data compared to when no shift is applied. However, we note that $3\arcsec$ is within the pointing error for the GBT data.

\subsubsection{Combination Tests} % (fold)
\label{appsub:combination_tests}

Following Chapter 3 of \citet{Stanimirovic1999PhDT}, we run two tests on the $uv$-amplitudes where the spatial coverage of the VLA and GBT data overlap to 1) ensure a bias is not added from using an incorrect single-dish beam model, and 2) derive a relative calibration factor to obtain a consistent flux-density scale. We define the $uv$ overlap region as all points between the 9.\arcmin8 beam size of the GBT data and 16.5\arcmin -- for the shortest baseline of 44 m in the VLA data. This gives $\sim200$ overlap points per channel at the grid size of the VLA data ($3\arcsec \times 3\arcsec$). To get the GBT amplitudes used below, we deconvolve its Fourier transform by dividing by the Fourier transform of the GBT beam. The GBT amplitudes in the overlap region are then multiplied by the ratio between the beam areas to account for the difference in resolution \citep{Stanimirovic1999PhDT}.

\begin{figure}
\includegraphics[width=0.5\textwidth]{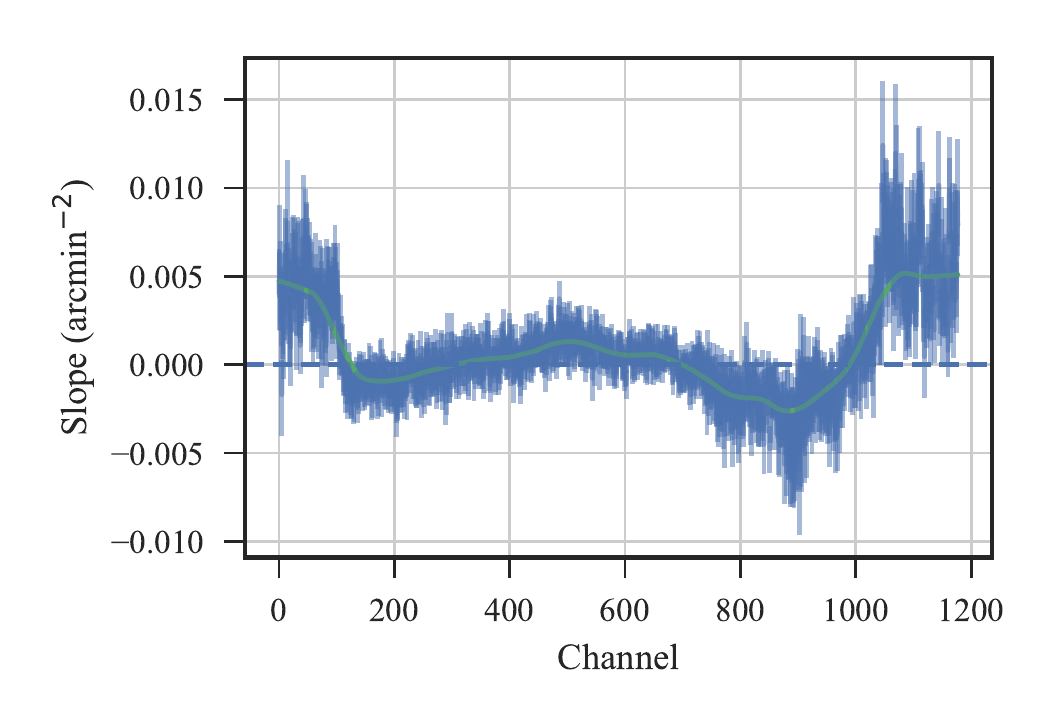}
\caption{\label{fig:feather_slopes} Derived slopes of the $uv$-distance versus the ratio between the VLA and GBT amplitudes in the overlap region. A slope consistent with zero suggests the beam model is correct; an incorrect beam model will incorrectly bias the amplitudes, giving a non-zero slope.}
\end{figure}

In \S3.2.1 of \citet{Stanimirovic1999PhDT}, an approximate relation between the scale factor and the $uv$-distance $k$ is given:
\begin{equation}
  \label{eq:scale_factor_slope}
  f_{\rm cal} = \left[ 1 + \frac{\Delta\theta (2\theta_0 + \Delta\theta)}{4 {\rm ln}2}k^2 \right],
\end{equation}
where $\theta_0$ is the true FWHM single-dish beam size and $\Delta\theta$ is the deviation from the true beam size and assumed to be small. This predicts a linear relation between $f_{\rm cal}$ and $k^2$ when an incorrect beam size is assumed. The correct beam size should have a slope of zero. We perform this fit for each velocity channel in the data using a robust Theil-Sen estimator to find the slope. A robust fitting method is required since the combined effects of the noise and different emission structure in the channels will naturally produce significant outliers. The Theil-Sen method computes the slope from the median of all slopes between each pair of data points, as is defined in the {\sc scipy} implementation\footnote{\url{https://docs.scipy.org/doc/scipy/reference/generated/scipy.stats.theilslopes.html\#scipy.stats.theilslopes}}. The slopes for each channel are shown in Figure \ref{fig:feather_slopes} using a GBT beam size of $9.\arcmin8$. This beam size maximizes the number of channels with slopes consistent with zero. The variations in the slope with different channels -- notably the first 100 and last 200 channels -- are driven by the lack of emission structure and correspondingly have larger uncertainties.  The channels dominated by signal in the $uv$-overlap range are from 200 to 800, and since this region has slopes largely consistent with zero, we adopt a GBT beam size of $9.\arcmin8$ for feathering.

With the correct GBT beam size established, we now address the scaling factor between the VLA and GBT amplitudes. We follow a similar procedure to the one described in \S3.2 in \citet{Stanimirovic1999PhDT}. The emission structure in both the VLA and GBT smoothly tapers near the map edges and no additional edge tapering is required. We first fit a line between the GBT and VLA amplitudes, similar to the method used by {\sc immerge} in {\sc miriad}. The issue with the linear fitting approach is consistently dealing with outliers. We adopt the Theil-Sen method rather than the L1-minimization used in {\sc immerge} since the former tends to be insensitive to outliers in both the $x$ and $y$ directions, while the latter is only robust against outliers in the $y$ direction \citep{wilcox_stats}. We fit a relation to the amplitudes in groups of five velocity channels to increase the number of points. Despite the robust nature of the Theil-Sen fit, we did not find consistent results across the channel groups. The issue is the severity of the outliers and their fraction relative to the entire data set: if there are too many extreme outliers, the Theil-Sen fit still has some sensitivity to the outliers. Increasing the number of channels simultaneously fit did not show improvement, and fitting the entire set of amplitudes is prohibitive for the Theil-Sen method since it computes the slope for every pair of data points.

\begin{figure}
\includegraphics[width=0.5\textwidth]{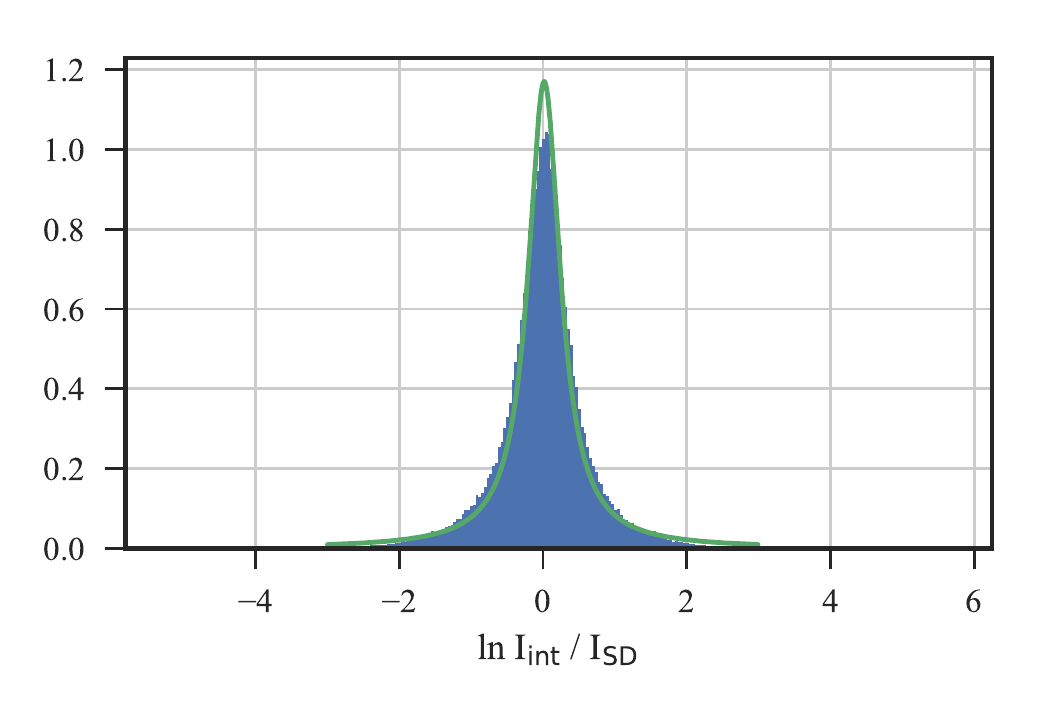}
\caption{\label{fig:feather_ratios} The log of the VLA-to-GBT amplitude ratios in the $uv$-overlap region across all channels. The histogram of the ratios is normalized to one and the green line is the best-fit Cauchy distribution. The peak of the distribution is at $1.02\pm0.06$, consistent with a scale factor of one.}
\end{figure}

Using a distribution-based method, we are able to provide more reliable constraints on the scaling factor. Motivated by \S3.2 in \citet{Stanimirovic1999PhDT}, we first examine the distributions of the GBT amplitudes, the VLA amplitudes, and their ratio across all channels. The GBT and VLA amplitudes reasonably follow a log-normal distribution\footnote{Note that this does not imply a log-normal distribution for the true signal. The log-normal shape results from the mixing between signal, noise, and other telescope effects.}. Since the ratio of two normal random variables follows a Cauchy (or Lorentzian) distribution, the log of the amplitude ratios can be fit to this form. The scaling factor is then the location of the distribution's peak. Figure \ref{fig:feather_ratios} shows the distribution of the amplitudes across all channels (blue) and the best-fit Cauchy distribution (green). This approach has the significant advantage that outliers are included in the expected model (the distribution tails) and do not require special treatment. The Cauchy distribution is also preferable to the Rayleigh distribution used in \citet{Stanimirovic1999PhDT}, which gives more weight to the outliers in the right-tail.  Adopting a maximum-likelihood approach, we find a scaling factor of $1.02\pm0.06$ using the ratios from all channels. As this is consistent with 1, {\it we do not apply a scaling factor to the GBT data before feathering}. We also perform the fitting on groups of five channels and find scaling factors consistent with one between channels 200 to 800, where most of the emission is contained.

% subsection combination_tests (end)

% subsection image_combination (end)

% section imaging_approach (end)

\section{Choice of Velocity Surface for the Rotational Model} % (fold)
\label{app:which_velocity_is_the_correct_rotational_velocity_}

Galactic rotation curves, with a few exceptions, are typically derived from velocity surfaces. A key question explored in several publications --- notably \citet{deBlok2008AJ....136.2648D} --- is: which velocity surface optimally traces galactic rotation? Commonly used methods include the centroid (\vcent), the velocity at the peak intensity (\vpeak), and Gauss-Hermite fitting.  We create velocity surfaces of the former two methods for the VLA-only and combined \hi cubes and fit a rotation curve using {\sc diskfit} with the same parameter settings.  We did not fit a rotation model for the Gauss-Hermite velocity surface for two reasons. First, using the VLA-only cube, the difference between the Gauss-Hermite centre and \vpeak was much smaller than the difference to the \vcent surface and will yield similar results.  The second reason is that the Gauss-Hermite fitting did not perform well with the enhanced line wings for the combined data.

We find that the rotation models using the VLA-only \vpeak and \vcent falls within the uncertainty of the combined VLA \& GBT \vcent surface, whose rotation curve is presented in \S\ref{sub:rotation_curve}.  However, the \vcent surface from the combined data has a shallower rotation curve.  The reason for this discrepancy, and the agreement of the other rotation models, is due to the bias from the line wings.  Without the component from the GBT, the VLA-only \vcent is not as biased by line wings and largely matches the \vpeak.

% section which_velocity_is_the_correct_rotational_velocity_ (end)

\section{Rotation Velocities} % (fold)
\label{app:rotation_velocities}

We provide a table of the calculated rotational velocities from \diskfit\, presented in \S\ref{sub:rotation_curve}.

\begin{table*}
\caption{Circular rotation velocities derived by \diskfit \,(\S\ref{sub:rotation_curve}). The rotation model is fit to the peak velocity (\vpeak) surface of the VLA+GBT data. \label{tab:rotvels}}
\begin{tabular}{rr}
Radius (\arcsec) & Circ. Velocity (\kms) \\\hline \\
9 & $1.58 \pm 6.24$ \\
27 & $11.88 \pm 5.26$ \\
45 & $22.71 \pm 5.25$ \\
63 & $30.53 \pm 5.12$ \\
81 & $29.95 \pm 4.47$ \\
99 & $34.20 \pm 4.03$ \\
117 & $36.76 \pm 3.99$ \\
135 & $44.61 \pm 3.79$ \\
153 & $47.80 \pm 3.49$ \\
171 & $51.67 \pm 3.08$ \\
189 & $52.73 \pm 2.96$ \\
207 & $53.24 \pm 3.06$ \\
225 & $55.30 \pm 3.22$ \\
243 & $57.58 \pm 3.30$ \\
261 & $58.70 \pm 3.42$ \\
279 & $59.47 \pm 3.50$ \\
297 & $60.30 \pm 3.59$ \\
315 & $64.32 \pm 3.78$ \\
333 & $ 67.66 \pm 3.75$ \\
351 & $ 69.07 \pm 4.03$ \\
369 & $ 72.05 \pm 4.49$ \\
387 & $ 75.89 \pm 4.82$ \\
405 & $ 74.09 \pm 4.91$ \\
423 & $ 73.44 \pm 3.29$ \\
441 & $ 77.06 \pm 2.76$ \\
459 & $ 77.06 \pm 2.80$ \\
477 & $ 77.82 \pm 2.60$ \\
495 & $ 79.01 \pm 2.48$ \\
513 & $ 80.96 \pm 2.59$ \\
531 & $ 80.34 \pm 2.50$ \\
549 & $ 80.35 \pm 2.65$ \\
567 & $ 82.90 \pm 2.66$ \\
585 & $ 86.23 \pm 2.66$ \\
603 & $ 86.92 \pm 2.59$ \\
621 & $ 86.13 \pm 2.63$ \\
639 & $ 87.43 \pm 2.57$ \\
657 & $ 87.89 \pm 2.63$ \\
675 & $ 87.82 \pm 2.35$ \\
693 & $ 91.08 \pm 2.44$ \\
711 & $ 90.37 \pm 2.25$ \\
729 & $ 88.22 \pm 2.09$ \\
747 & $ 90.67 \pm 2.36$ \\
765 & $ 92.32 \pm 2.48$ \\
783 & $ 93.31 \pm 2.26$ \\
801 & $ 94.15 \pm 2.15$ \\
819 & $ 93.40 \pm 2.37$ \\
837 & $ 94.02 \pm 2.44$ \\
855 & $ 94.91 \pm 2.50$ \\
873 & $ 95.57 \pm 2.48$ \\
891 & $ 95.15 \pm 2.22$ \\
\end{tabular}
\begin{tabular}{rr}
Radius (\arcsec) &  Circ. Velocity (\kms) \\\hline
909 & $ 93.19 \pm 2.46$ \\
927 & $ 94.52 \pm 2.63$ \\
945 & $ 95.00 \pm 2.77$ \\
963 & $ 95.72 \pm 2.90$ \\
981 & $ 96.89 \pm 2.81$ \\
999 & $ 98.58 \pm 2.73$ \\
1017 & $ 98.09 \pm 2.61$ \\
1035 & $ 99.87 \pm 2.53$ \\
1053 & $ 99.10 \pm 2.41$ \\
1071 & $ 99.01 \pm 2.40$ \\
1089 & $ 97.58 \pm 2.21$ \\
1107 & $ 98.60 \pm 2.26$ \\
1125 & $ 99.61 \pm 2.22$ \\
1143 & $ 99.61 \pm 2.49$ \\
1161 & $102.02 \pm 2.40$ \\
1179 & $101.84 \pm 2.52$ \\
1197 & $102.76 \pm 2.38$ \\
1215 & $102.66 \pm 2.62$ \\
1233 & $102.57 \pm 2.68$ \\
1251 & $104.08 \pm 2.58$ \\
1269 & $103.24 \pm 2.39$ \\
1287 & $103.17 \pm 2.42$ \\
1305 & $103.33 \pm 2.50$ \\
1323 & $103.41 \pm 2.64$ \\
1341 & $103.91 \pm 2.56$ \\
1359 & $102.91 \pm 2.63$ \\
1377 & $105.50 \pm 2.47$ \\
1395 & $104.05 \pm 2.47$ \\
1413 & $106.03 \pm 2.47$ \\
1431 & $104.89 \pm 2.57$ \\
1449 & $106.68 \pm 2.48$ \\
1467 & $105.53 \pm 2.36$ \\
1485 & $105.80 \pm 2.33$ \\
1503 & $104.03 \pm 2.23$ \\
1521 & $105.58 \pm 2.23$ \\
1539 & $105.69 \pm 2.26$ \\
1557 & $106.32 \pm 2.17$ \\
1575 & $105.97 \pm 2.32$ \\
1593 & $105.58 \pm 2.20$ \\
1611 & $106.18 \pm 2.26$ \\
1629 & $106.08 \pm 2.28$ \\
1647 & $106.24 \pm 2.36$ \\
1665 & $106.39 \pm 2.37$ \\
1683 & $106.47 \pm 2.43$ \\
1701 & $107.19 \pm 2.44$ \\
1719 & $105.88 \pm 2.58$ \\
1737 & $106.58 \pm 2.52$ \\
1755 & $106.09 \pm 2.56$ \\
1773 & $106.30 \pm 2.35$ \\
1791 & $107.23 \pm 2.61$ \\
1809 & $104.01 \pm 2.82$ \\
1827 & $104.50 \pm 2.59$ \\\hline \\
\end{tabular}
\end{table*}

% section rotation_velocity (end)

\section{Modeling super-profiles} % (fold)
\label{app:modelling_super_profiles}

\begin{figure*}
\mbox{
\subfigure{
\includegraphics[width=0.5\textwidth]{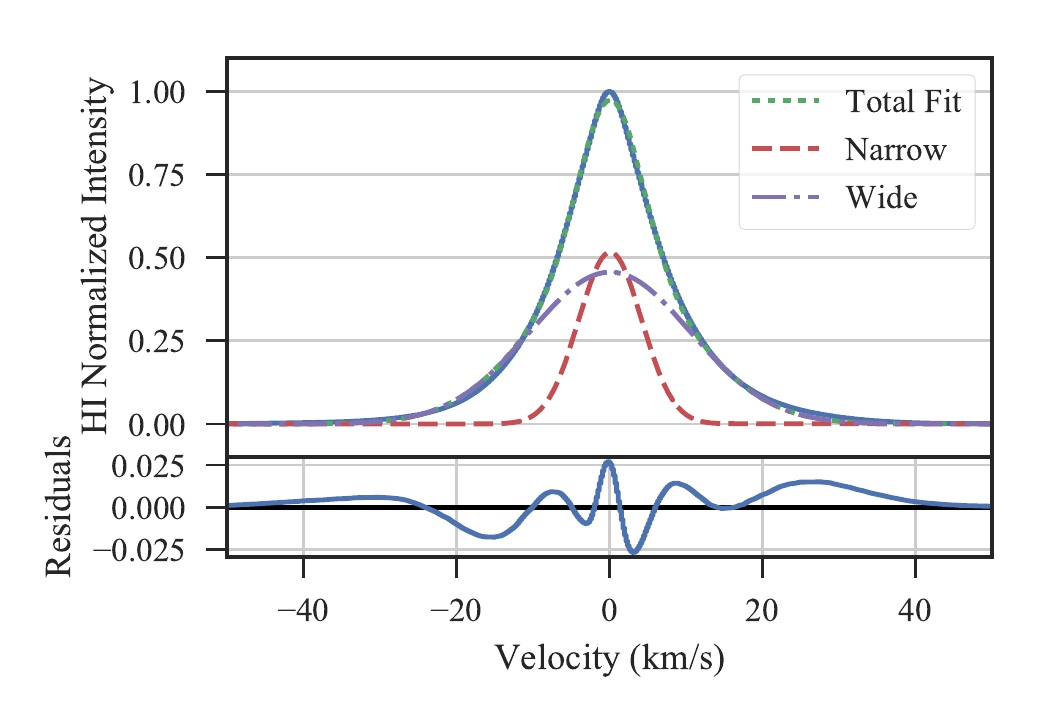}
}\quad
\subfigure{
\includegraphics[width=0.5\textwidth]{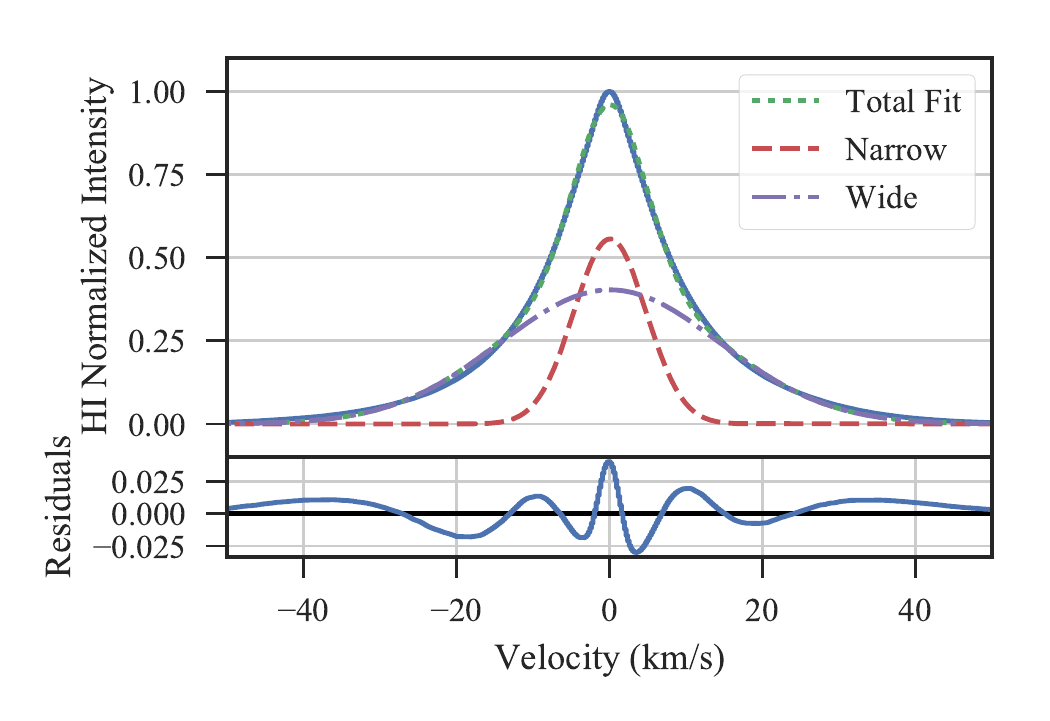}
}
}
\caption{\label{fig:peaksub_2gauss} Two Gaussian fits to the \vpeak stacked profiles shown in Figure \ref{fig:peakvel_profiles}. The left panel is a stack of the VLA-only data and the right uses the combined data.  The model residuals are shown in the bottom panels.}
\end{figure*}

We demonstrate an alternative modeling approach for the super-profiles: fitting two stacked Gaussian profiles. Figure \ref{fig:peaksub_2gauss} shows the \vpeak stacked profiles presented in \S\ref{sub:stacking_spectra} with a fitted two-Gaussian model and the model residuals. The model parameters are given below. Note that we do not list uncertainties due to the large covariance between parameters, which we discuss below.

These two stacked spectra drove our decision to use the HWHM scaling from \citet{Stilp2013ApJ...765..136S}, since the central peak is non-Gaussian and the resulting two-Gaussian fit does not provide a better description for the data.  Of particular concern is that the model components cannot account for the central peak. The sum of the amplitudes in both fits is $\sim0.95$. The narrow and wide amplitudes are 0.51 and 0.55 for the narrow component, and 0.45 and 0.4 for the wide component, for the VLA and combined profiles, respectively.  The difference in the components is about the same as the missing peak intensity, making comparisons of their ratios uncertain.

We find widths of 4.1 and 10.5 \kms for the VLA-only profile, and widths of 4.7 and 14.4 \kms for the combined profile.  The HWHM widths of 5.9 and 6.6 \kms for these two profiles are $\sim50\%$ larger than the narrow component. This is consistent with comparing the common galaxies examined with these two methods in \citet{Stilp2013ApJ...765..136S} and \citet{Ianjasm2012AJ....144...96I}.

Since Gaussian components do not form an orthogonal set, there are large covariances between the parameters.  \citet{Stilp2013ApJ...765..136S} explore this issue in depth and demonstrate that it is this covariance that leads to the overall good fits presented in \citet{Ianjasm2012AJ....144...96I}.  \citet{Stilp2013ApJ...765..136S} show that the individual components from the two-Gaussian fit do not account for the overall profile shape.  They demonstrate this by scaling all profiles to a common width, based on the different models, and find that the shapes from the broad and narrow components have significantly more scatter compared to the HWHM estimate.

While the two-Gaussian fits provide an adequate representation of the profiles, there are model-dependencies that impact the connection to the underlying physical parameters, to the extent that the super-profiles can give. Thus we use the HWHM modeling throughout this work.

% section super_profiles (end)

\section{Issues with the Second Moment for Estimating the Line Width in Local Group Galaxies}
\label{app:second_moment}

\begin{figure}
\includegraphics[width=0.5\textwidth]{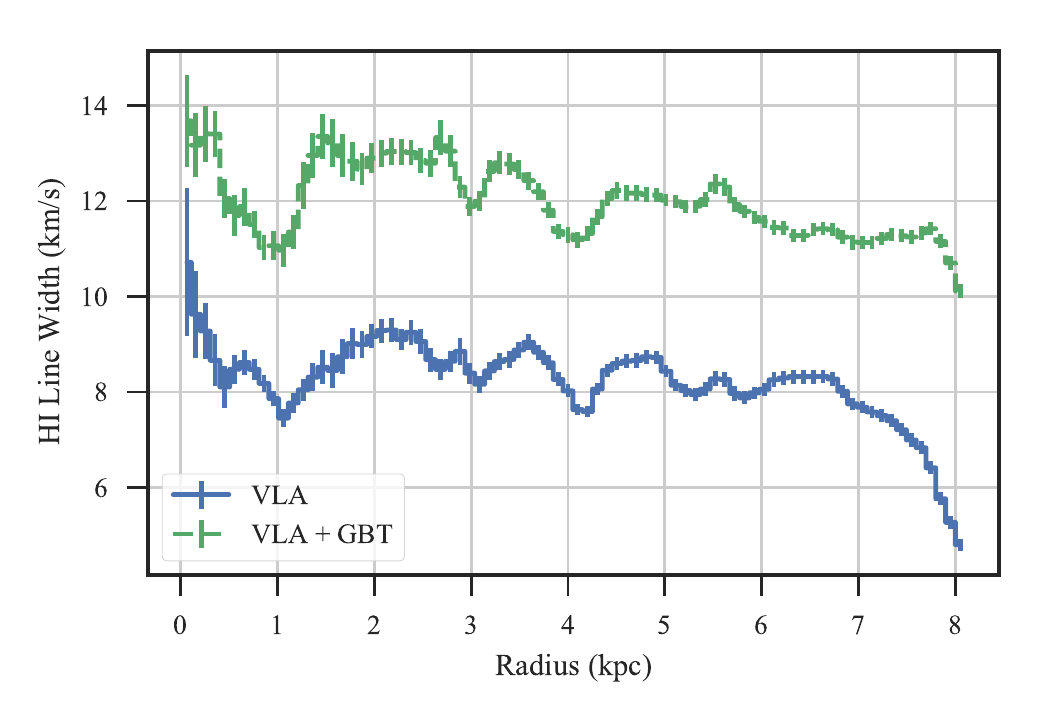}
\caption{\label{fig:veldisp_mom2_comparison} The azimuthally-averaged line width estimated from the second moment in 100 pc bins. The combined VLA+GBT data (green dot dashed) are 30\% larger than the VLA-only values (blue solid).  From the stacking analysis (Figure \ref{fig:peakvel_profiles}, however, we find that adding the GBT component tends to only affect the shape of the line wings.  The large increase from the inclusion of the GBT data highlights that estimates from the second moment are very sensitive to the line wing structure.}
\end{figure}

In Figure \ref{fig:veldisp_mom2_comparison}, we show that there is a stark difference between the averaged second moment line widths with and without the total power component added -- the values increase by $\sim30\%$.  This highlights the impact of extended line wings when using the second moment to estimate the line width, making it difficult to directly connect with the underlying physical conditions.  The large discrepancy seen here is a product of two factors. The first is the limit of small angular scales in the VLA data for a relatively nearby system.  Figure \ref{fig:total_rot_profiles} shows that the combined VLA and GBT data recover significantly more of the large-scale disk structure in the southern half than the VLA data only.  This issue is less important for more distant systems \citep[e.g., ][]{Walter2008AJ....136.2563W}.  The second factor is the extensive extra-planar component in the warped disk of M33.  As we suggest from the skewness maps (\S\ref{sub:exploring_spectral_complexities}), the warped disk component influences the line shape near the edges of our map, increasing the line widths estimated from the second moment.  The extent of this factor depends on the galaxy's environment. However, these average line widths are consistent with the range found in other \hi studies of nearby galaxies \citep{Tamburro2009AJ....137.4424T,Mogotsi2016AJ....151...15M}, showing that this effect is not strong enough to make M33 an outlier, or that line widths from the second moment in other galaxies are similarly affected.

\label{lastpage}
\end{document}